\documentclass[aps,pre,reprint,twocolumn,nofootinbib]{revtex4-1}

\usepackage{subfigure}
\usepackage{amsmath}    
\usepackage{amssymb}
\usepackage{graphicx}   
\usepackage{verbatim}   
\usepackage{color}      
\usepackage{subfigure}  
\usepackage{hyperref}   

\usepackage{array}
\newcommand{\PreserveBackslash}[1]{\let\temp=\\#1\let\\=\temp}
\newcolumntype{C}[1]{>{\PreserveBackslash\centering}p{#1}}
\newcolumntype{R}[1]{>{\PreserveBackslash\raggedleft}p{#1}}
\newcolumntype{L}[1]{>{\PreserveBackslash\raggedright}p{#1}}

\DeclareMathOperator*{\argmax}{arg\,max}

\begin{document}

\title{A random matrix perspective of cultural structure: groups or redundancies?}

\author{Alexandru-Ionu\c{t} B\u{a}beanu}
\affiliation{Lorentz Institute for Theoretical Physics, Leiden University, The Netherlands}
\affiliation{Rotterdam School of Management, Erasmus University, The Netherlands}
\date{\today}

\begin{abstract}
Recent studies have highlighted interesting structural properties of empirical cultural states:
collections of cultural traits sequences of real individuals.
Matrices of similarity between individuals may be constructed from these states, 
allowing for further structural insights to be gained using concepts from random matrix theory,
approach first exploited in this study.
For generating random matrices that are appropriate as a structureless reference, we propose a null model that enforces, on average, the empirical occurrence frequency of each possible trait.
With respect to this null model, the empirical matrices show deviating eigenvalues, 
which may be signatures of subtle cultural groups.
However, they can conceivably also be artifacts of arbitrary redundancies between cultural variables. 
We first study this possibility in a highly simplified setting,
using a toy model that enforces a certain level of redundancy in a minimally-biased way, 
in parallel with another toy model that enforces group structure.
By analyzing and comparing cultural states generated with these toy models, we show that a deviating eigenvalue can indeed be a redundancy signature, 
which can be distinguished from a grouping signature by evaluating the uniformity of the entries of the respective eigenvector,
as well as the uniformity-based compatibility with the null model. 
For empirical data, the eigenvector uniformities of all deviating eigenvalues are shown to be compatible with the null model, 
apparently suggesting that we are not dealing with genuine group structure. 
However, we demonstrate that some deviating eigenvalues might actually be due to authentic groups that are internally non-uniform. 
A generic procedure for distinguishing such groups from redundancy artifacts requires further research.

\end{abstract}

\maketitle

\section{Introduction}\label{Intr}

Understanding the complex behavior of social systems greatly benefits from constructively combining the increasing amount of empirical data with a variety of quantitative, theoretical approaches, 
often originating in the natural sciences \cite{Urry, Lazer}.
Although much of this interdisciplinary research focuses on the network and connectivity aspects of social systems \cite{Kadushin}, 
efforts are also being made for understanding a complementary aspect: the formation and dynamics of opinions, preferences, attitudes and beliefs, more generically referred to as ``cultural traits'' \cite{Castellano}.
In particular, recent studies have placed a stronger emphasis on using empirical data about the cultural traits of real individuals \cite{Valori, Stivala, Babeanu_1, Babeanu_2, Babeanu_3}.
Such data is typically recorded within a short period of time from a random sample of people in a population, via a social survey with a large number of questions/items, 
so that a vector (or sequence) of cultural traits can be constructed for every individual, where each trait is an answer to one of the questions.
The collection of all cultural vectors constructed from one empirical source is called an empirical ``cultural state'', or an empirical ``set of cultural vectors'', 
since it can be used to empirically specify the initial conditions of an Axelrod-type model of cultural dynamics~\cite{Axelrod}. 
Using previously developed tools~\cite{Valori, Stivala} that relied on models of cultural and opinion dynamics,
Ref.~\cite{Babeanu_1} showed that empirical cultural states are characterized by properties that are highly robust across different datasets.
These properties have been further explored~\cite{Babeanu_2, Babeanu_3} but not entirely understood. 
The generic empirical structure appears to be largely captured by the matrix of cultural similarities between individuals, which are computed for all pairs of cultural vectors.
This opens the possibility to further investigate the empirical structure by means of a random matrix approach.

Random matrix theory~\cite{Mehta, Edelman} has been successfully used for a variety of applications, 
among which the analysis of financial systems~\cite{Potters} is an important highlight.
The framework deals with various properties of random matrices, under certain distributional assumptions.
The associated statistical ensembles of matrices are used to compute the expected values (or even the probability distributions) of interesting, matrix dependent quantities.
These theoretical expectations can be compared to empirical counterparts evaluated on matrices that encode information about the real world systems that are being studied. 
Statistically significant deviations of the empirical quantities are then interpreted as interesting, non-trivial structural properties of the respective systems.  
The focus is on the eigenvalue spectrum of the empirical matrix, which very often consists of correlations between time series associated, for instance, 
to the dynamics of stocks \cite{MacMahon} or the activity of neurons \cite{Almog}. 
In such cases, the appropriate assumptions of randomness are captured by the the Marchenko-Pastur~\cite{Marchenko} law, which gives a limiting distribution for the spectrum.
The empirical eigenmodes whose eigenvalues are significantly larger than what is expected based on the Marchenko-Pastur law
are interpreted as joint dynamical patterns in terms of which the non-trivial behavior of the system can be understood,
while the other are interpreted as noise components. 
Recently, Ref.~\cite{Patil} extended this approach to similarity matrices constructed from categorical data,
where each entry of the matrix is a similarity between two time series of discrete symbols.
For instance, for one of the datasets in Ref.~\cite{Patil}, 
each sequence of symbols corresponds to an electoral constituency of India, with different symbols associated to different winning parties and successive time steps associated to successive elections. 
	  
This study extends the approach to spectra of empirical matrices of cultural similarities, 
constructed from data previously used in Refs.~\cite{Valori, Stivala, Babeanu_1, Babeanu_2, Babeanu_3}.
Instead of relying on analytic formulas for estimating and filtering the noise, numerical methods are extensively used. 
This allows for a detailed investigation of three null models (Sec.~\ref{Empir_1}), 
among which the uniformly random generation is the simplest and conceptually closest to analytic approaches behind the Marchenko-Pastur distribution~\cite{Patil}.
As a second null model, we make use of trait shuffling, which is known to be important for understanding empirical cultural states, independently from spectral decomposition and random matrix notions~\cite{Valori, Stivala, Babeanu_1, Babeanu_2, Babeanu_3},
since it reproduces exactly the empirical trait occurrence frequencies.
We propose an additional null model which also reproduces these empirical trait frequencies on average, 
while also incorporating some mathematically desirable properties of the uniform random generation. 
We name this procedure "restricted random generation".
These null models are compared in terms of how well they reproduce the upper boundary of the noisy spectral region (``the bulk''), as well as the position of the highest eigenvalue -- 
a strong outlier which can be understood as a ``global mode'', which for similarity matrices is guaranteed to be present even under the uniformly random scenario~\cite{Patil}.
As shown in Sec.~\ref{Empir_1}, the restricted random generation turns out to be more appropriate and is thus selected for further analysis.  
Based on restricted randomness, we numerically evaluate the probability distribution of the upper noise boundary,
showing that there are several empirical eigenvalues significantly above this boundary.
These ``deviating eigenmodes'' capture the structure of empirical data, since they are incompatible with the null hypothesis behind restricted randomness.
Hence, this manuscript will often refer to them as ``structural modes''.

It is tempting to interpret the structural modes as manifestations of cultural groups, in a manner similar to time series analysis \cite{MacMahon},
suggesting that individuals fall under several classes, categories or clusters that are inherent to the system.
This is particularly intriguing, given that Ref.~\cite{Babeanu_2} provides indirect evidence for cultural structure being governed by a small number of cultural prototypes supposedly induced by universal ``rationalities''. 
However, it is important to keep in mind that the empirical data also shows pairwise correlations between cultural variables (or ``features''), that are at least partly due to arbitrary, 
dataset-dependent redundancies (semantic overlaps) between the corresponding survey items, as previously pointed out~\cite{Valori, Babeanu_1, Stivala}.
Since these correlations are not retained by restricted randomness nor by shuffling, it is possible that deviating eigenmodes are a direct consequence of arbitrary redundancies. 

The question of whether deviating eigenmodes are signatures of authentic groups of individuals or just artifacts of arbitrary redundancies between variables motivates the rest of the study.  
First, we explicitly show that it is mathematically meaningful to differentiate between a ``redundancies scenario'' and the ``groups scenarios'',
which is not obvious, first because both scenarios may induce pairwise correlations between features, second because these features are discrete variables with small numbers of possible values. 
This is done in Sec.~\ref{Theor_1} by studying, in a highly simplified, abstract setting, consisting of only binary features, two probabilistic models for generating (sets of) cultural vectors.
The first model, labeled ``FCI'' (Sec.~\ref{FCI}), explicitly enforces a certain pairwise coupling between all features, in a manner that gives rise to a certain level of correlation, 
without introducing unintended assumptions or biases in the underlying probability distribution.
This is ensured by a maximum-entropy derivation~\cite{Jaynes}, which leads to a statistical ensemble that is mathematically equivalent to the canonical ensemble of the Ising model on a fully-connected lattice~\cite{Colonna-Romano}, 
where each feature corresponds to one lattice site and each cultural vector corresponds to a spin configuration.
The second model, labeled ``S2G'' (Sec.~\ref{S2G}), explicitly enforces a binary group structure, whose strength can be analytically tuned to match the first model in terms of the level of feature-feature correlation. 

For any given level of feature-feature correlation, the FCI and S2G models are used (Sec.~\ref{Theor_2}) for generating cultural states and associated similarity matrices.
The latter generally exhibit one structural mode, associated to the subleading eigenvalue, whose strength increases with the correlation level.
However, for any given correlation level, the average value and significance of the subleading eigenvalue turns out to be exactly the same for the FCI and S2G models,
so the subleading eigenvalue does not discriminate between the two scenarios.
This explicitly demonstrates that the presence of deviating eigenvalues alone does not automatically imply the presence of authentic group structure. 
We show that the essential difference between FCI and S2G is captured by the entries of the eigenvector associated to the subleading eigenvalue. 
In particular, the uniformity of these entries, quantified by the ``eigenvector entropy'' (Sec.~\ref{Theor_2}), shows a clearly different behavior as a function of correlation for the two models:
S2G exhibits systematically higher uniformities than FCI.
Moreover, the dependence of the second-highest eigenvector entropy on the correlation level reproduces well the symmetry-breaking phase transitions that characterize the two models.
In each case, the eigenvector entropy suddenly jumps from a regime of compatibility to a regime of incompatibility with the null model
exactly when the probability distribution associated to the respective model becomes bi-modal. 
The critical correlation associated to this transition is almost one order of magnitude smaller for S2G than for FCI. 
This justifies the use of eigenvector entropy as an indicator of group structure in empirical data, as a complement to eigenvalue information. 

Along these lines, Sec.~\ref{Empir_2} presents an enhanced analysis of empirical data, showing how the eigenmodes are distributed in terms of eigenvalue and eigenvector entropy, 
in comparison to expectations based on restricted randomness. 
Interestingly, all deviating eigenvalues are actually associated to non-deviating eigenvector entropies,
suggesting that all structural modes are artifacts of arbitrary redundancies between cultural variables. 
However, such a conclusion is conditional on how representative the S2G model is for authentic group structure that empirical data might capture. 

As explained in Sec.~\ref{Non_Unif_Group}, groups generated with S2G actually have internal uniformity built in, to an extent that conflicts with basic intuition/expectations about real world groups. 
This is illustrated by means of a contrast to another, third toy model, inherited from previous work~\cite{Babeanu_2}, namely the ``Mixed prototype generation'' (MPG).
MPG can generate internally non-uniform groups that are more compatible with intuition, while also enjoying some support from social science theories and an some degree of empirical validation.
Sec.~\ref{Mixed_Protos} explicitly shows that structural modes of MPG states, like those of empirical state, systematically fail to exhibit eigenvector uniformity:
the quantity departs from null model expectation only when the correlation level attains a very high value, relative to S2G and event to FCI.
Following a complementary, more data-driven approach, Sec.~\ref{Feat_Elim} shows that two of the four structural modes identified for one empirical dataset, although exhibit uniformities that are compatible with the null model, are unlikely to be due to survey-specific redundancies between variables.
This takes advantage of redundancies being very obvious in the feature-feature correlation matrix for that particular dataset. 
Sec.~\ref{Non_Unif_Group} thus suggests that eigenvector uniformity provides a criterion that is too strong for the purpose of validating structural modes as authentic signatures of real world cultural groups, so the results in Sec.~\ref{Empir_2} should no be used for drawing strong conclusions.

For now, the existence of cultural groups cannot be rejected nor confirmed. 
Still, this study makes important steps in that direction, by developing important methodology and 
providing insights which will likely find use beyond the analysis of cultural states.
As discussed in Sec.~\ref{Disc}, 
more research is needed for developing a reliable way of distinguishing structural modes induced by internally non-uniform groups from those induced by redundancies. 
The study is concluded in Sec.~\ref{Conc}.
 
\section{Eigenvalue distributions for empirical data and null models}\label{Empir_1}

In this section, the eigenvalue spectra of empirical matrices of cultural similarities are evaluated. 
At the same time, three null models are evaluated and compared.
Each null model is used to numerically generate similarity matrices, by randomly sampling from the associated statistical ensemble, which enforces, to a certain extent, the empirical information that is expected to not be of interest 
-- this is information that, on a priori grounds, clearly has more to do with arbitrary survey design choices than with any authentic cultural structure. 
One of these models, namely the ``restricted random'' model, which is first introduced here, is chosen as a good benchmark with respect to which interesting structure is to be measured, as explained below.
Before presenting the actual results, some mathematical clarifications are given with respect to the computation of similarity matrices, the spectral decomposition procedure and the definitions of the null models. 

A cultural similarity matrix is a square, $N \times N$ matrix obtained from $N$ cultural vectors,
which are all defined with respect the same set of $F$ cultural features (variables or dimensions). 
Each feature can take one of $q^k$ possible, discrete values, called ``cultural traits'', where $k$ labels the features, according to some order that is arbitrary, but consistent across all vectors.
Moreover, each feature can be either nominal, marked as $f_{\text{nom}}^k = 1$, or ordinal, marked as $f_{\text{nom}}^k = 0$, which affects how its similarity contribution is defined. 
Each entry $s_{ij}$ of the similarity matrix is then computed according to:
\begin{widetext}
\begin{equation}
  \label{CultSim}
  s_{ij} = \frac{1}{F}\sum_{k=1}^{F}\left[ f_{\text{nom}}^k \delta(x_i^k, x_j^k) + (1-f_{\text{nom}}^k) \left( 1 - \frac{|x_i^k - x_j^k|}{q^k-1} \right) \right],
\end{equation}
\end{widetext}
encoding the similarity between vectors $i$ and $j$, where $\delta$ stands for the Kronecker delta function and $x_i^k$ and $x_j^k$ denote the traits recorded with respect feature $k$ in vectors $i$ and $j$ respectively
 -- for the ordinal case, it is important that $x_i^k$ and $x_j^k$ take discrete, rational values between $1$ and $q^k$, while for the nominal case they only need to take symbolic values from any (feature-specific) alphabet. 
Note that the similarity measure in Eq.~\eqref{CultSim} is an arithmetic average of the similarity contributions of the $F$ cultural features,
in agreement with Refs.~\cite{Valori, Stivala, Babeanu_1, Babeanu_2, Babeanu_3} -- although in these studies most concepts are presented in terms of cultural distances $d_{ij}$, 
these have a trivial relationship to cultural similarities: $d_{ij} = 1 - s_{ij}$.
For an empirical matrix, each vector $i$ corresponds to one individual in the real world, 
each feature $k$ to one question or item in the questionnaire used to collect the data,
so that the realized trait $x_i^k$, which lies at the intersection between vector $i$ and feature $k$, corresponds to the answer/rating given by individual $i$ to question/item $k$.
For a matrix generated based on a null model, the $N$ vectors are generated according to the specified random procedure, while retaining (at least) the empirical data format, 
namely the type $f_{\text{nom}}^k$ and range $q^k$ of each feature $k$.
Note that, in contrast to the empirical symbolic sequences used in Ref.~\cite{Patil}, 
cultural vectors have no axis of time, so everything is equivalent up to a reordering of the cultural features, as long as this is done consistently for all cultural vectors. 
This is irrelevant for any of the mathematical operations involved by the analysis here, but it is relevant for the interpretation: 
cultural vectors capture no time-evolution, and should be interpreted as instantaneous, multidimensional opinion profiles, rather than as dynamical, one-dimensional dynamical profiles.	

\begin{figure*}
\centering
	\subfigure[\label{Spec:empir}]{\includegraphics[width=8.5cm]{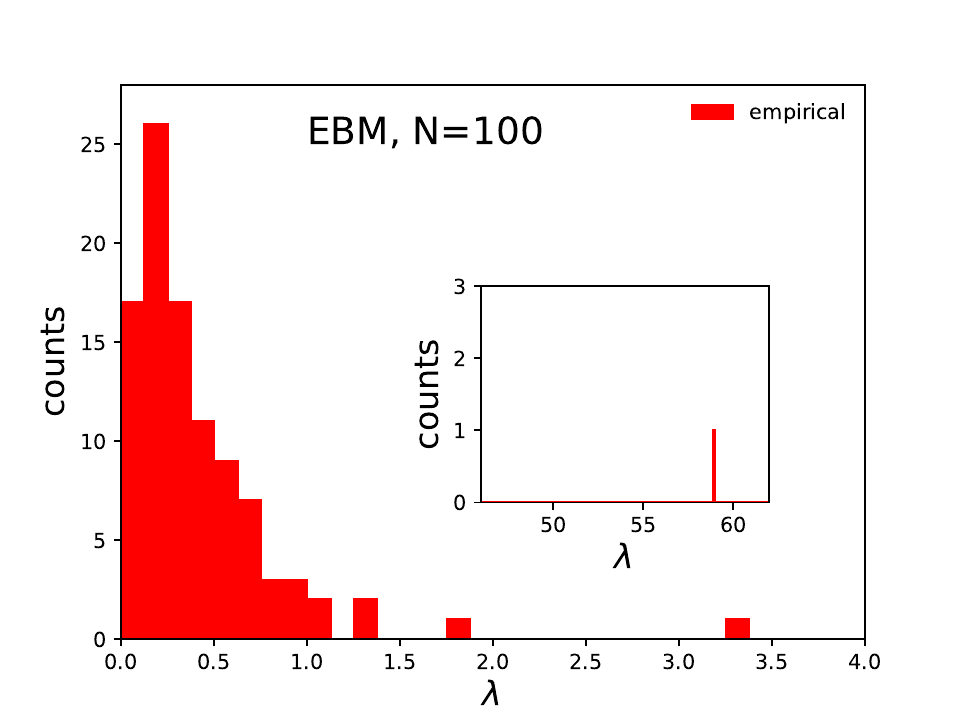}}  \hspace{0.6cm}
	\subfigure[\label{Spec:u-rand}]{\includegraphics[width=8.5cm]{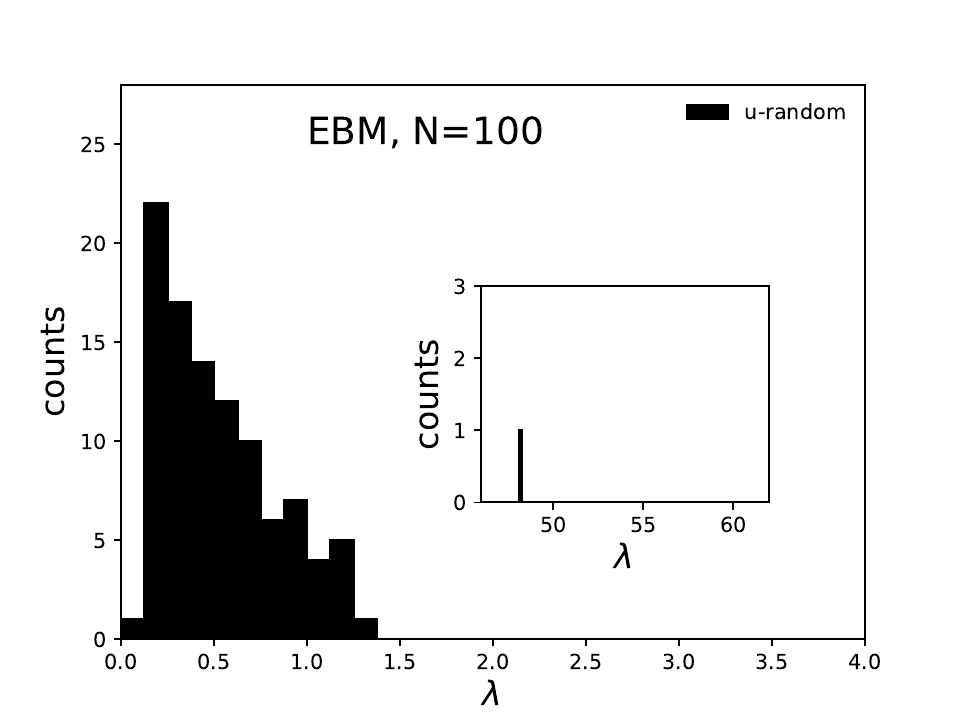}} \\ \vspace{-0.3cm}
	\subfigure[\label{Spec:shuf}]{\includegraphics[width=8.5cm]{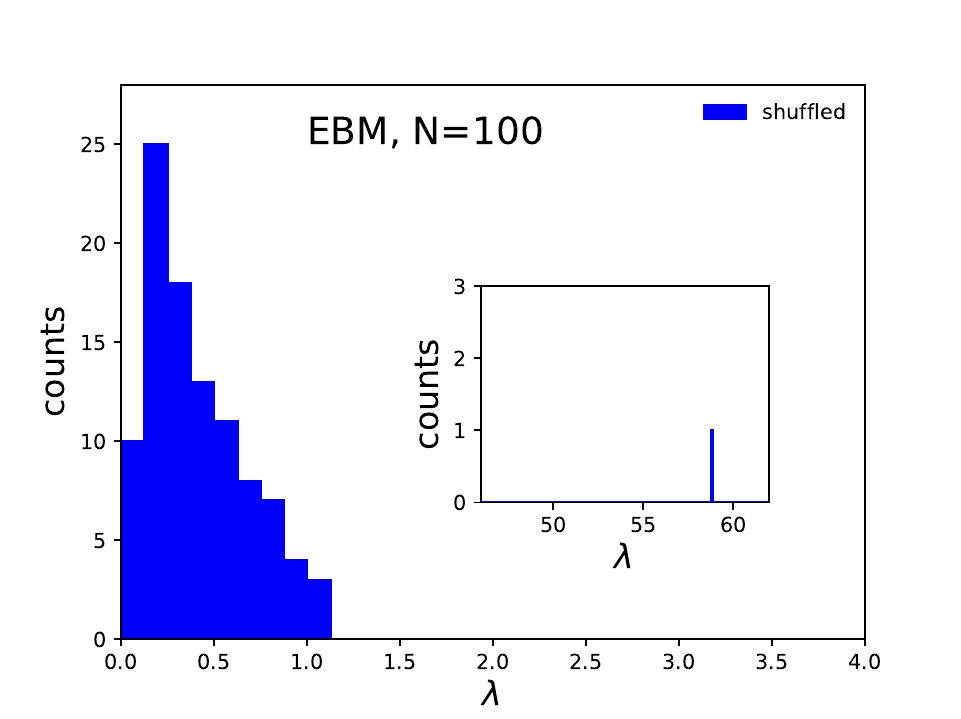}} \hspace{0.6cm}
	\subfigure[\label{Spec:r-rand}]{\includegraphics[width=8.5cm]{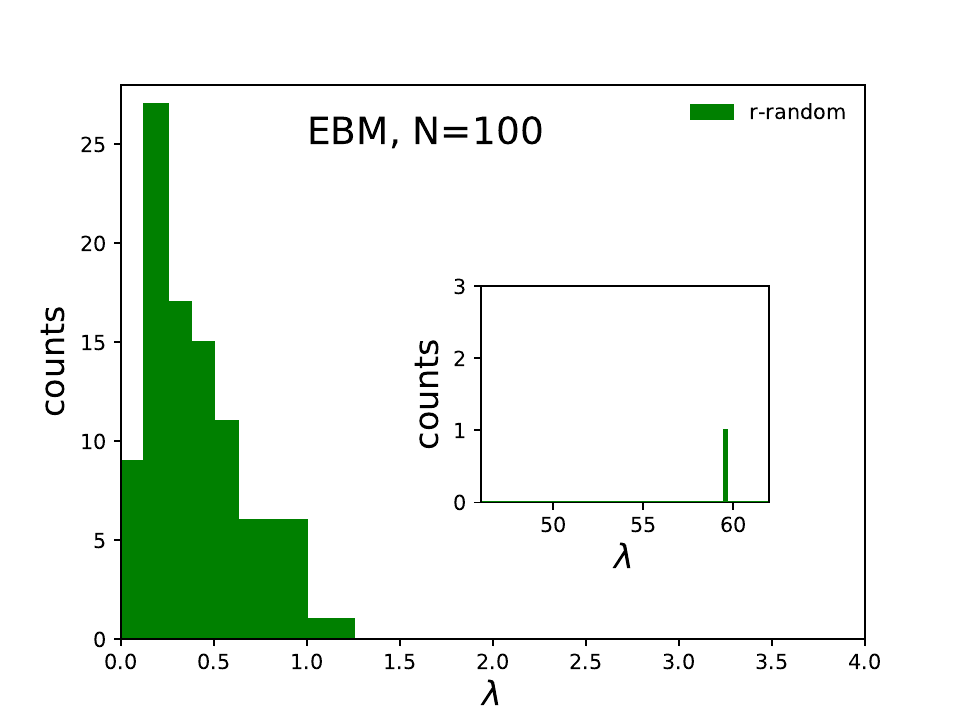}}  \vspace{-0.3cm}
	\caption{Eigenvalue spectra of cultural similarity matrices.
	The first panel correspond to an empirical cultural state~\subref{Spec:empir} with $N=100$ vectors constructed from Eurobarometer (EBM) data, 
	while the other three correspond to associated cultural states generated with the uniform random~\subref{Spec:u-rand}, 
	the shuffled~\subref{Spec:shuf} and the restricted random~\subref{Spec:r-rand} null models, 
	using partial information from the empirical cultural state and the same $N=100$. 
	For each panel, the inset shows the leading eigenvalue of the respective spectrum. 
	For comparison purposes, the axis ranges and bin widths are the same across the four panels, for the main plots as well as for the insets. 
}
\label{Spec}
\end{figure*} 

From Eq.~\eqref{CultSim} it follows that such a similarity matrix is real and symmetric, from which it follows, according to the spectral theorem,
that it has $N$ real eigenvalues with $N$ associated orthonormal eigenvectors with real entries. 
This implies that the matrix can be decomposed in the following way: 
\begin{equation}
	\label{MatDec}
	s_{ij} = \sum_{l=1}^N \lambda_l v_l^i v_l^j,
\end{equation}
where ``$\lambda_l$'' and ``$v_l$'' are used to denote the $l$th highest eigenvalue and, respectively, the eigenvector associated to it, 
while $v_l^i$ is the $i$th entry of eigenvector $v_l$. 
Throughout this study, special attention is payed to $\lambda_1$ and $\lambda_2$, the highest and second highest eigenvalues of various similarity matrices,
also denoted as the ``leading'' and ``subleading'' eigenvalues respectively.
In parallel, ``$\lambda$'' is used to denote any generic eigenvalue.
More notation will be introduced below, as needed.  

It is remarkable that all eigenvalues of any similarity matrix computed according to Eq.~\eqref{CultSim} are bound to the positive real axis: $\lambda_l \ge 0, \forall l$. 
Thus, in this study, no eigenvalue-related figure axis needs to be concerned with values smaller than $0.0$.  
This positive semidefiniteness property is rigorously shown to hold in Sec.~\ref{App_Proof_PSD}.
The property may be relevant beyond the analysis of cultural states,
since it holds for any similarity matrix belonging to what one may call the ``Hamming-Manhattan'' class:
a matrix whose elements $s_{ij}$ satisfy $s_{ij} = 1-d_{ij}$, where every $d_{ij}$ is a combined, Hamming-Manhattan distance,
with nominal and ordinal features corresponding to the Hamming and Manhattan contributions respectively.

All similarity matrices used in this study are based on sets of $N = 100$ cultural vectors, regardless of whether they are empirical, 
generated with one of the three null models introduced below or with one of the toy models introduced in Sec.~\ref{Theor_1}.
Moreover, all these matrices satisfy $s_{ii} = 1.0, \forall i$, as a consequence of Eq.~\eqref{CultSim}, meaning that the trace is always: $\sum_i {s_{ii}} = N = 100$. 
Since diagonalization preserves the trace, the eigenvalues are also bound to add up to $\sum_l \lambda_l = N = 100$ for every matrix.  

Fig.~\ref{Spec:empir} shows the eigenvalue spectrum of an empirical similarity matrix computed based on $N=100$ cultural vectors extracted from Eurobarometer (EBM) data, 
which records attitudes and opinions of European Union citizens on various topics concerning technology, the environment and certain policy issues~\cite{EBM}.
The data is formatted according to the procedure described in Ref.~\cite{Babeanu_1}, which makes $F=144$ cultural features available.
The vertical axis gives the number of eigenvalues occurring in each bin along the horizontal axis. 
The inset focuses on the higher $\lambda$ region of the horizontal axis where the leading eigenvalue $\lambda_1$ is located.
The high value of $\lambda_1$ is expected based on purely mathematical grounds~\cite{Patil}, due to the overall positivity of any such similarity matrix. 
In most cases, all entries of the eigenvector associated to $\lambda_1$ have the same sign and very similar absolute values, meaning that, according to Eq.~\eqref{MatDec}, 
the $\lambda_1 v_1^i v_1^j$ captures a large, highly uniform, positive component of the matrix entries $s_{ij}$.
The $\lambda_1$ eigenmode thus accounts for the overall tendency towards similarity of the entire system, which is partly due to how similarity is defined and partly (see below) due to feature-level non-uniformities. 
For this reason, the $\lambda_1$ mode will also be referred as the ``global mode'', term which originates from time-series analysis~\cite{MacMahon} based on correlation matrices, 
for which a global mode may or may not be present, depending on the system.
Using exactly the same format as Fig.~\ref{Spec:empir}, each of the other three panels of Fig.~\ref{Spec} shows the spectrum of a similarity matrix generated from each of the three null models:
``uniform randomness'', ``shuffling'' and ``restricted randomness''.  

First, Fig.~\ref{Spec:u-rand} shows the spectrum of a similarity matrix generated via uniform randomness (abbreviated as ``u-random'').
Specifically, for every vector, each trait is chosen independently at random from the traits available at the level of the respective feature, 
with equal probability attached each possible trait. 
This means that uniform randomness retains minimal information from the empirical cultural state used for Fig.~\ref{Spec:empir}: only the number of features, the type and the number of traits of each feature. 
Note that the leading eigenvalue of this matrix is comparable to that of the empirical matrix.  
Ref.~\cite{Patil} showed that the analytic, limiting distribution given by the Marchenko-Pastur formula has a shape that is qualitatively similar to the bulk of the u-random spectrum.
Quantitatively however, the analytic and numerical distributions become truly similar only if an important parameter controlling the former is left free and fit to the numerical results, 
instead of being directly set to $F/N$, which can be done when dealing with matrices of correlations between $N$ time series with $F$ numerical entries each.
Moreover, the Marchenko-Pastur formula completely fails to describe the leading eigenvalue.

Second, Fig.~\ref{Spec:shuf} shows the eigenvalue spectrum of a similarity matrix generated via shuffling.
Specifically, with respect to every feature, the traits realized in the empirical state are randomly permuted among the vectors, such that every permutation is equally likely. 
This is done independently for every feature, so that all types of correlations between features are destroyed.
The procedure preserves exactly the number of times each trait is empirically realized, 
in addition to preserving the data format of the empirical state in Fig.~\ref{Spec:empir}.
Note that, by construction, the assignment of traits to vectors is not entirely independent across vectors,
implying that the number of vectors $N$ resulting from shuffling has to be exactly the same as the number of empirical vectors used.

\begin{figure*}
\centering
	\subfigure[\label{DistVal:2}]{\includegraphics[width=8.5cm]{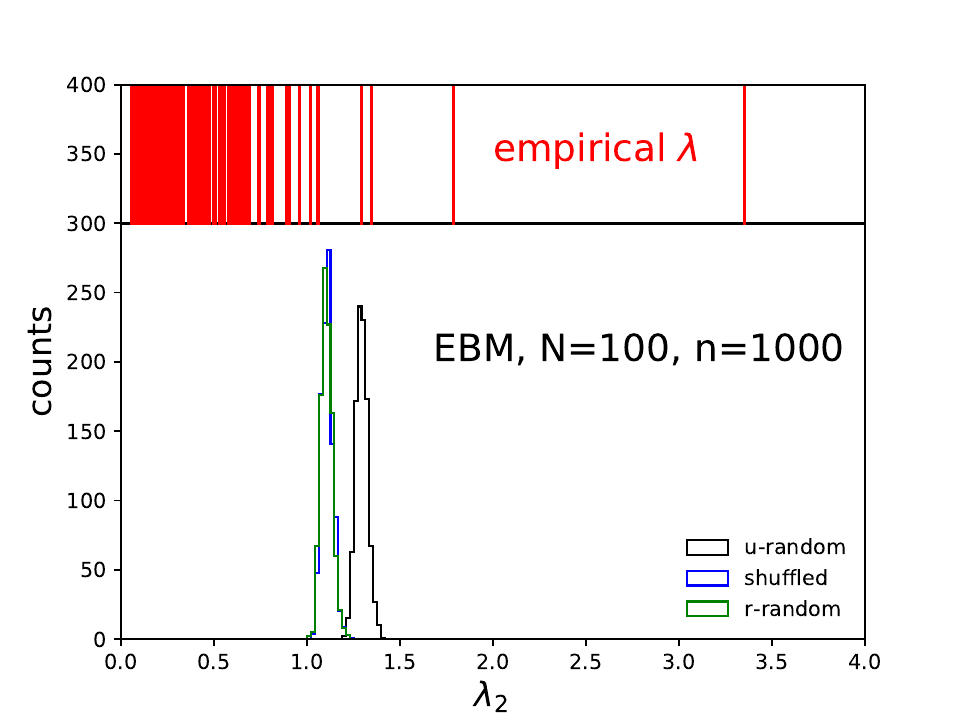}} \hspace{0.6cm}
	\subfigure[\label{DistVal:1}]{\includegraphics[width=8.5cm]{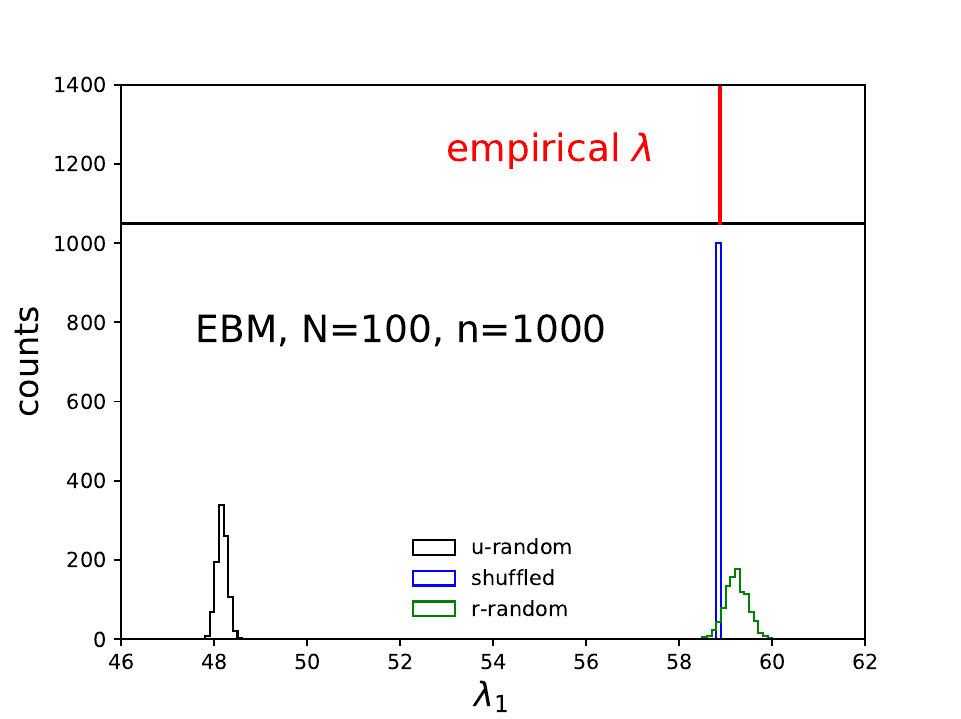}} 
	\caption{Leading and subleading eigenvalue distributions for random matrices. 
	The figure shows the subleading eigenvalue $\lambda_2$ distribution~\subref{DistVal:2} and the leading eigenvalue $\lambda_1$ distribution~\subref{DistVal:1}, 
	for the three null models (legends), implementing uniform randomness (black), shuffling (blue) and restricted randomness (green), 
	in comparison to the empirical eigenvalues, whose positions are marked by the vertical (red) lines in the upper bands. 
	For each distribution, $n = 1000$ similarity matrices are numerically generated from the respective null model. 
	Everything is based on the same set of $N=100$ vectors constructed from Eurobarometer (EBM) data used in Fig.~\ref{Spec}.
}
\label{DistVal}
\end{figure*} 

\begin{figure}
\centering
	\includegraphics[width=8.5cm]{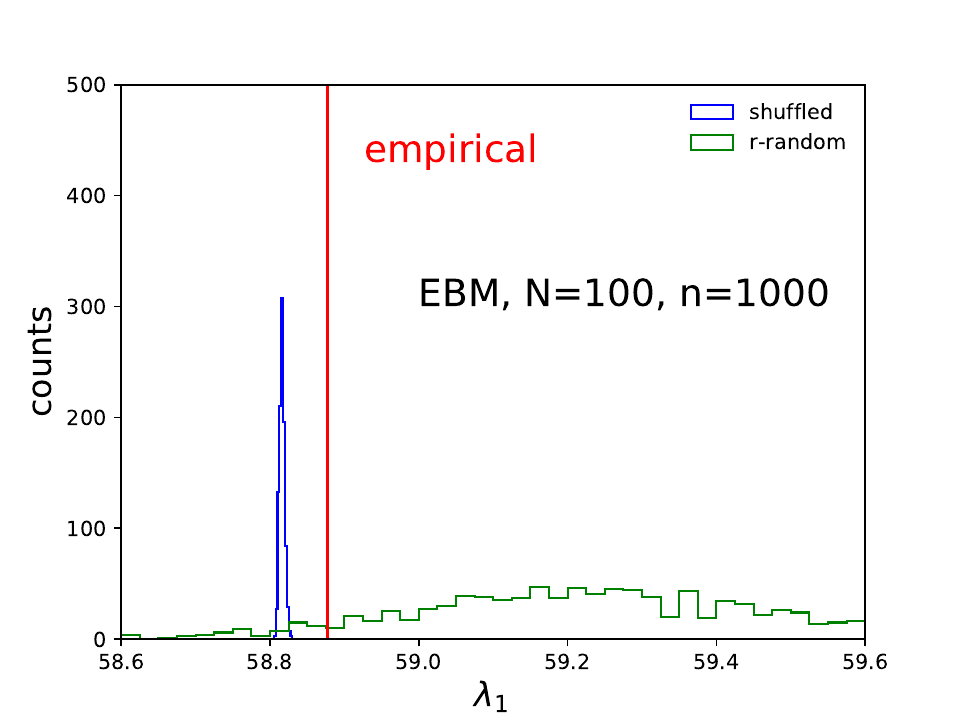}
	\caption{Detailed comparison in terms of the leading eigenvalue. 
	The figure shows in detail the distributions of the leading eigenvalue $\lambda_1$ for the shuffled (blue) and restricted random (green) null models, 
	in comparison to the empirical value (vertical red line).
	For each distribution, $n = 1000$ similarity matrices are numerically generated from the respective null model. 
	Everything is based on the same set of $N=100$ vectors constructed from Eurobarometer (EBM) data used in Fig.~\ref{Spec}.
	For visual purposes, the bin size of the restricted random histogram is ten times smaller than for the shuffled histogram.
}
\label{DistVal_1_Det}
\end{figure} 

\begin{figure*}
\centering
	\subfigure[\label{DistVal_2:GSS}]{\includegraphics[width=8.5cm]{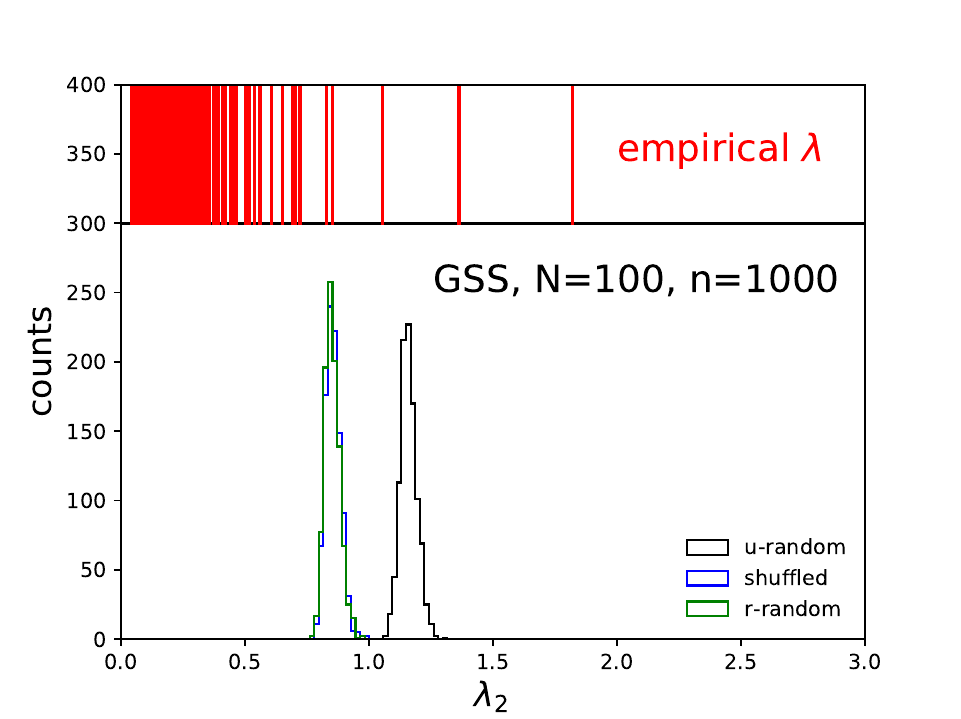}} \hspace{0.6cm}
	\subfigure[\label{DistVal_2:JS}]{\includegraphics[width=8.5cm]{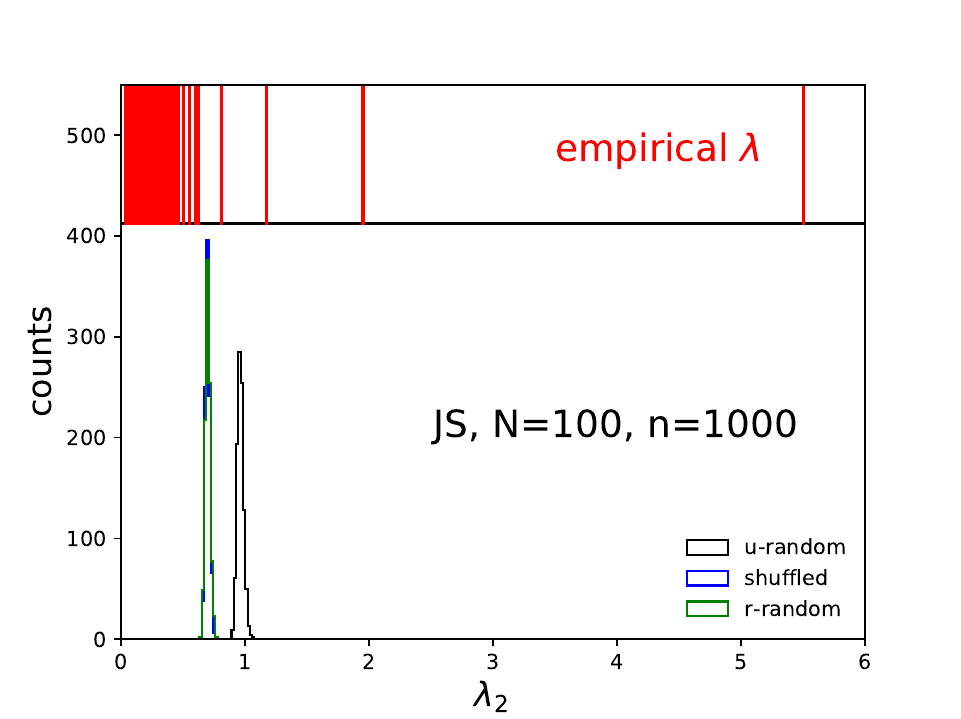}}
	\caption{Empirical structure in other datasets. 
	The figure shows the subleading eigenvalue $\lambda_2$ distribution for the three null models (legends), implementing uniform randomness (black), shuffling (blue) and restricted randomness (green),
	in comparison to the empirical eigenvalues, whose positions are marked by the vertical (red) lines in the upper band, 
	based on General Social Survey data~\subref{DistVal_2:GSS} and for Jester data~\subref{DistVal_2:JS}.
	In each case, $N = 100$ cultural vectors are constructed from the respective dataset. 
	For each null model, $n = 1000$ random matrices of $N=100$ vectors are generated for drawing the associated distribution. 	
}
\label{DistVal_2}
\end{figure*} 

Third, Fig.~\ref{Spec:r-rand} shows the spectrum of a similarity matrix generated via restricted randomness (abbreviated as ``r-random''). 
Specifically, for every vector, each trait is chosen independently at random from the traits available at the level of the respective feature, 
with different probabilities attached to the possible traits, these probabilities being directly proportional to the empirical occurrence frequencies of the respective traits.
This means that, like the shuffling procedure, restricted randomness also reproduces the empirical trait frequencies, but on average. 
Moreover, it also retains the independent generation specific to uniform randomness,
which allows for an arbitrary number $N$ of cultural vectors to be generated, regardless of how large this number is for empirical data.
The independent generation should also make the analytic tractability of the model easier. 
Although neither of these two advantages are directly exploited in this study, they suggest that restricted randomness is conceptually more appropriate than either uniform randomness or shuffling, 
as it incorporates the desirable properties of both. 
 
The rough shape of the eigenvalue histogram is quite similar across the four panels of Fig.~\ref{Spec}, 
which means that empirical data contains a large amount of noise, which can be described reasonably well by any of the three null models. 
Interesting discrepancies are present in terms of the leading eigenvalue: 
the empirical value is very similar to the shuffled and r-randomn values, while higher than the u-random value. 
This shows that the overall tendency towards similarity is smaller in the uniformly-random case than in the other three cases.
This is due to shuffling and restricted randomness reproducing the feature-level non-uniformities, which are not reproduced by, leading to an enhanced global mode.

Very important are the empirical outliers in Fig~\ref{Spec:empir}, which encode empirical structure that is independent of feature-level non-uniformities.
The two higher outliers are larger than the bulk boundary as predicted by any of the three null models, while the other two appear compatible with the random bulk predicted by uniform randomness. 
This highlights the importance of choosing the appropriate null model, since this determines the position of the boundary between noise modes and structural modes along the $\lambda$ axis, 
which in turn decides how many empirical eigenmodes are to be regarded as structurally relevant on the higher $\lambda$ side of this boundary. 
It appears that the position of this boundary is somewhat different for the three null models, but this is hard to evaluate only based on Fig.~\ref{Spec},
due to limitations inherent in the binning. 

Fig.~\ref{DistVal:2} overcomes these limitations by showing the subleading eigenvalue distribution for the three null models, 
in parallel with the leading eigenvalue distributions in Fig.~\ref{DistVal:1}, where the colors associated to the three null models are the same as those in Fig.~\ref{Spec}.
For comparison, the empirical eigenvalues are shown by the vertical (red) lines in the upper bands of Fig.~\ref{DistVal}.
Each $\lambda_1$ and $\lambda_2$ distribution is produced numerically by sampling $n=1000$ sets of cultural vectors from the statistical ensemble of the respective null model. 
It appears that shuffling and r-random show almost the same $\lambda_2$ distribution,
while for u-random this is located at higher values.
Since $\lambda_2$ sets the boundary for the random bulk, 
more empirical eigenmodes are to be regarded as structurally relevant with respect to a null model based on shuffling or restricted randomness, rather then on uniform randomness.
Choosing between shuffling and r-random appears appropriate, since they are consistent with empirical data in terms of the leading eigenvalue, as noted before, 
now confirmed in a more statistically reliable way by Fig.~\ref{DistVal:1}. 
Such a choice is compatible with the idea of focusing on the empirical structure that is present independently of feature-level non-uniformities, 
which are expected to strongly depend on how the associated questions and the possible answers are formulated and much less on authentic properties of the real social system from which the data is extracted. 
With respect to either the shuffled or the r-random $\lambda_2$ distribution, all four empirical outliers noted in Fig.~\ref{Spec:empir} appear statistically significant, 
with a departure of at least two standard deviations from the mean.

On the other hand, based on Fig.~\ref{DistVal:1}, the empirical leading eigenvalue also appears statistically compatible with both shuffling and restricted randomness, 
but closer to the mean of the former. 
This, however, deserves a closer inspection, due to the limitations inherent in the binning of Fig.~\ref{DistVal:1}.
Fig.~\ref{DistVal_1_Det} focuses on the shuffled and r-random $\lambda_1$ distributions, 
giving a better impression of how well either null model predicts the empirical leading eigenvalue based on partial information about trait frequencies.
It appears that, due to the sharpness of the shuffled $\lambda_1$ distribution, the empirical value is actually not statistically compatible with it, 
while it is clearly compatible with the r-random distribution. 
For this reason, we choose restricted randomness as the appropriate null model. 
Note that, for visual purposes, the bins are chosen to be much smaller for the shuffled than for the r-random distribution -- both histograms contain $n=1000$ entries, 
one for each random matrix sampled from the respective ensemble. 

Finally, it is worth repeating the analysis on empirical cultural states constructed from two more datasets, 
namely the General Social Survey~\cite{GSS} (GSS) -- Fig.~\ref{DistVal_2:GSS} -- and Jester~\cite{JS} (JS) -- Fig.~\ref{DistVal_2:JS}.
Both datasets are also formatted according to the procedure described in Ref.~\cite{Babeanu_1}, leading to $F=122$ features for GSS and to $F=128$ features for JS.
The two figures follow the format of Fig.~\ref{DistVal:2}, since this emphasizes the empirical outliers and their departure from the subleading eigenvalue distributions of the three null models --
although, at this point, the choice has already been made in favor of restricted randomness, the other two distributions are also shown for consistency. 
Both the GSS and JS eigenvalue spectra show outliers that are significantly larger than what is expected based on the r-random null model:
three such outliers are present for GSS and four for JS. 
The deviating eigenvalues are, on average, larger for JS than for EBM, and higher for EBM than for GSS. 
-- note that the axis ranges of Figs.~\ref{DistVal_2:GSS},~\ref{DistVal_2:JS} and~\ref{DistVal:2} are not the same.

Based on the results above, one can say that the empirical structure captured by matrices of cultural similarity is generally recognizable via eigenvalues 
that are significantly larger than what is expected based on a null hypothesis accounting for empirical trait frequencies: 
they are significantly higher than the subleading eigenvalue and much lower than the leading eigenvalue expected from this null hypothesis. 
For the rest of this study, the eigenpairs (eigenvector-value pairs) associated to these deviating eigenvalues will often be referred to as ``structural modes''.

\section{Two interpretations of structural modes}\label{Theor_1}

This section explores possible ways of interpreting the structural modes of culture described above. 
To begin with, certain aspects of linear algebra are emphasized, in relation to the diagonalization of similarity matrices. 
These provide theoretical some theoretical justification for an interpretation of structural modes as group modes, like in the context of correlation matrices. 
Then, two scenarios are formulated: first, that structural modes are just the effect of redundancies between cultural features, thus only retaining information about how the associated questions/items
are chosen; second, that structural modes are an effect of genuine groups or grouping tendencies among the individuals, thus retaining information about the social system from which the data is extracted. 
This leads to probabilistic formulations of the two scenarios in a very simplistic setting: 
the redundancy scenario is realized as the ``fully-connected Ising'' (FCI) model in Sec.~\ref{FCI}, 
while the groups scenario is realized as the ``symmetric two-groups'' (S2G) model in Sec.~\ref{S2G}.
Finally, in Sec.~\ref{FCIvS2G}, the mathematical properties of the two models are studied in order to check that they behave as expected and to better emphasize their differences.

It is instructive to first consider some elementary, but important mathematical properties of the eigenvalues $\lambda_l$ and the associated eigenvectors $v_l$ satisfying Eq.~\eqref{MatDec}. 
For the sake of clarity, the following explanations make use of the term ``individual'' as a replacement for ``cultural vector'', 
although most of the concepts presented are also valid, at least mathematically, for similarity matrices constructed from randomly generated cultural vectors, based on any probabilistic model.

Since the eigenvectors $v_l$ have only real entries and form an orthonormal basis, one can write any real vector $w$ with $N$ entries as a linear combinations of the eigenvectors:
\begin{equation}
	\label{VecDec}
	w = \sum_{l=1}^N \alpha_l v_l, 
\end{equation}
with real coefficients $\alpha_l$. 
The rest of this argument is restricted to unit vectors $w$, which satisfy $\sum_{i=1}^N w_i^2 = 1$, 
which can be translated as $\sum_{l=1}^N \alpha_l^2 = 1$ in terms of the eigenvectors' coefficients. 
This encompasses all the eigenvectors $w=v_l, \forall l$ as special cases.
Moreover, let us define the following scalar quantity:
\begin{equation}
	\label{Sandwich}
	S = \sum_{i=1}^N \sum_{j=1}^N w_i s_{ij} w_j,
\end{equation}
as the double contraction of the similarity matrix $s$ with the vector $w$.
By means of Eq.~\ref{Sandwich} and Eq.~\ref{VecDec}, for any vector $w$ (including the special cases when this entirely matches one of the eigenvectors $v_l$) 
every entry of $w$ becomes associated to one of the individuals based on which the similarity matrix $s$ is computed.
Thus, $w$ can be seen as a (normalized) linear combination of the $N$ individuals.
$S$ can be then interpreted as the self-similarity of any normalized linear combination $w$, 
since every pairwise similarity $s_{ij}$ is multiplied by the numbers $w_i$ and $w_j$ attached to individuals $i$ and $j$.
For any normalized $w$, one can show that: 
\begin{equation}
	\label{SelfSim}
	S = 1 + 2 \sum_{i=1}^{N-1} \sum_{j=1+1}^{N} w_i s_{ij} w_j,
\end{equation}
which immediately follows from the fact that $s_{ii} = 1, \forall i$, which is a direct consequence of how the similarity is defined in Eq.~\eqref{CultSim}. 
Note that $S=1$ whenever $w$ gives a strength of $1$ to one individual and $0$ to all the other, which supports the interpretation of $S$ as a self similarity.
It is also important to note, from Eq.~\eqref{SelfSim}, that $S$ is larger when $w$ is such that pairs of entries $(i,j)$ with the same sign correspond to higher values of $s_{ij}$ and higher values of $|w_i w_j|$, 
while pairs with opposite signs correspond to lower values of $s_{ij}$ and lower values of $|w_i w_j|$.

The largest self-similarity $S$ is attained when the linear combination $w$, among all unit vectors, 
takes the form of the eigenvector $v_1$ with the largest associated eigenvalue $\lambda_1$, corresponding to $\alpha_{l} = \delta_l^{1}, \forall l$.
This largest self-similarity value is actually equal to the largest eigenvalue: $S = \lambda_1$. 
This is shown by plugging Eq.~\eqref{MatDec} and Eq.~\eqref{VecDec} into \eqref{Sandwich} and using the normalization condition, leading to: 
\begin{equation}
	S = \sum_{l=1}^N \alpha_l^2 \lambda_l.
\end{equation}
More generally, one can see here that each eigenvector $v_l$ with the $l$th highest eigenvalue $\lambda_l$, corresponding to $\alpha_{l'} = \delta_{l'}^{l}, \forall l'$,
is such that it gives the largest possible value of $S = \lambda_l$, while also being normalized and orthogonal to all eigenvectors $v_{l'}$ with $\lambda_{l'} > \lambda_{l}$
When confronting this with the insights provided by Eq.~\eqref{SelfSim}, 
one realizes that any subset of individuals with strong, internal similarities is captured by one of the eigenmodes, 
whose eigenvalue is larger if the overall level of internal similarity is higher.
Moreover, the eigenvector entries of these strongly similar elements will have the same sign and the highest absolute values.

By combining the above with the findings of Sec.~\ref{Empir_1}, a more complete interpretation is obtained for structural modes: 
they are the normalized linear combinations of the individuals, orthogonal to each other and to the global mode, 
with the highest possible self-similarities, of which the lowest is significantly higher than what is expected from restricted randomness.
Each of the structural modes could indicate the presence of a group of highly similar individuals, which is why in the context of time-series analysis they are often called ``group modes''~\cite{MacMahon}. 
Although it is not clear how a linear combination of individuals (or of cultural vectors) should be expressed in terms of cultural traits and features, this is not important for this study and does not affect the above arguments.

An alternative interpretation of structural modes comes from realizing that social surveys are imperfect, in the sense that one cannot guarantee the absence of semantic overlaps (redundancies or similarities) between the variables that are used.
These translate to correlations between cultural features, which have been noticed in previous studies~\cite{Valori, Stivala, Babeanu_1} and which are specific to the design of each dataset. 
It is conceivable that feature redundancies, if strong enough, could induce artifactual structural modes themselves. 
For example, if a large fraction of the associated items or questions are designed such that they are mostly sensitive to the same underlying degree of freedom,  
the similarity between individuals responding to any of these items in a certain way will be high, since these individuals will likely respond to all the other similar items in the same way. 
It appears likely that this behavior would be captured by a structural mode. 
If this is the mechanism behind the structural modes shown in Sec.~\ref{Empir_1}, 
it means that they do not provide information about the inherent organization of real-world culture, 
but just about the design of the ``instrument'' used to ``measure'' culture.
Although redundancies between features would manifest as correlations between those features, 
authentic groups can also induce such correlations, so this aspect cannot be directly exploited for differentiating redundancies from groups.

There are thus fundamental reasons that make it very difficult to understand the extent to which structural modes of culture are due to details of the experimental setting and the extent to which they are due to authentic properties of the underlying system.
This study makes a first step in this direction, by formulating the two scenarios as mathematical, probabilistic models capable of generating (sets of) cultural vectors that are governed either by a coupling between cultural features (Sec.~\ref{FCI}) or by a grouping tendency (Sec.~\ref{S2G}).  
These models are designed to work without any empirical input, in the same, simplest conceivable setting, consisting of $F$ binary features 
-- it does not matter whether these features are regarded as ordinal or nominal, since the two types of similarity contributions are equivalent if there only $q=2$ traits available, as can be seen from Eq.~\eqref{CultSim}. 
For each feature, the two traits are marked as ``$-1$'' and ``$+1$'' -- although the former should be mapped to ``$0$'' when computing similarities between vectors, if features are assumed ordinal.
Each of the two models defines a statistical ensemble (and an associated cultural space distribution, in the language of Refs.~\cite{Babeanu_1, Babeanu_2}),
according to which cultural vectors can be independently drawn, in a random, but non-uniform way.
For both statistical ensembles, each feature-level probability distribution is uniform -- the two traits have an equal probability of $0.5$ attached.
Note that, although both models are probabilistic in nature, neither of them is intended as a null model, 
since neither makes use of information from empirical data nor is it intended for direct, quantitative comparisons to empirical data,
nor to be realistic to any extent.
They are toy-models, used for illustrating conceptual differences and ambiguities between redundancies and groups in the context of cultural states.
Nonetheless, they do provide an arena for studying and developing certain mathematical tools in a highly controlled setting, 
tools that can are later used for studying empirical data. 

\subsection{The first scenario: redundancies}\label{FCI}

This section explains the ``fully-connected Ising'' (FCI) model, in the context of generating (sets of) cultural vectors in a stochastic way.
The purpose of this probabilistic model is to enforce a certain level of redundancy for all pairs of cultural features, controllable via one parameter, but as little as possible in addition.
This can be done by properly choosing the probability distribution $p$ taking as support the set of possible cultural vectors with $F$ binary features, or, in other words,
the set of possible spin configurations $\vec{S}$ with $F$ lattice sites.
Note that the support of this distribution has $2^F$ elements, which is the number of sites/points of the ``cultural space'', according to the formalism in Ref.~\cite{Babeanu_1}.

One needs to choose the maximally-random (thus minimally biased) probability distribution $p$ that entails a certain level of feature-feature correlation.
This is found by maximizing the Shannon entropy (Eq.~\eqref{entropy}) subject to two constraints: 
one enforcing the normalization of the probability distribution (Eq.~\eqref{constr1}), 
the other enforcing the overall level of pairwise coupling between cultural features (Eq.~\eqref{constr2_1}).  
This procedure is a realization of maximum-entropy inference introduced in Ref.~\cite{Jaynes},
and is described in detail in Sec.~\ref{App_Theor_FCI}. 
The resulting probability distribution can be expressed as:
\begin{widetext}
\begin{equation}
	\label{prob_mu_F}
	p(\mu, F, F_{+}) = \frac{1}{Z(\mu)} \frac{F!}{F_{+}!(F-F_{+})!} \exp{\left[\frac{\mu}{2} \left( (2F_{+}-F)^2 - F \right) \right]}.
\end{equation}
\end{widetext}
This gives the (total) probability attached to all cultural vectors with $F_{+}$ out of $F$ traits marked as ``+'' or ``$+1$'',
where $\mu$ is the parameter controlling the overall level of coupling between features.
Moreover, $Z(\mu)$ is a normalization factor, namely the partition function in Eq.~\eqref{partFunc_mu_F}. 
Note that $\sum_{F_{+}=0}^{F} p(\mu, F, F_{+}) = 1.0$, 
since the expression combines the probability of different possible configurations with the same $F_{+}$, which, due to symmetry reasons are equally likely. 
There are $F!/\left(F_{+}!(F-F_{+})!\right)$ such configurations (the physical ``density of states'', where a ``state'' would correspond to a possible cultural vector rather than a what we call here a ``cultural state'') for each $F_{+}$.

The model is mathematically equivalent to the Ising model of magnetism on a fully connected lattice~\cite{Colonna-Romano}, described in the canonical ensemble,
with the parameter $\mu$ replacing the ratio between spin-spin coupling and temperature, which controls for the overall level of alignment between spins.
This parallel does not come as a surprise: for any statistical physics ensemble defined by the averages of certain, externally controlled/measured (physical) quantities, 
the mathematical derivation can be formulated in terms of maximum-entropy inference~\cite{Jaynes}, 
which ultimately provides a statistical, information-theoretic justification of minimum-bias as a replacement for assumptions like ``ergodicity''. 
Due to this parallel, the nomenclature related to spins is sometimes used instead of that related to cultural features. 

Based on Eq.~\eqref{prob_mu_F}, one can derive the expression for the correlation between any two features:
\begin{widetext}
\begin{equation}
	\label{C_mu}
	C(\mu, F) = \frac{1}{Z(\mu)} \sum_{F_{+} = 0}^{F} \frac{(F-2)! \left( (2F_{+}-F)^2 - F \right)}{F_{+}!(F-F_{+})!} \exp{\left[\frac{\mu}{2} \left( (2F_{+}-F)^2 - F \right) \right]},
\end{equation}
\end{widetext}
based on the entire statistical ensemble.
The details of this derivations are also given in Sec.~\ref{App_Theor_FCI}.

In Eq.~\eqref{prob_mu_F} and Eq.~\eqref{C_mu}, the coupling parameter is positive: $\mu \in [0,\infty)$.
Physically, this corresponds to ferromagnetism, meaning that alignment between spins is favored, a tendency which is enhanced with increasing $\mu$.
Using Eq.~\eqref{prob_mu_F}, one can check that, for vanishing coupling $\mu=0$, 
the probability of choosing a configuration with a given $F_{+}$ is directly proportional to the number of such configurations, 
which is specified by the binomial coefficient preceding the exponential. 
As $\mu$ is increased, 
more emphasis is given to configurations with unequal numbers of $-1$ and $+1$ traits,
at the expense of configurations that are more balanced.
Using Eq.~\eqref{C_mu}, one can also check that the correlation $C(\mu, F)$ increases with increasing coupling $\mu$, as expected, 
and that $C(0.0, F) = 0.0$ for any $F$. 
 
\subsection{The second scenario: groups}\label{S2G}

This section explains the ``symmetric two-groups'' (S2G) model, in the context of generating (sets of) cultural vectors in a stochastic way. 
This probabilistic model enforces an organization of cultural vectors in terms of two, equally sized groups, with high similarities within groups and low similarities between groups. 
The model defines a probability distribution $p$ taking as support the same set of possible cultural vectors as in Sec.~\ref{FCI}:
the cultural space defined by $F$ binary features, with $2^F$ configuration.
One of the groups is ``centered'' around the configuration with a $-1$ trait with respect to each feature, 
while the other group is centered around the opposite configuration, having a $+1$ trait with respect to each feature.
The model is designed such that all features contribute equally to the group structure.
As a consequence, this induces a certain level of correlation for all pairs of cultural features.  
The strength of these correlations is controlled by the same free parameter that controls the strength of the group structure.

According to the S2G model, every cultural vector that is generated is first randomly assigned to one of the two groups, with equal probabilities.
These two groups are denoted as the ``$-1$'' group and the ``$+1$'' group. 
Then, at the level of every feature, the trait is randomly and independently chosen among the two possibilities, but with unequal probabilities:
the trait with the same sign as the group is chosen with probability $1-2\nu$,
while the trait with the opposite sign is chosen with probability $2\nu$.
Here, $\nu \in [0,0.25]$ is the free model parameter controlling the strength of the group structure: 
lower $\nu$ values imply stronger group structure and stronger correlations between features, as made more explicit by Eq.~\eqref{C_nu}.
From this procedure, it follows that, at the level of every feature, each generated trait falls under one of the following situations:
\begin{itemize}
	\item with probability $0.5 - \nu$, it is attached to a vector belonging to group $-1$ and has a value of $-1$;
	\item with probability $\nu$, it is attached to a vector belonging to group $-1$ and has a value of $+1$; 
	\item with probability $0.5 - \nu$, it is attached to a vector belonging to group $+1$ and has a value of $+1$; 
	\item with probability $\nu$, it is attached to a vector belonging to group $+1$ and has a value of $-1$. 
\end{itemize}
Note that the probabilities of the four cases add up to $1.0$, that the combined probability of either value is $0.5$ and that the probability of either group is also $0.5$.

For this model, the probability that a generated configuration has $F_{+}$ traits $+1$ is:
\begin{widetext}
\begin{equation}
	\label{prob_nu_F}
	p(\nu, F, F_{+}) = \frac{1}{2} \frac{F!}{F_{+}!(F-F_{+})!} (2\nu)^{F_{+}} (1 - 2\nu)^{F_{+}} \left[ (2\nu)^{F - 2F_{+}} + (1 - 2\nu)^{F - 2F_{+}} \right],
\end{equation}
\end{widetext}
while the correlation between any two features is: 
\begin{equation}
	\label{C_nu}
	C(\nu) = 1 - 8\nu + 16\nu^2.
\end{equation}
The mathematical derivations of Eq.~\eqref{prob_nu_F} and Eq.~\eqref{C_nu} are given in Sec.~\ref{App_Theor_S2G}.
Note that the correlation in Eq.~\eqref{C_nu} behaves as expected, namely: 
$C(0.0) = 1$ (when the two groups are maximally dissimilar the correlation is maximal) and 
$C(0.25) = 0.0$ (when the two groups are indistinguishable the correlation is zero).
Finally, Eq.~\ref{C_nu} can be written in the form of a quadratic equation, whose solution reads:
\begin{equation}
	\label{nu_C}
	\nu(C) = \frac{1 - \sqrt{C}}{4},
\end{equation}
after having taken into account that $\nu \in [0,0.25]$.
Note that the alternative, $1 + \sqrt{C}$ solution given by the quadratic formula would be valid for the $\nu \in [0.25,0.5]$ interval, 
which is not used here, since it is entirely equivalent (up to an inversion) with the $\nu \in [0,0.25]$ interval, 
while being relevant only if group $-1$ is allowed to be biased towards $+1$ traits instead of towards $-1$ traits, and viceversa, which is not the case here.

\subsection{Mathematical comparison of the two scenarios}\label{FCIvS2G}

\begin{figure*}
\centering
	\subfigure[\label{FFCor:FCI}]{\includegraphics[width=8.5cm]{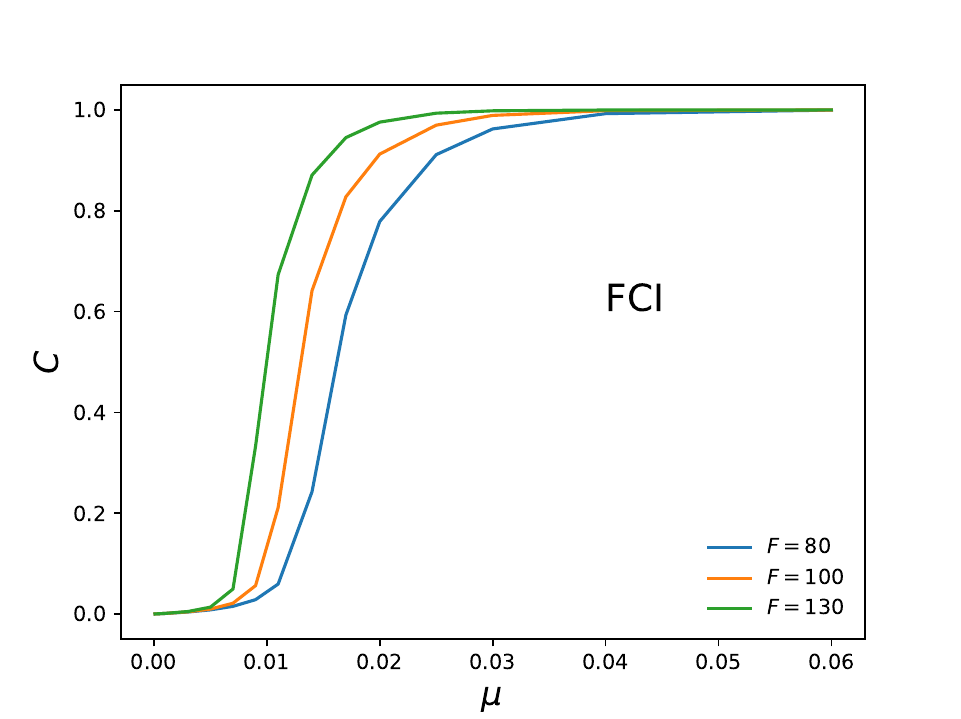}} \hspace{0.6cm}
	\subfigure[\label{FFCor:S2G}]{\includegraphics[width=8.5cm]{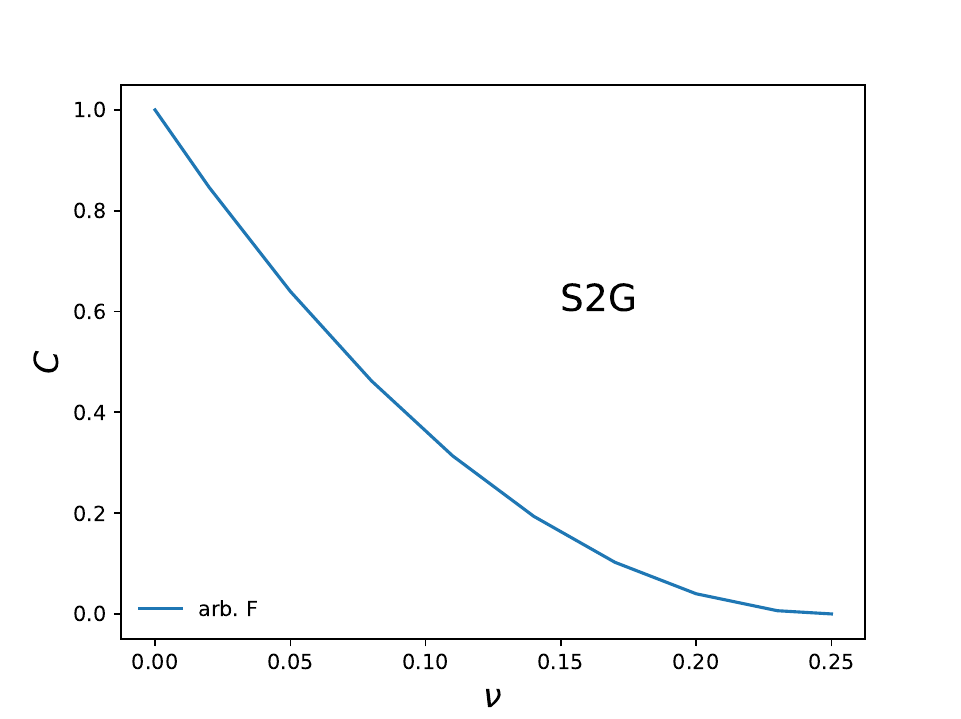}} 
	\caption{Correlation behaviour. 
	The figure shows the dependence of the pairwise feature correlation $C$, 
	first~\subref{FFCor:FCI} on the feature-feature coupling strength parameter $\mu$ controlling the fully-connected Ising model (FCI),
	second~\subref{FFCor:S2G} on the group strength parameter $\nu$ controlling the symmetric two-groups model (S2G).
	In the case of FCI, different curves (legend) are shown for different values of the number of features $F$,
	while in the case of S2G, a single curve is shown, which is valid for any value of $F$.
}
\label{FFCor}
\end{figure*}

\begin{figure}
\centering
	\includegraphics[width=8.5cm]{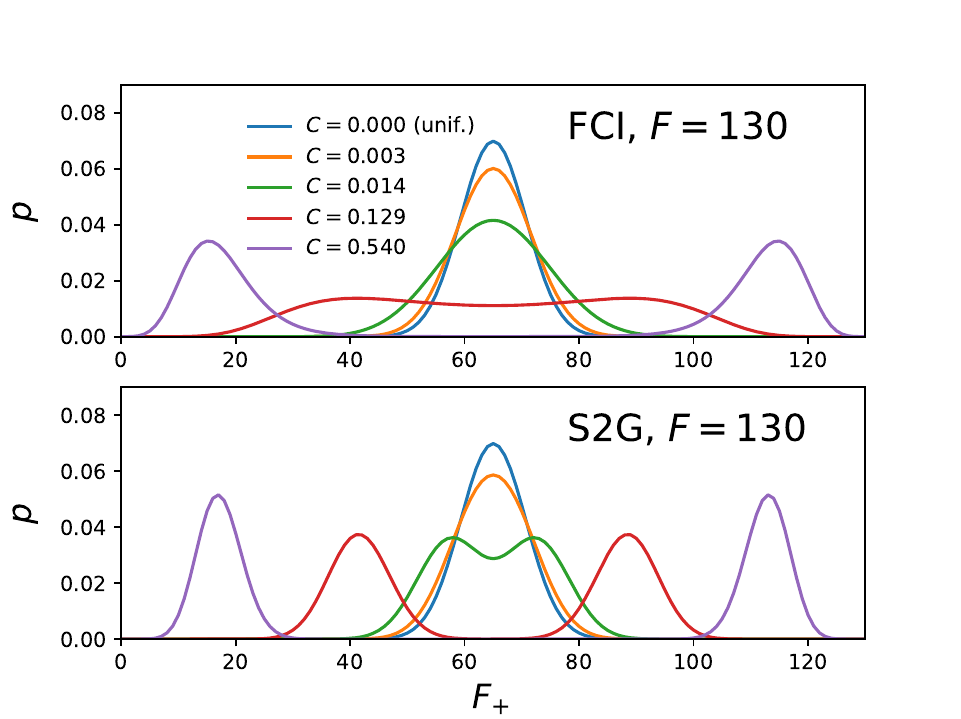}
	\caption{Shape of probability distribution. 
	The figure shows the probability associated to configurations with $F_{+}$ traits $+1$, for $F=130$ features,
	for different values of the feature-feature correlation level $C$ (legend),
	for the fully-connected Ising model (FCI, top) and for the symmetric two-groups model (S2G, bottom).
}
\label{ProbDist}
\end{figure}

\begin{figure*}
\centering
	\subfigure[\label{SBPT:FCI_1}]{\includegraphics[width=8.5cm]{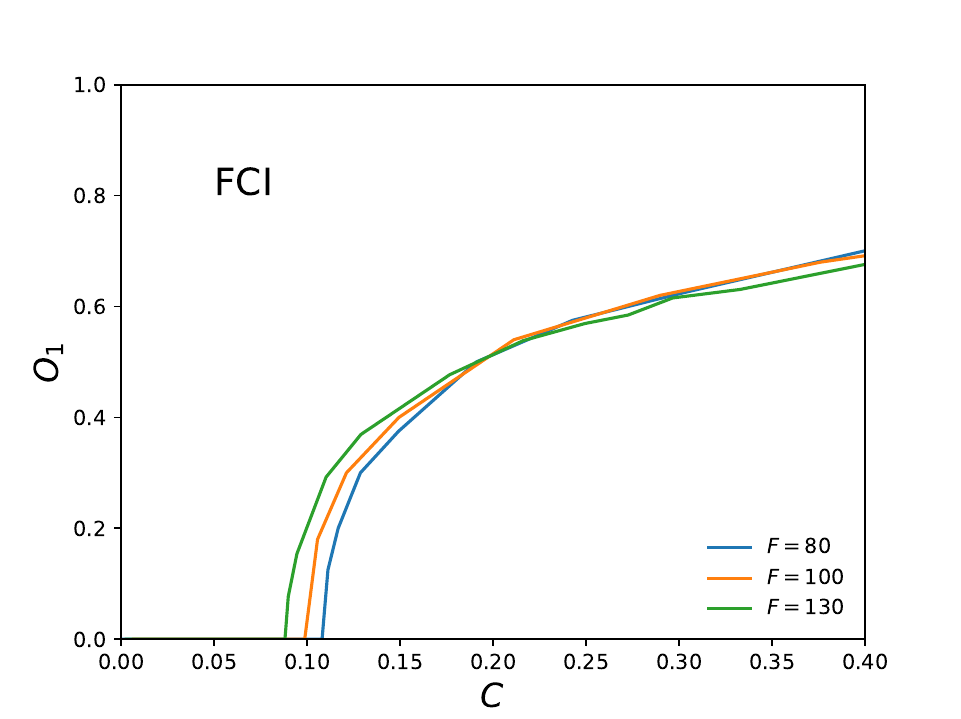}} \hspace{0.6cm}
	\subfigure[\label{SBPT:S2G_1}]{\includegraphics[width=8.5cm]{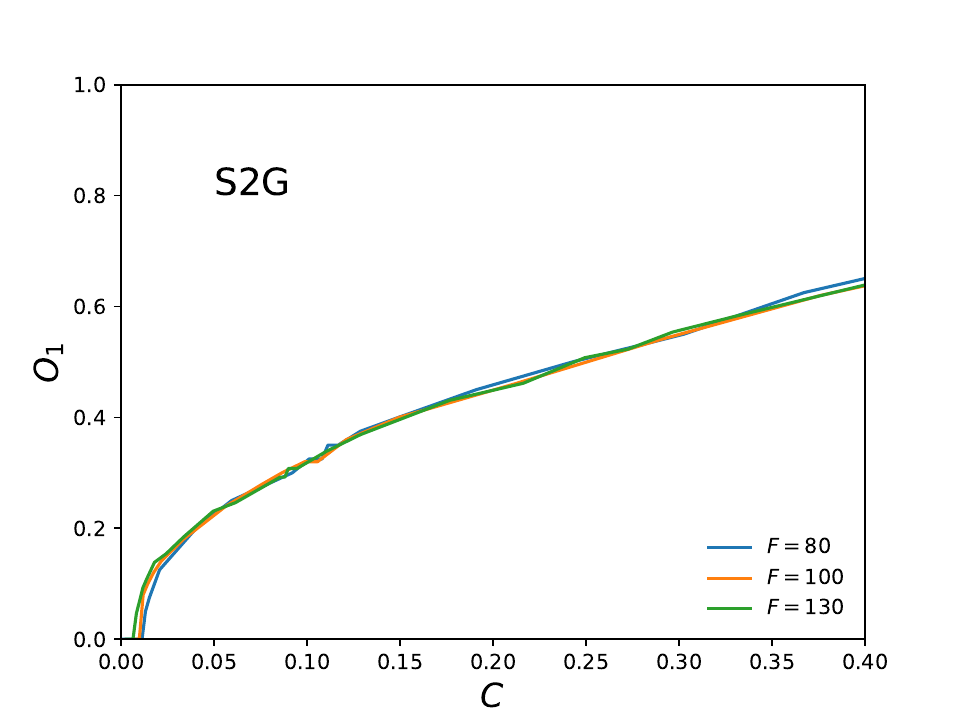}} \\ \vspace{-0.3cm}
	\subfigure[\label{SBPT:FCI_2}]{\includegraphics[width=8.5cm]{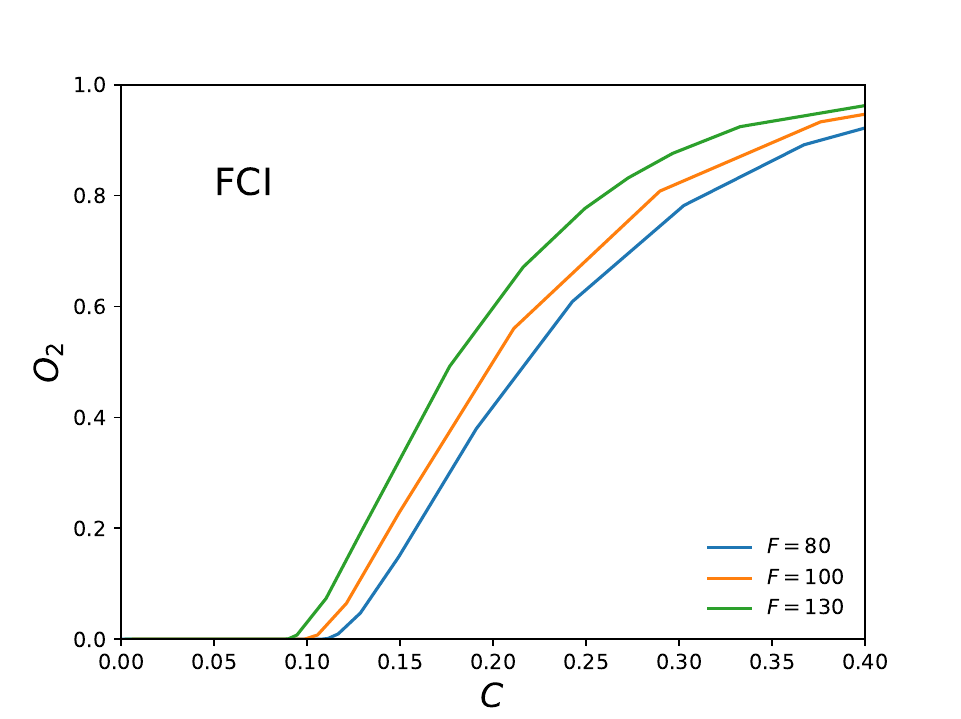}} \hspace{0.6cm}
	\subfigure[\label{SBPT:S2G_2}]{\includegraphics[width=8.5cm]{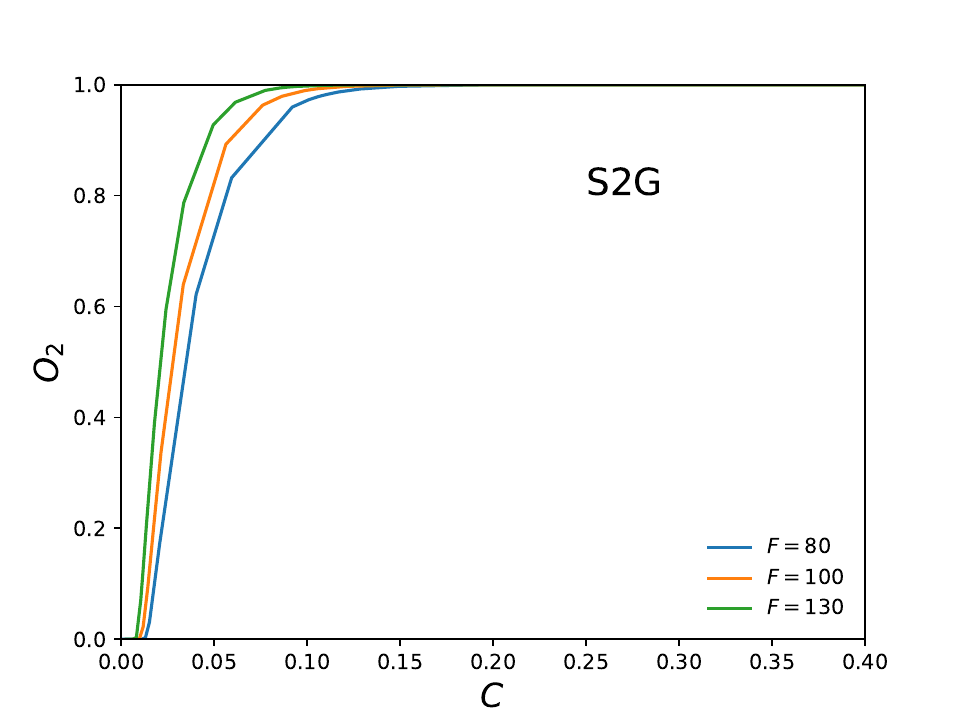}}
	\caption{Symmetry breaking phase transitions. 
	The figure shows the behavior of the normalized probability peak departure $O_1$ (top) and of the normalized peak height $O_2$ (bottom), 
	as functions of the model parameters, 
	for the fully-connected Ising (FCI, top) and the symmetric two-groups (S2G, bottom) models.
	Different curves corresponds to different values of the $F$ parameter, controlling the number of features (legends).
	Both the $\mu$ parameter of the FCI model and the $\nu$ parameter of the S2G model are remapped to the associated correlation value $C$, 
	which is shown along the horizontal axis of each plot. 
}
\label{SBPT}
\end{figure*}

This section deals with the comparison between the FCI and the S2G models, 
in terms of properties that can be extracted directly from the equations in Sec.~\ref{FCI} and Sec.~\ref{S2G}, 
without the need of randomly sampling from the the two statistical ensembles.
Specifically, we focus on the behavior of the feature-feature correlation (Fig.~\ref{FFCor}), 
the shape of the probability distribution (Fig.~\ref{ProbDist}) and the symmetry breaking phase transition (Fig.~\ref{SBPT}) associated to each model.

Fig.~\ref{FFCor} shows the behavior of the correlation between any two cultural features for the two models. 
First, Fig.~\ref{FFCor:FCI} shows how the correlation entailed by the FCI model depends on the model parameter $\mu$ controlling the pairwise couplings between features,
based on Eq.~\eqref{C_mu}.
Different curves correspond to different values of $F$.
Note that the correlation increases from $C=0.0$ to $C=1.0$ as the coupling $\mu$ is increased,
but it also increases as the number of features $F$ is increased.
Second, Fig.~\ref{FFCor:S2G} shows how the correlation entailed by the S2G model depends on the model parameter $\nu$ controlling the group strength,
based on Eq.~\eqref{C_nu}.
Here, the correlation decreases from $C=1.0$ to $C=0.0$ as $\nu$ is increased, which is consistent with the fact that, by construction,
lower values of $\nu$ correspond to a stronger group structure. 
Note that the $C(\nu)$ behavior is independent of $F$, which is obvious from Eq.~\eqref{C_nu}.

All the following comparisons are based on a matching of the two models in terms of the correlation level $C$.
Specific values of $\mu$ are chosen, based on which the correlation level entailed by the FCI model $C(\mu, F)$ is computed via Eq.~\eqref{C_mu}, for a given $F$.
Then, the corresponding $\nu(C)$ of S2G entailing the same correlation is calculated based on Eq.~\eqref{nu_C}.
This creates a correspondence between parameter $\mu$ of FCI and parameter $\nu$ of S2G by means of the correlation $C$.
Since $C$ is a number extracted from the full statistical ensemble under a specific parameterization, it can be regarded as a model parameter, 
namely as a replacement or remapping of $\mu$ (in the case of FCI) and of $\nu$ (in the case of S2G), which allows for a side-by-side comparison of the two models in terms of other quantities. 

This $\mu$-to-$C$-to-$\nu$ mapping is first exploited by Fig.~\ref{ProbDist}, which shows the probability distributions associated to the FCI and S2G models,
as described by Eq.~\eqref{prob_mu_F} and Eq.~\eqref{prob_nu_F} respectively. 
In either case, the distribution is shown for the same values of the correlation $C$ that are listed by the legend at the top. 
These $C$ values correspond to the values of the $\mu$ and $\nu$ parameters that are listed in Table~\ref{TabPars}.
The calculations are based on a value of $F=130$, which is comparable to the $F$ values associated to the empirical cultural states used in Sec.~\ref{Empir_1} and Sec.~\ref{Empir_2}.

\begin{table}
\begin{center}
  \begin{tabular}{ c | c | c }
    \hline
    correlation level $C$ & FCI parameter $\mu$ & S2G paramter $\nu$  \\ \hline \hline
    0.000	& 0.000	& 0.250 \\ 
    0.003	& 0.002	& 0.237 \\ 
    0.014	& 0.005	& 0.221 \\ 
    0.129	& 0.008	& 0.160 \\ 
    0.540	& 0.010	& 0.066 \\ 
  \end{tabular}
	\caption{Parameter mapping. 
	The table shows the correspondence between the correlation values $C$ shown in Fig.~\ref{ProbDist}, 
	the associated $\mu$ values used for generating the FCI probability curves and 
	the associated $\nu$ values used for generating the S2G probability curves. 
	This correspondence is valid when $F=130$ features are used for the FCI and S2G models.}
	\label{TabPars}
\end{center}
\end{table}

Note that, in the limit of vanishing correlation $C$, the distributions of both models converge to the uniform probability distribution, 
which assigns to every value of $F_{+}$ a probability that is equal to the fraction of possible configurations with that many ``+'' traits .
This uniform distribution is characterized by the existence of one maximum at the center of the $F_{+}$ axis. 
As the correlation $C$ increases, the shape of the distribution becomes wider, with two equal maxima arising on either side of the $F_{+}$ axis, whose separation also increases with increasing $C$.
Thus, both models exhibit a symmetry breaking phase transition.

However, a close inspection of Fig.~\ref{ProbDist} reveals that the symmetry breaking happens later (higher values of $C$) for the FCI model than for the S2G model, 
meaning that there is a non-vanishing $C$ interval for which FCI exhibits a unimodal behavior, while the S2G exhibits a bimodal behavior, interval which contains the $C=0.014$ value. 
This $C$ interval is of crucial interest for this study, since it corresponds to the correlation regime for which the symmetric group structure built into the S2G model is visible in the shape of the probability distribution, 
while the feature-feature coupling built into the FCI model is not strong enough to induce a qualitatively similar shape. 
Still, even for $C$ values that are high enough for the FCI distribution to also show maxima, the exact shapes of the two distributions are also different, 
with the S2G maxima being stronger than the FCI ones (visible for $C=0.129$ $C=0.540$).
This is a visual confirmation that the two statistical ensembles are indeed different
and that the S2G ensemble has a smaller Shannon entropy than the FCI ensemble, for any, non-vanishing value of $C$, 
thus being more biased, more constrained and encoding more structure, which should manifest itself at the level of higher-order correlations (involving more than two spins/features). 

A more complete picture of the phase transitions exhibited by the two models is provided by Fig.~\ref{SBPT}. 
This shows the dependence of two mathematical properties of the probability distributions in Fig.~\ref{ProbDist} on the model parameters.
The first property, denoted here by $O_1(\gamma,F) \in [0,1]$, is a normalized departure of either probability peak from the center of the (horizontal) $F_{+}$ axis.
The second property, denoted here by $O_2(\gamma,F) \in [0,1]$, is a normalized height of either probability peak compared to the probability at the center of the (horizontal) $F_{+}$ axis. 
Note that $\gamma$ is a placeholder for either the $\mu$ parameter or the $\nu$ parameter, depending, respectively, on whether the FCI or the S2G model is used.
Both quantities are zero when symmetry breaking is not present and are positive when symmetry breaking is present, giving higher values for better defined probability peaks.  
They can thus be used as ``order parameters'' characterizing the phase transition, although they are evaluated in a priori, based on the expression of the probability distribution, 
rather than based on configurations sampled from the associated ensemble. 
Mathematically, the first quantity is defined as:
\begin{equation}
	O_1(\gamma, F) = \frac{[0.5 F] - F_{+}^{\text{*}}(\gamma,F)}{[0.5 F]},
\end{equation}
while the second quantity is defined as:
\begin{equation}
	O_2(\gamma, F) = \frac{p^{*}(\gamma,F) - p(\gamma, F, [0.5 F])}{p^{*}(\gamma,F)},
\end{equation}
where the square brackets stand for the ``integer part'' operation.
Moreover, $F_+^{\text{*}}(\gamma,F)$ is the (integer) position along the $F_{+}$ axis of the first (lower-$F_{+}$) peak and $p^{*}(\gamma,F)$ is the height of this peak.
At the same time, $p(\gamma, F, [0.5 F])$ is evaluated according to either Eq.~\eqref{prob_mu_F} or Eq.~\eqref{prob_nu_F}, 
depending on whether the quantity is evaluated for the FCI model ($\gamma$ is replaced by $\mu$) or for the S2G model ($\gamma$ is replaced by $\nu$).
The value of $F_{+}^{\text{*}}(\gamma,F)$ is extracted by iteratively exploring the lower half of the $F_{+}$ axis, 
while evaluating $p(\gamma, F, F_{+})$ according to either Eq.~\eqref{prob_mu_F} or Eq.~\eqref{prob_nu_F}.
On the other hand, $p^{*}(\gamma,F)$ is essentially an abbreviation for $p(\gamma, F, F_{+}^{\text{*}}(\gamma,F))$.

The four panels of Fig.~\ref{SBPT} show 
the behaviour of $O_1$ for the FCI model (Fig.~\ref{SBPT:FCI_1}), 
the behaviour of $O_1$ for the S2G model (Fig.~\ref{SBPT:S2G_1}),
the behaviour of $O_2$ for the FCI model (Fig.~\ref{SBPT:FCI_2}) and
the behaviour of $O_2$ for the S2G model (Fig.~\ref{SBPT:S2G_2}).
The dependence of either quantity on the $\mu$ parameter (for FCI) and on the $\nu$ parameter (for S2G) is translated in terms of the corresponding correlation value $C$, 
via Eq.~\eqref{C_mu} and Eq.~\eqref{C_nu} respectively. 
Note that the two quantities agree in terms of the correlation value for which the transition occurs, 
for both the FCI (Fig.~\ref{SBPT:FCI_1} vs Fig.~\ref{SBPT:FCI_2}) and the S2G (Fig.~\ref{SBPT:S2G_1} vs Fig.~\ref{SBPT:S2G_2}), 
for any number of features $F$.
It is clear that the transition point comes closer to $C = 0.0$ with increasing $F$ for both models. 
Finally, Fig.~\ref{SBPT} shows that, independently of $F$, the transition point of S2G is located at lower values of $C$ than that of FCI.

\section{Discriminating between the two interpretations}\label{Theor_2}

This section investigates, from a spectral analysis and random matrix perspective, quantities that may differentiate between the two structural scenarios introduced in Sec.~\ref{Theor_1}:  
feature-feature redundancies vs group structure.
To this end, sets of cultural vectors are numerically sampled from the two ensembles and similarity matrices are computed, based on Eq.~\eqref{CultSim}.
Since both the FCI and S2G ensembles are such that the (marginal) feature-level probability distributions are uniform, 
restricted randomness (see Sec.~\ref{Empir_1}) is equivalent to uniform randomness as a null model (at least if the number of cultural vectors $N$ is reasonably high) 
with respect to which structure is to be evaluated.
Thus, for simplicity, uniform randomness (u-random) is used as a null model in this section.
All comparisons made here make use of matching the feature-feature coupling parameter $\mu$ of FCI and the group strength parameter $\nu$ of S2G in terms of the correlation level $C$, 
as described in Sec.~\ref{FCIvS2G}. 
Moreover, the number of features and the number of cultural vectors are $F=130$ and $N=100$ for all the FCI, S2G and u-random cultural states generated and used for the figures of this section. 

\begin{figure}
\centering
	\includegraphics[width=8.5cm]{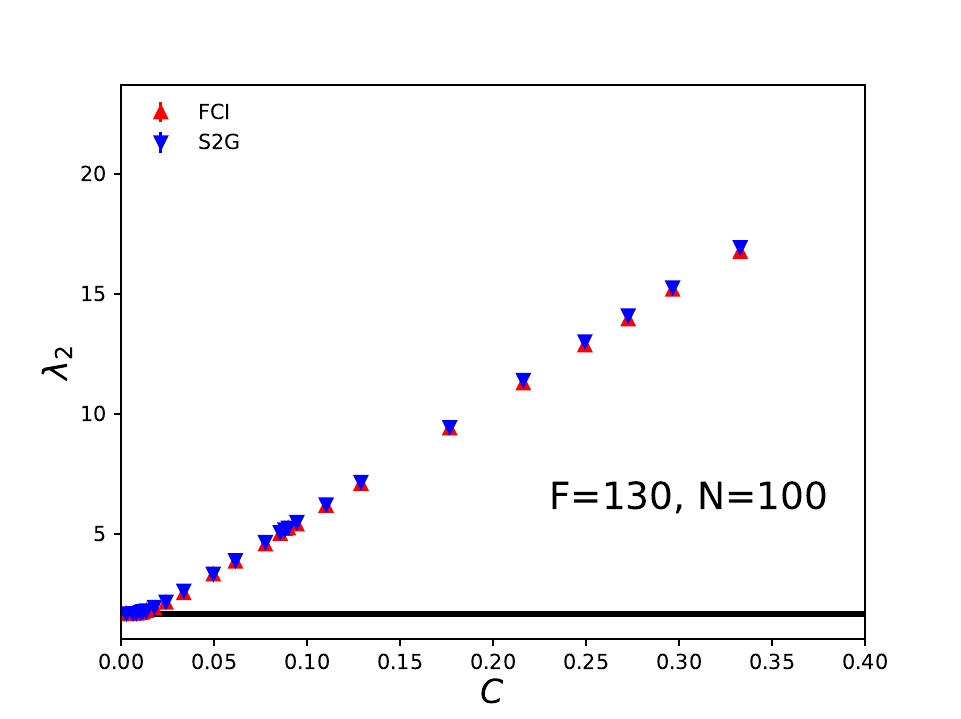}
	\caption{Behavior of subleading eigenvalue ($\lambda_2$). 
	The figure shows how $\lambda_2$ depends on the correlation level $C$ for the fully-connected Ising (FCI, red, upward triangles) and for the symmetric two-groups (S2G, blue, downward triangles) models. 
	For each $C$ value, for each of the two models, an averaging is performed over 80 sets of cultural vectors independently sampled from the respective ensemble --
	the vertical bar associated to each point shows the interval spanned by one standard mean error on each side of the mean. 
	The black, horizontal lines show, for comparison, the mean $\lambda_2$ expected based on uniform randomness, along with the width of the $\lambda_2$ distribution -- one standard deviation on each side -- 
	where the calculations are based on 60 sets of cultural vectors generated via uniform randomness 
	-- these lines do not imply that, for uniform randomness, the correlation $C$ (which actually vanishes by construction) can be arbitrarily large. 
}
\label{Val2}
\end{figure} 

\begin{figure}
\centering
	\includegraphics[width=8.5cm]{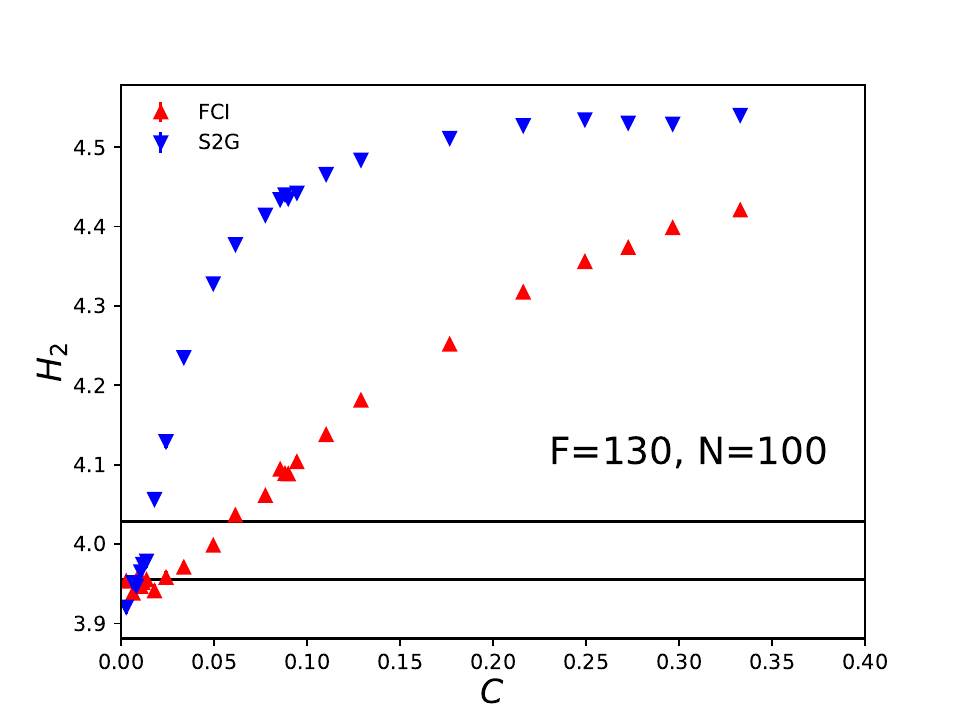}
	\caption{Behavior of uniformity $H_2$ associated to subleading eigenvalue.
	The figure shows how $H_2$ depends on the correlation level $C$ for the fully-connected Ising (FCI, red) and for the symmetric two-groups (S2G, blue) models. 
	For each $C$ value, for each of the two models, an averaging is performed over 80 sets of cultural vectors independently sampled from the respective ensemble --
	the vertical bar associated to each point shows the interval spanned by one standard mean error on each side of the mean. 
	The black, horizontal lines show, for comparison, the mean $H_2$ expected based on uniform randomness, along with the width of the $H_2$ distribution -- one standard deviation on each side -- 
	where the calculations are based on 60 sets of cultural vectors generated via uniform randomness 
	-- these lines do not imply that, for uniform randomness, the correlation $C$ (which actually vanishes by construction) can be arbitarily large. 
}
\label{Ent2}
\end{figure}

\begin{figure}
\centering
	\includegraphics[width=8.5cm]{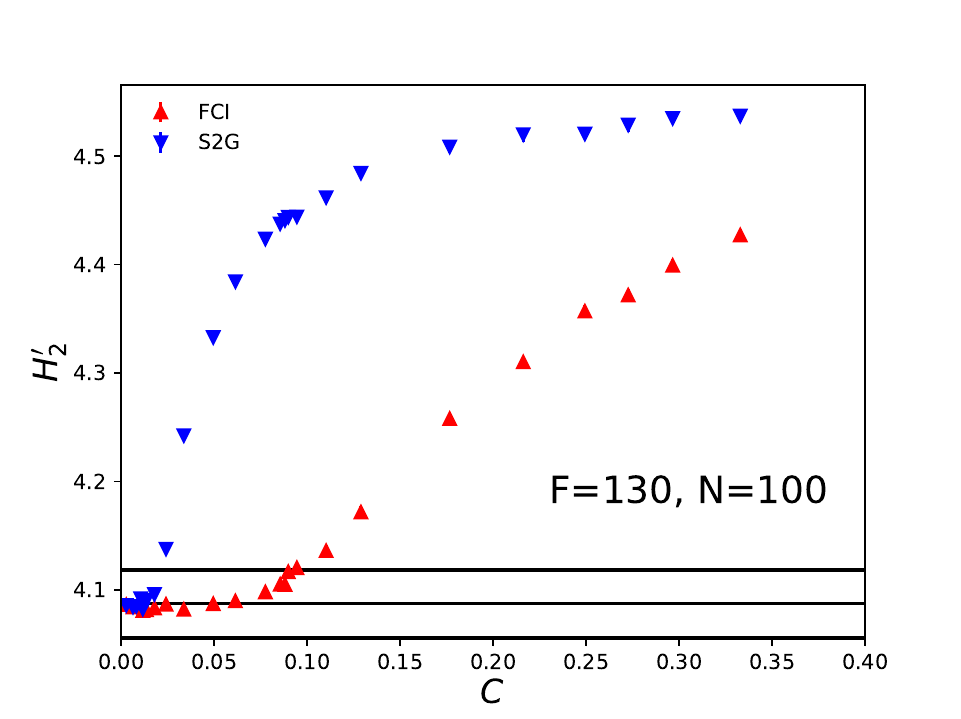}
	\caption{Behavior of subleading uniformity $H_2'$.
	The figure shows how $H_2'$ depends on the correlation level $C$ for the fully-connected Ising (FCI, red) and for the symmetric two-groups (S2G, blue) models. 
	For each $C$ value, for each of the two models, an averaging is performed over 80 sets of cultural vectors independently sampled from the respective ensemble --
	the vertical bar associated to each point shows the interval spanned by one standard mean error on each side of the mean. 
	The black, horizontal lines show, for comparison, the mean $H_2'$ expected based on uniform randomness, along with the width of the $H_2'$ distribution -- one standard deviation on each side --
	where the calculations are based on 60 sets of cultural vectors generated via uniform randomness 
	-- these lines do not imply that, for uniform randomness, the correlation $C$ (which actually vanishes by construction) can be arbitrarily large. 
}
\label{Ent_by_Ent2}
\end{figure}

The most obvious quantity that could conceivably discriminate between the FCI and the S2G models is the subleading eigenvalue $\lambda_2$, 
or the extent to which this goes above the uncertainty range predicted by uniform randomness. 
Fig.~\ref{Val2} shows the dependence of $\lambda_2$ on the correlation level $C$ for FCI (red) and S2G (blue), 
while the horizontal black lines show the u-random uncertainty range (the mean value and 1 standard deviation on each side of the mean), 
as a compact replacement of the distributions shown in Fig.~\ref{DistVal:2}, Fig.~\ref{DistVal_2:GSS} and Fig.~\ref{DistVal_2:JS} 
-- as mentioned in the figure caption, these lines are not meant to give any information about the correlation level of the u-random null model, 
nor about realized correlations based on specific sets of vectors sampled from the ensemble.
Surprisingly, $\lambda_2$ does not distinguish between the FCI and the S2G models, for any given correlation level $C$, since the average $\lambda_2$ values clearly overlap. 
At the same time, $\lambda_2$ (for both models) does depart significantly from the null model expectations.
This explicitly shows that empirical structural modes such as those identified in Sec.~\ref{Empir_1} can actually be triggered by feature-feature redundancies alone, 
at least in certain cases (those for which the simplistic setting behind the FCI and S2G models is reasonably representative).
Thus, empirical eigenvalues that significantly depart from what is expected based on the null hypothesis do not automatically indicate groups. 
In the light of Sec.~\ref{Theor_1}, Fig.~\ref{Val2} also implies that the subleading eigenmodes of matrices produced via FCI are associated, on average,
to the same self-similarity as those of matrices produced via S2G, for a given correlation level. 
This appears counter-intuitive, since the low-$C$ presence of symmetry breaking for S2G makes it much easier to identify two, well separated groups, one for each side of the $F_{+}$ axis of Fig.~\ref{ProbDist}. 
However, a closer inspection of the probability distributions in Fig.~\ref{ProbDist} reveals that
FCI is more likely to produce, even in the absence of symmetry breaking, cultural vectors that are at one extreme or the other (almost fully populated with $+1$ traits or with $-1$ traits).
These extremal configurations are much more representative, or ``central'', for the configurations that are possible on the respective side of the $F_{+}$ axis, thus compensating for the softer separation at $F_{+} = 0.5 F$. 
Also note that the values of $C$ used in Fig.~\ref{Val2} are the same for FCI and S2G and the same as those used in Fig.~\ref{Ent2} and Fig.~\ref{Ent_by_Ent2} described below.
For each FCI and S2G point in any of these plots, explicit averaging over the sampled sets of cultural vectors is only performed with respect to the quantity associated to the vertical axes.
For the correlation level $C$, associated to the horizontal axes, we simply use the analytically-computed, ensemble-level value, for the given parameterization of the model (Eq.~\eqref{C_mu} and Eq.~\eqref{C_nu}).

Sec.~\ref{App_lambda_1-3} shows, in a manner similar to Fig.~\ref{Val2}, the behavior of the largest and and third largest eigenvalues -- $\lambda_1$ and $\lambda_3$ respectively -- 
for the FCI and S2G models, in comparison to the u-random null model. 
The analysis there makes it clear that the $\lambda_1$ and $\lambda_3$ are both compatible with the null hypothesis. 
Thus, all or most of the structural information of cultural states generated from either the FCI or the S2G model is captured by the $(\lambda_2, v_2)$ eigenpair.
Since $\lambda_2$ cannot discriminate between the two scenarios, this means that all or most discriminating power is encoded in the associated eigenvector $v_2$, 
which is the focus of the rest of this section. 

Based on Sec.~\ref{FCIvS2G} and in particular on Fig.~\ref{ProbDist}, one can say that, 
for the interesting correlation interval where FCI does not exhibit symmetry breaking while S2G does, 
configurations that are on one side of the $F_{+}$ axis and are generated with S2G exhibit relatively equal fractions of traits of a certain sign,
compared to those that are generated with FCI.
The S2G configurations should thus also display relatively equal contributions to the structural mode $(\lambda_2,v_2)$, 
so the associated $v_2$ entries should be much more similar for S2G than for FCI. 
Given the symmetric nature of both models, it follows that the absolute values of all the $v_2$ entries should be much closer to each other for S2G cultural states than for FCI ones -- in either case, the entries associated to different sides of the $F_{+}$ axis would (typically) have different signs.
This reasoning suggests that the difference between FCI and S2G would be captured by a quantity that evaluates the overall ``uniformity'' of the $v_2$ eigenvector,
based on the absolute values of its entries.
Since these are normalized via $\sum_{i=1}^N |v_l^i|^2 = 1$ for any eigenvector $v_l$, the Shannon entropy is a natural quantity for evaluating the uniformity. 
This leads to the definition of ``eigenvector entropy'' $H_l$ associated to to the $l$th highest eigenvalue $\lambda_l$, as a measure of uniformity:
\begin{equation}
	\label{EigVecEnt}
		H_l = - \sum_{i=1}^N |v_l^i|^2 \log |v_l^i|^2
\end{equation}
where $v_l^i$ is the $i$th entry of the eigenvector associated to $\lambda_l$ -- note that this quantity was also used in Ref.~\cite{Patil}, which cites Ref.~\cite{Jones}.

Fig.~\ref{Ent2} shows the behavior of the eigenvector entropy $H_2$ associated to the second highest eigenvalue $\lambda_2$, in a format very similar to that of Fig.~\ref{Val2}.
This confirms that $H_2$ discriminates well between the two models, with S2G showing clearly higher $H_2$ values than FCI as long the correlation level does not come arbitrarily close to $C = 0.0$.
Moreover, comparing the two profiles with the u-random one-$\sigma$ band reveals that the structure of S2G becomes incompatible with the null-hypothesis for much lower correlation values than the structure of FCI.
However, for either model, the $H_2(C)$ curve does not show the sudden increase that one would expect based on the phase transitions described in Sec.~\ref{FCIvS2G}, 
in the manner they are exhibited by the more theoretical $O_1(C)$ and $O_2(C)$ curves in Fig.~\ref{SBPT}.
 
The smoothness of the $H_2(C)$ curves is actually related to the fact that, for the low-$C$ regime, where $\lambda_2$ is highly compatible with the null hypothesis, 
$H_2$ is typically not the second highest eigenvector entropy, although it is associated to the second highest eigenvalue.
This suggests a definition of $H_l'$ as the $l$th highest eigenvector entropy, independently of the associated eigenvalue.
Fig.~\ref{Ent_by_Ent2} is a modification of Fig.~\ref{Ent2}, with $H_2'$ used as a replacement for $H_2$ for the vertical axis, affecting all the FCI, S2G and u-random calculations. 
Note that, unlike in Fig.~\ref{Ent2}, the sudden changes in Fig.~\ref{SBPT} are now reflected in Fig.~\ref{Ent_by_Ent2}.
Moreover, the transition points at $F=130$ in Fig.~\ref{SBPT} seem to be well reproduced in Fig.~\ref{Ent_by_Ent2}, 
while the FCI and S2G shapes of the $H_2'(C)$ curves are quite similar to those of $O_2(C)$, which are related to the height of the probability distribution peaks.
Finally for higher $C$ values, each $H_2'(C)$ curve in Fig.~\ref{Ent_by_Ent2} is almost identical to the associated $H_2(C)$ in Fig.~\ref{Ent2}, 
so strong structure makes it very likely that the eigenvector of the second highest eigenvalue has the second highest entropy, 
and $H_2'$ is effectively equivalent to $H_2$.

The considerations above strongly suggest that a significant departure of the eigenvector entropy from the null model expectation is a good indication that the eigenvector encodes information about a group or a grouping tendency:
this departure is present for S2G but absent for FCI for the interesting $C$ inteval for which S2G exhibits symmetry breaking and FCI does not.
However, the criterion breaks down when $C$ is higher than the critical FCI value,
illustrating that, in an empirical setting, very strong redundancies would not be distinguishable from groups at all.

\section{Revisiting the empirical data}\label{Empir_2}

The findings of Sec.~\ref{Theor_2} point out the importance of the eigenvector entropy, in addition to the eigenvalue, for deciding whether a structural mode qualifies as an authentic group mode or not. 
As a consequence, in this section, the two quantities are being used together for a second, more detailed inspection of the empirical data in Sec.~\ref{Empir_1}.  
The empirical similarity matrices are computed based on the same three sets of $N=100$ cultural vectors used in Sec.~\ref{Empir_1},
constructed from Eurobarometer (EBM), General Social Survey (GSS) and Jester (JS) data. 

\begin{figure}
\centering
	\includegraphics[width=8.5cm]{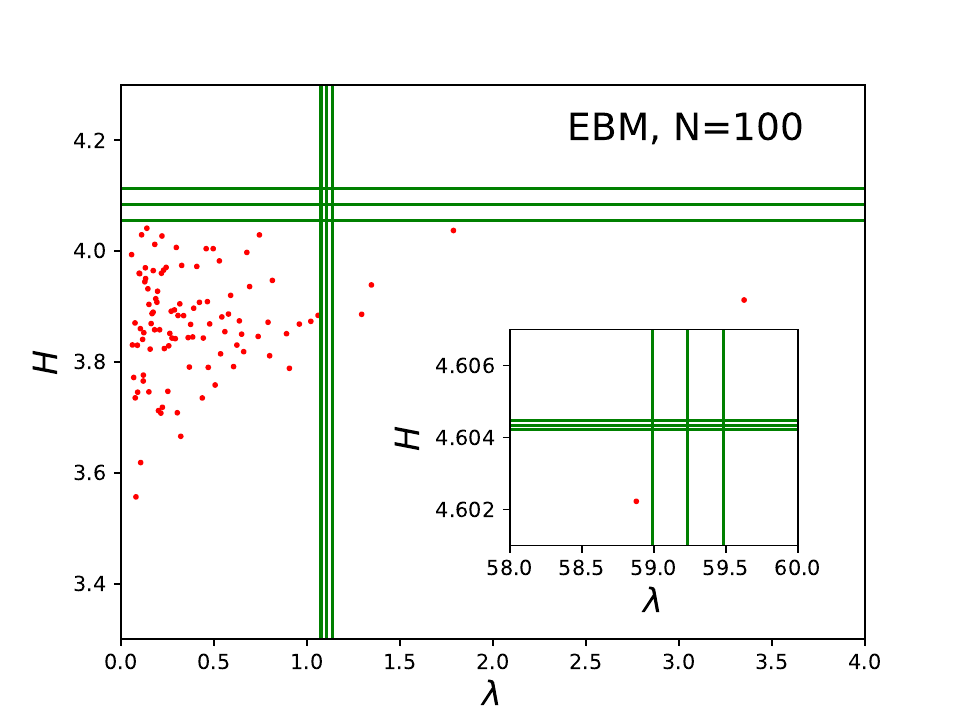}
	\caption{Eigenvalues and eigenvector entropies for empirical data.
	Every point corresponds to an empirical eigenpair, with the eigenvalue $\lambda$ shown along the horizontal axis and the eigenvector entropy $H$ shown along the vertical axis.
	The inset focuses on the leading eigenvalue, which also corresponds to the highest eigenvector entropy.
	The vertical lines in the main plot and in the inset show, respectively, the widths (one-standard deviation on each side of the mean) of the subleading and leading eigenvalue distributions, based on restricted randomness. 
	The horizontal lines in the main plot and in the inset show, respectively, the widths (one-standard deviation on each side of the mean) of the second highest and highest eigenvector entropy distributions, based on restricted randomness. 
	The vertical lines are not intended to provide any information about the eigenvector entropies associated to the respective eigenvalues, 
	while the horizontal lines are not intended to provide any information about the eigenvalues associated to the respective eigenvector entropies. 
	The figure is based on the same, Eurobarometer (EBM) data with $N=100$ cultural vectors used in Figs.~\ref{Spec},~\ref{DistVal} and~\ref{DistVal_1_Det}.
}
\label{EntVal_EBM}
\end{figure} 

Fig.~\ref{EntVal_EBM} shows a scatter of the empirical eigenpairs of the EBM matrix, 
where the horizontal axis is associated to the eigenvalue $\lambda$, 
while the vertical axis is associated to the eigenvector entropy $H$.
The global mode eigenpair is highlighted by the inset. 
In the main plot, the vertical lines show the average and the 1-$\sigma$ band of what one may expect for the subleading eigenvalue $\lambda_2$, based on the r-random null model, 
which reproduces, on average, the empirical trait frequencies (see Sec.~\ref{Empir_1}).
In the inset, the vertical lines show the same type of information for the leading eigenvalue $\lambda_1$.
The horizontal lines in the main plot and the inset show the average and the 1-$\sigma$ band of what one may expect for, respectively,
the subleading entropy $H_2'$ and the leading entropy $H_1'$, based on the r-random null model.
Note that, as anticipated in Sec.~\ref{Theor_2}, the subleading entropy is usually not associated to the subleading eigenvalue, 
while the leading entropy appears to always be associated to the leading eigenvalue. 

\begin{figure*}
\centering
	\subfigure[\label{EntVal_other:GSS}]{\includegraphics[width=8.5cm]{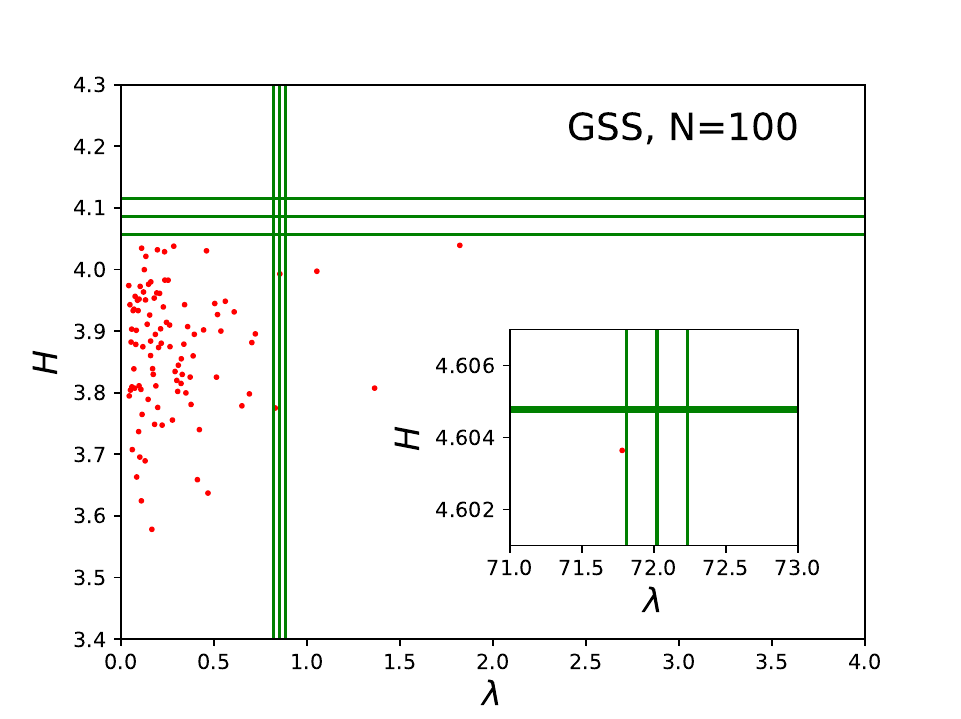}}
 	\subfigure[\label{EntVal_other:JS}]{\includegraphics[width=8.5cm]{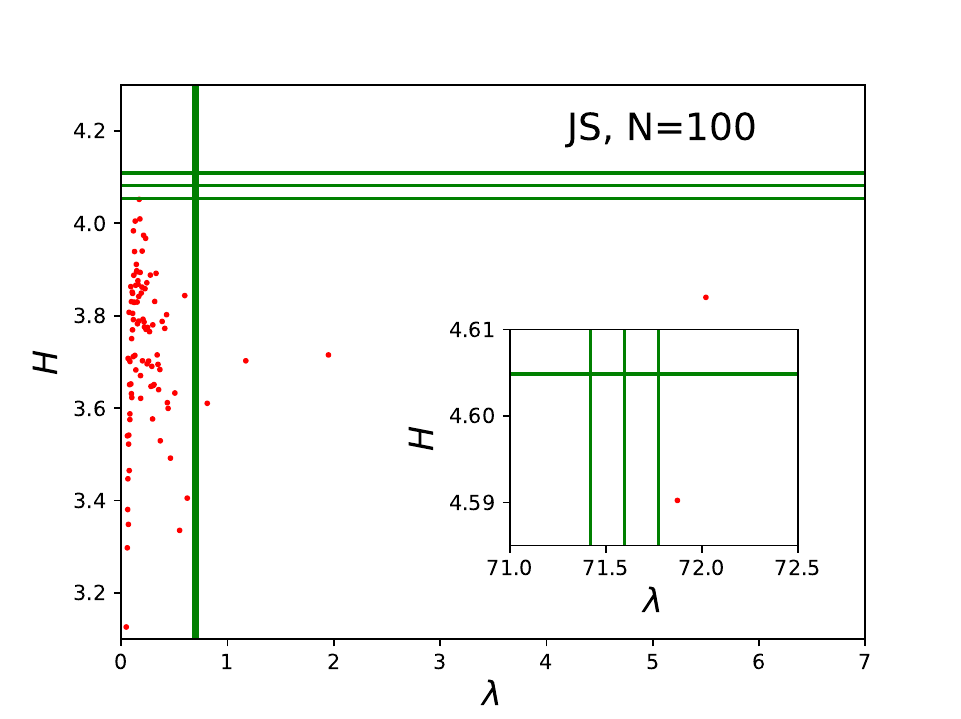}}
	\caption{Eigenvalues and eigenvector entropies in other empirical datasets.
	Fig.~\subref{EntVal_other:GSS} is based on the General Social Survey (GSS) data with $N=100$ cultural vectors used for Fig.~\ref{EntVal_other:GSS},
	while Fig.~\subref{EntVal_other:JS} is based on the Jester (JS) data with $N=100$ cultural vectors used in Fis.~\ref{EntVal_other:JS}.
    Each plot makes use of the same type of eigenpair analysis as Fig.~\ref{EntVal_EBM}.
}
\label{EntVal_other}
\end{figure*} 

The four structural modes identified based on Fig.~\ref{DistVal:2} are also visible in the main plot of Fig.~\ref{EntVal_EBM}, 
to the right of the vertical r-random band.
Importantly, all their eigenvector entropies are below the horizontal r-random band, suggesting that neither of them qualifies as a group mode.
Actually, all the bulk EBM eigenpairs are also below the r-random band, and thus compatible with the null hypothesis in terms of the uniformity of eigenvector entries. 
Also note that the leading eigenvector entropy is significantly smaller than what the null model predicts, 
but this difference is much smaller than the difference between the leading eigenvector entropy and the subleading one. 
This means that the contributions of different cultural vectors to the global mode are less equal than expected based on randomness, but much more equal than the contributions to any of the structural modes. 

The analysis in Fig.~\ref{EntVal_EBM} is also applied to the other datasets and the results are presented in Fig.~\ref{EntVal_other}, 
with Fig.~\ref{EntVal_other:GSS} showing the results for GSS data and Fig.~\ref{EntVal_other:JS} showing the results for JS data.
In both cases, the results are similar to those of EBM data: the structural modes do not show a higher eigenvector uniformity than what is expected based on the null model, 
nor do any of the smaller-$\lambda$ modes, while the eigenvector uniformity of the global mode is smaller than what is expected based on the null model, but much higher than what is expected or realized for the structural modes and the random modes.
In the light of Sec.~\ref{Theor_1} and Sec.~\ref{Theor_2}, these results suggests that structural modes of empirical matrices of cultural similarity are not due to authentic group structure, 
but due to arbitrary semantic redundancies between the questions or items used for the dataset.
However, such a conclusion would be premature, as it implicitly uses strong assumptions about cultural groups.
This aspect is further explored in Sec.~\ref{Non_Unif_Group}.

\section{Internally non-uniform groups}\label{Non_Unif_Group}

This section elaborates on another interpretation of the above results: empirical structural modes (or at least some of them) are actually signatures of authentic, system specific groups that somehow do not exhibit significant eigenvector uniformity;
this is an alternative to the interpretation resulting from Sec.~\ref{Empir_2}: empirical structural modes are due to arbitrary, instrument-specific feature-feature redundancies. 
The latter interpretation is based on the observation that the eigenvector uniformities of these modes are not higher than null model expectations,
thus not satisfying the criterion emerging from Sec.~\ref{Theor_1} and Sec.~\ref{Theor_2}.
In turn, this criterion relies on comparisons between the S2G and FCI models and is potentially sensitive to assumptions about cultural groups
that might be inherent in the S2G model, while not necessarily valid for real-world cultural groups.

In fact, the high subleading eigenvector uniformity characterizing S2G appears to be a direct consequence of the following property of the model:
generated vectors associated to either group share a typical separation from what one may call the ``center'' of the group -- the hypothetical vector that would best represent that grouping tendency, which in this case would be either the full ``$-1$'' or the full ``$+1$'' vector.
Most of the generated vectors fall within a relatively narrow region around that typical separation.
The separation bands associated to the two groups are obvious upon inspecting the bottom of Fig.~\ref{ProbDist},
in the form of the two peaks of the probability distribution, present for a wide range of the correlation level $C$
-- in this representation, the two group centers correspond to the two extremes of the $F_{+}$ axis: $F_{+} = 0$ and $F_{+} = F$.
As explained in Sec.~\ref{Theor_2}, such a well defined separation band implies that the eigenvector capturing the respective group 
(effectively capturing both groups, in the case of S2G, due to the highly symmetric setting)
typically has a large number of entries with relatively similar (absolute) strengths and no entries of significantly higher strengths.

One can thus argue that, by construction, S2G gives rise to groups that are ``internally uniform'',
property which seems inadequate for real world cultural groups:
it is hard to imagine a reason why individuals would be effectively forbidden from coming arbitrarily close to or from going arbitrarily far from the center of the group in cultural space.
More concretely, if the group is a manifestation of a political ideology, 
there seems to be no reason why there should exist a preferred number of topics/items in terms of which individuals under the influence of that ideology would agree with its most representative opinion profile. 

Instead, it is very plausible that a real cultural group exhibits a high variability of the extent to which different individuals identify with it,
so that one encounters non-vanishing numbers of individuals that are very central or very peripheral.
Such ``internally non-uniform'' groups would likely not exhibit statistically significant eigenvector uniformities,
so the eigenvector entropy criterion developed and used above would fail to recognize them as authentic.
In order to illustrate this scenario in a more quantitative manner, Sec.~\ref{Mixed_Protos} makes use of another toy model, 
called ``mixed prototype generation'' (MPG), inherited from previous work~\cite{Babeanu_2}, 
where it was shown to be capable of generating cultural states that reproduced other, important empirical properties.
This model explicitly randomizes the strengths of the couplings between generated vectors and central group profiles, or ``prototypes'',
so that the distribution of vectors' separations from these ``prototpyes'' is rather flat, without separation peaks/bands like those of S2G.
The ``mixing'' relates to the fact that every generated vector is a quasi-unique combination of all the prototypes, although typically dominated by either of them. 
While structurally different from both S2G and FCI, MPG can also be used in the binary-feature setting employed in Sec.~\ref{Theor_1} and Sec.~\ref{Theor_2} (although in a manner that is somewhat less elegant mathematically).
This is exploited by Sec.~\ref{Mixed_Protos}, which explicitly shows that cultural groups with strongly-significant eigenvalues but non-significant eigenvector uniformities may exist, 
thus providing an important, theoretical indication that empirical structural modes highlighted in Sec.~\ref{Empir_1} and Sec.~\ref{Empir_2} might still be signatures of authentic cultural groups that are internally non-uniform.

Another, more empirically-based indication of this possibility is provided by Sec.~\ref{Feat_Elim}. 
This takes advantage of the block-diagonal form of the feature-feature correlation matrix of one of the datasets used in this study,
which allows for easy identification and elimination of blocks of obviously redundant features, while checking the robustness of the structural modes under this operation. 
It turns out that some of the structural modes retain their eigenvalue significance after eliminating all the obvious feature redundancies,
suggesting that these robust modes are actually due to authentic but internally non-uniform cultural groups. 

\subsection{The mixed prototypes scenario}\label{Mixed_Protos}

\begin{figure*}
\centering
	\subfigure[\label{EntVal_MPG:70}]{\includegraphics[width=8.5cm]{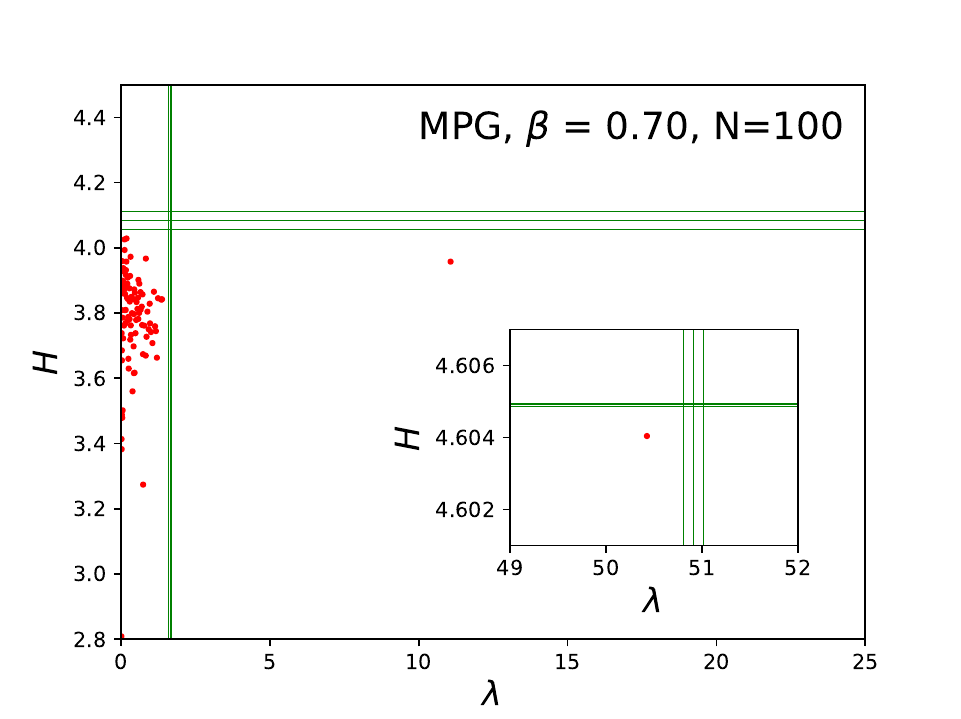}}
 	\subfigure[\label{EntVal_MPG:80}]{\includegraphics[width=8.5cm]{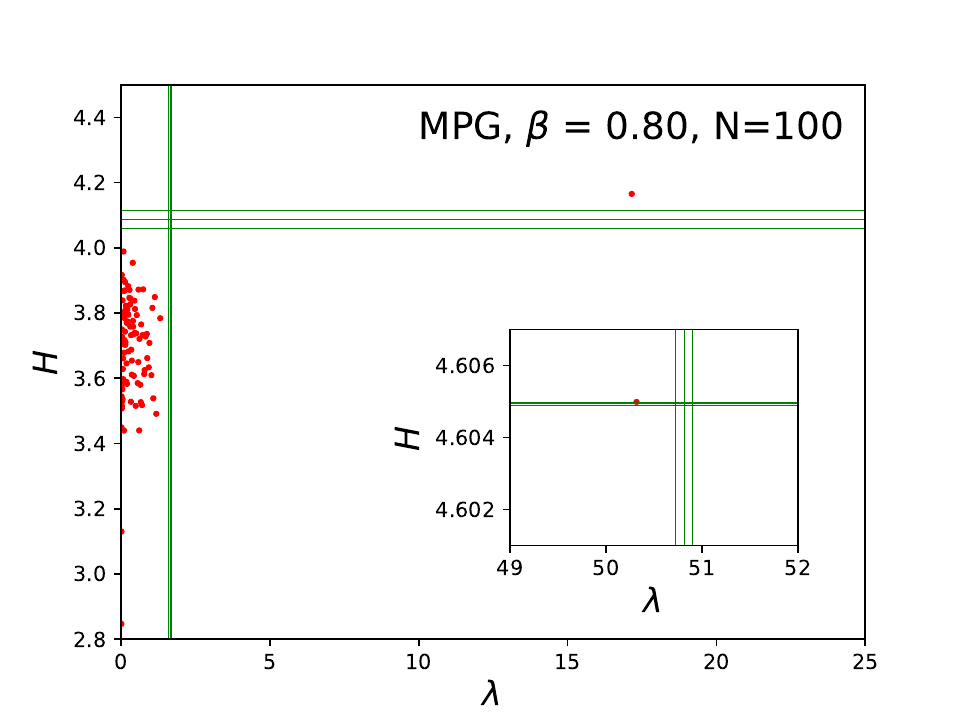}}
	\caption{Eigenvalues and eigenvector entropies for cultural states generated via mixed prototypes.
    Each plot corresponds to one MPG state of $N=100$ cultural vectors generated with mixing parameter 
    $\beta=0.70$ (Fig.~\subref{EntVal_MPG:70}) and 
	$\beta=0.80$ (Fig.~\subref{EntVal_MPG:80}) respectively.
    Each plot makes use of the same type of eigenpair analysis as Fig.~\ref{EntVal_EBM}.
}
\label{EntVal_MPG}
\end{figure*}

\begin{figure}
\centering
	\includegraphics[width=8.5cm]{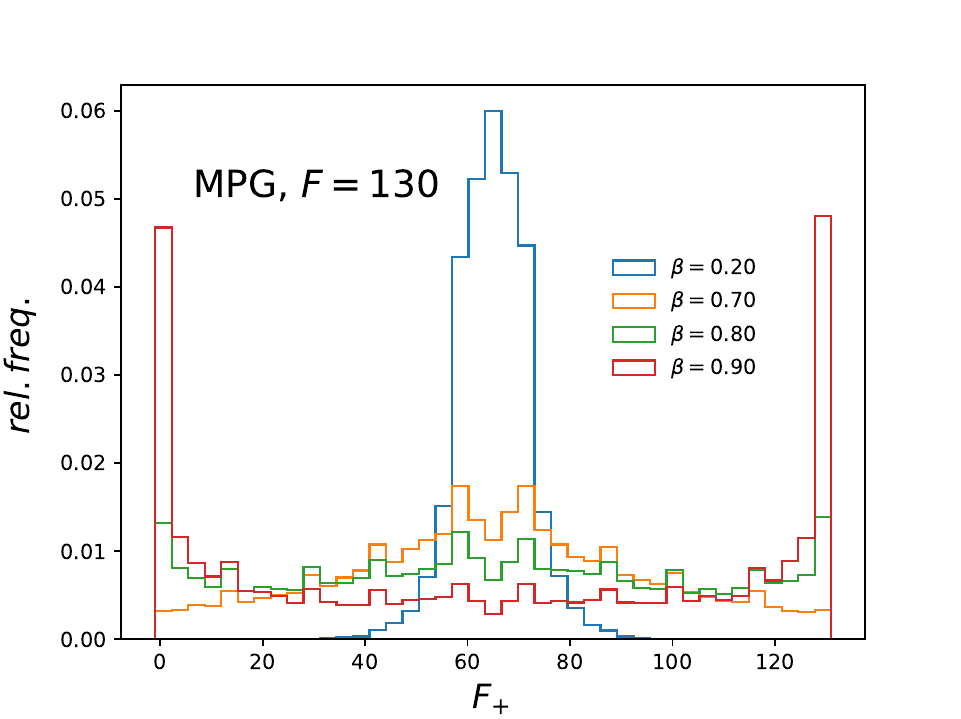}
	\caption{Approximate shape of probability distribution. 
	The figure shows the relative frequency associated to vectors with $F_{+}$ traits $+1$, 
	obtained via mixed prototpe generation (MPG) with two, maximally separated prototypes with $F_{+} = 0$ and $F_{+} = F$,
	for several values of the mixing parameter $\beta$ (legend),
    for $F=130$ binary features.
    The estimation is done based on $N=20000$ numerically sampled vectors for each $\beta$ value.
    Note that the histogram use 41 bins that cover a probability support with $F+1=131$ discrete points, 
    so each bin includes 3 or 4 of these points.
}
\label{ProbDist_MPG}
\end{figure}

This section focuses on cultural states generated with the ``mixed prototype generation'' (MPG) procedure proposed in Ref.~\cite{Babeanu_2}. 
Besides its interesting social science foundation and demonstrated structural realism,
MPG is capable of generating cultural states characterized by groups that are internally non-uniform, 
which is why it is employed here.
After providing a review of the central assumptions and technical aspects behind MPG in the following few paragraphs, 
we present, via Fig.~\ref{EntVal_MPG} and Table~\ref{TabPars_MPG}, relevant random matrix results obtained for MPG cultural states. 
In parallel, Fig.~\ref{ProbDist_MPG} illustrates how MPG vectors are distributed with respect to the group prototypes,
while providing further understanding of the essential structural differences between MPG, S2G and FCI. 

MPG is designed~\cite{Babeanu_2} to generate cultural states that reflect the existence of a certain number $K$ of underlying ideologies that are effectively recognizable, in different extents and ways, at the level of every individual in the population.
These ideologies are formally represented by abstract cultural vectors, or prototypes, so that each prototype is the ``ideal'' opinion profile of one ideology.
Every concrete cultural vector is generated as a random, quasi-unique mixture of the $K$ prototypes and a uniformly random, pure noise component. 
The associated $K+1$ contributions are deliberately unequal, so there is an ordering of the prototypes in terms of the number of traits that are copied from each of them.
This ordering is different across different generated vectors, so that every vector has a dominant prototype, 
while the number of traits generated from pure noise is by construction always smaller or equal to that of the lowest-contributing prototype.

The mixing contributions are specified by randomly picking, for each generated vector, a set of $K+1$ weights
$\{w_1(\beta),...,w_{K+1}(\beta)\}$ satisfying $\sum_{i=1}^{K+1} w_i(\beta) = 1$, which are assigned to the $K$ prototypes and to the pure noise component. 
The latter is bound to always receive the smallest weight, while the prototype that receives the highest weight is understood as the dominant prototype of that vector.
For each vector, roughly $w_l(\beta)F$ features are randomly assigned to contribution $l$ and the associated traits are generated accordingly, by either being copied from the respective prototype or being randomly generated, depending on the type of contribution that $l$ stands for.
Here, $\beta \in (0,1)$ is a free model parameter controlling the overall/expected strength of the groups and mixing (stronger groups and weaker mixing for higher $\beta$),
together with the shape of the joint probability distribution from which the weights are effectively sampled.
This is essentially a $\beta$-dependent probability distribution taking as support the volume of the regular $(K)$-simplex spanned by a vector $\vec{w}$ taking the weights as its entries. 
MPG employes a pragmatic, computational sampling procedure that does not explicitly state nor or use the associated joint distribution. 
While this procedure is explained in Ref.~\cite{Babeanu_2}, it is worth providing here an intuition of the role that $\beta$ plays:
the joint distribution places most emphasis on the vertices of the simplex when $\beta \rightarrow 1$,
while placing most emphasis to the center of the simplex when $\beta \rightarrow 0$.
The former extreme corresponds to having one weight of almost $1.0$, while the other are almost $0.0$ (very strong groups and boundaries, very weak mixing).
The latter extreme corresponds to having all weights almost equal to $1/(K+1)$ (very weak groups and boundaries, very strong mixing).

In practice, for an intermediate $\beta$ value, a generated MPG vector effectively falls under one of $K$ possible types, or groups, depending on its dominant prototype.
However, the flexibility of the weights associated to different prototypes make the boundaries between groups rather soft.
Moreover, within each group, the vectors exhibit significant variability in terms or how close they actually are to the prototype, 
variability associated to the ($\beta$-dependent) marginal distribution of the largest weight.
In turn, this gives rise to the internally non-uniform nature of MPG groups, containing vectors that are arbitrarily close to or far from the ``core''. 

The above description of MPG is conditional on cultural prototypes already being fully specified in terms of the traits they pick for every feature.  
In Ref.~\cite{Babeanu_2}, the $K$ prototypes themselves are also randomly generated during a preliminary step,
according to a procedure that uses another parameter $\alpha$ controling for the separation between prototypes.
Here, instead, $K=2$ prototypes are manually defined and kept fixed, in a manner that allows for direct comparisons with the S2G and FCI models introduced and studied in Sec.~\ref{Theor_1} and Sec.~\ref{Theor_2}.
Specifically, a cultural space consisting of $F=130$ binary features is used, 
where the two (maximally-dissimilar) prototypes are filled entirely with ``$-1$'' and ``$+1$'' traits respectively (the same labeling convention as in Sec.~\ref{Theor_1}), 
so that a binary, symmetric group structure is induced.
These prototypes coincide with the central/representative vectors of the two S2G groups, 
as well as with the unique spin configurations corresponding to the two extremes of the $F_{+}$ axis in Fig.~\ref{ProbDist}. 
Although S2G also has a binary, symmetric group structure, the S2G probability of generating vectors identical with the extreme configurations effectively vanishes, as long as the group strength parameter $\nu$ is not too low (group strength and correlation level not too high).
As suggested by the above explanations and shown by Fig.~\ref{ProbDist_MPG}, this is not the case for MPG. 
Just like for FCI and S2G, the binary, symmetric setting used for MPG gives rise to the expectation that only one structural mode would be present, 
with an associated $\lambda_2$ eigenvalue increasing with increasing $\beta$.
Fig.~\ref{EntVal_MPG} and Table~\ref{TabPars_MPG} confirm this expectation.

A multitude of values are selected for the $\beta$ parameter (listed in Table~\ref{TabPars_MPG}) and one cultural state is generated for each of these values. 
For each of these states, the enhanced eigenpair analysis from Sec.~\ref{Empir_2} is carried out,
showing the presence of only one structural mode, whose eigenvalue increases with $\beta$.
The two plots in Fig.~\ref{EntVal_MPG} illustrate this analysis for $\beta=0.70$ and $\beta=0.80$.
For these values, the structural mode is, respectively, below and above the r-random expectation, in terms of eigenvector uniformity. 
Note that this expectation band is crossed for a very high (and extremely significant) subleading eigenvalue $\lambda_2 \approx 15$.
For comparison, the FCI and S2G models exhibit similar crossings already when $\lambda_2 \approx 5$ and $\lambda_2 \approx 1$ respectively, 
as revealed when combining the information in Fig~\ref{Val2} and Fig~\ref{Ent_by_Ent2}.
The comparison between MPG and S2G confirms that the the binary group structure induced by the former is very different from that induced by the latter, 
as the structural mode needs to be almost $\approx 15$ times stronger in order to exhibit significant eigenvector uniformity.
The comparison between MPG and FCI confirms that the eigenvector uniformity criterion is entirely inappropriate for validating 
groups like those induced by MPG, as the structural mode needs to be $\approx 3$ times stronger in order to exhibit significant eigenvector uniformity --
the criterion is much less likely to (correctly) identify structural modes based on mixed prototypes as authentic then (erroneously) identify structural modes based on feature-feature redundancies as authentic.

\begin{table}
\begin{center}
  \begin{tabular}{ C{1.4cm} || C{1.4cm} || C{1.4cm} | C{1.4cm} | C{1.4cm} }
    \hline
    $\beta$ & $\tilde{C}$ & $z(\lambda_2)$ & $z(H_2)$ & $z(H_2')$   \\ \hline \hline
    0.20        &   0.00    &   0.41    &   -4.05   &   -0.33   \\ 
    0.40        &   0.04    &   20.92   &   -13.36  &   0.18    \\
    0.60        &   0.12    &   79.07   &   -6.24   &   -0.07   \\
    {\bf 0.70}  &   0.21    &   182.83  &   -4.50   &   -1.98   \\
    0.75        &   0.26    &   242.82  &   -1.44   &   -1.44   \\
    {\bf 0.80}  &   0.34    &   303.79  &   2.90    &   2.90    \\
    0.85        &   0.44    &   410.63  &   4.90    &   4.90    \\
    0.90        &   0.57    &   468.78  &   8.25    &   8.25    \\
\end{tabular}
	\caption{Relevant estimates for mixed prototypes generation (MPG). 
	The first column shows values of the MPG $\beta$ parameter controlling the strength of prototype mixing. 
	The second column shows values of the estimated feature-feature correlation level $\tilde{C}$,
	which is numerically computed by generating $20000$ MPG vectors for each $\beta$ value.
	The last three columns show z-scores of the 
        subleading eigenvalue $z(\lambda_2)$, 
        the uniformity associated to the subleading eigenvalue $z(H_2)$ and
        the subleading uniformity $z(H_2')$.
    The $\lambda_2$, $H_2$ and $H_2'$ values are extracted from one MPG cultural state of $N=100$ vectors generated for each $\beta$ value,
    while the associated z-scores are computed with respect to the r-random null model, from which $n=1000$ random matrices are sampled for each $\beta$ value.
	All estimates are valid for $F=130$ binary features.
	The highlighted values of $\beta$ indicate direct correspondences with the plots in Fig.~\ref{EntVal_MPG}.}
	\label{TabPars_MPG}
\end{center}
\end{table}

Such insights become even clearer when inspecting Table~\ref{TabPars_MPG}, which shows how several relevant quantities depend on the MPG mixing parameter $\beta$.
In particular, the second column shows a numerical estimate $\tilde{C}$ of the feature-feature correlation level, based on $N=20000$ MPG vectors generated for each value of $\beta$.
The value of $\tilde{C}$ is obtained by averaging over the $F/2=65$ Pearson correlation values corresponding to all pairs of consecutive features 
-- although this estimator might be biased, it should be consistent (asymptotically approach the true value in the limit of $N\rightarrow\infty$); 
unlike FCI and S2G, MPG does not seem to allow for an exact, analytic, full-ensemble formula for computing $C$.
The following three columns show the z-scores of the subleading eigenvalue $\lambda_2$, the associated eigenvector entropy $H_2$ and the subleading eigenvector entropy $H_2'$. 
For each $\beta$ value, $\lambda_2$, $H_2$ and $H_2'$ are extracted from one MPG cultural state of $N=100$ vectors (the MPG state used in Fig.~\ref{EntVal_MPG}, when $\beta=0.70$ and $\beta=0.80$),
while the associated z-scores are computed with respect to the subleading eigenvalues (for $\lambda_2$) and subleading eigenvector entropies (for $H_2$ and $H_2'$) of $n=1000$ r-random matrices suitable for the respective MPG state (the same r-random matrices on which the uncertainty bands of Fig.~\ref{EntVal_MPG} are based, when $\beta=0.70$ and $\beta=0.80$).

One notices in Table~\ref{TabPars_MPG} the increase of the eigenvalue significance $z(\lambda_2)$ of the MPG structural mode with increasing $\beta$ and increasing correlation $\tilde{C}$. 
Only for $\beta = 0.75$ does the eigenvector entropy of the structural mode qualify as the subleading eigenvector entropy, while still smaller than the r-random expectation, since $z(H_2) = z(H_2') < 0$.
When $\beta=0.80$, the structural mode also becomes significant in terms of eigenvector entropy, with respect to the r-random expectation, as $z(H_2) = z(H_2') > 2.0$, while this significance further increases for higher $\beta$ values.
A transition takes place somewhere between $\beta = 0.75$ and $\beta=0.80$, so that for higher $\beta$ the two, symmetric groups induced by MPG, captured by the $\lambda_2$ structural mode start exhibiting internal uniformity that becomes statistically recognizable via eigenvector entropy.
Note that the correlation-level $\tilde{C} \approx 0.2$ corresponding to this transition is much higher than that associated to the S2G phase transition $C \approx 0.01$ (Fig.~\ref{SBPT}-right and Fig.~\ref{Ent_by_Ent2}-blue) and higher even than that associated to the FCI phase transition $C \approx 0.08$ (Fig.~\ref{SBPT}-left and Fig.~\ref{Ent_by_Ent2}-red).

With these results in mind, it is instructive to see how MPG vectors are distributed in terms of their number $F_{+}$ of ``$+1$'' traits. 
This is shown in Fig.~\ref{ProbDist_MPG}, for several $\beta$ values associated to MPG states described by Table~.\ref{TabPars_MPG}, 
two of which ($\beta=0.70$ and $\beta=0.80$) are also present in Fig.~\ref{EntVal_MPG}. 
As a reminder, each of the two extremes ($F_{+} = 0$ and $F_{+} = F$) of the horizontal axis is compatible with only one possible configuration of traits, 
which perfectly matches one of the two prototypes, so the location on the axis also determines the separation from the two prototypes: $\max(F_{+}, F-F_{+})$ is effectively the similarity with the dominant prototype of a vector sitting at $F_{+}$.
As expected, the three higher $\beta$ values, which exhibit a significant $\lambda_2$ (see $z(\lambda_2)$ column in Table~.\ref{TabPars_MPG}), 
also show $F_{+}$ distributions that are very broad and flat, when compared to the S2G (and even to the FCI) distributions in Fig.~\ref{ProbDist}.
Although such comparisons are somewhat obscured by MPG distributions being shown for different correlation levels than S2G (and FCI) distributions (a correlation-based correspondence like that inherent in Fig.~\ref{ProbDist} is hard to achieve, since analytic calculations are much harder for MPG), the trend is clear: 
MPG groups are internally much more flexible than those of S2G, in terms of the separation of the vectors from the group cores, 
which makes them exhibit low eigenvector uniformity.
Compared to S2G and even to FCI, the MPG correlation level needs to be much higher in order to exhibit a bi-modal-like $F_{+}$ distribution,
in the form of an accumulation of vectors very close to the two prototypes, which is visible for $\beta=0.80$ and obvious for $\beta=0.90$.
This accumulation is responsible for the significant eigenvector uniformity exhibited by states with $\beta=0.80$ or higher. 

On one hand, one might also notice the presence of two, small probability peaks close to the center of the $F_{+}$ axis in Fig.~\ref{ProbDist_MPG}, for $\beta=0.70$ and $\beta=0.80$.
This is likely a consequence of MPG being formulated in a somewhat arbitrary way (the lack of an explicit, analytic control of the joint weight distribution, the presence of the pure noise component), which is inherited from Ref.~\cite{Babeanu_2} -- future research on internally non-uniform groups and/or mixing prototypes would likely benefit from a revised version of MPG.  
In any case, these peaks cannot drive up the uniformity, since the eigenvector entries of associated configurations are relatively weak: there are many other configurations further to the extremes of the axis, which are closer to the group cores and to each other.
By contrast, the symmetry-breaking peaks of S2G and FCI (Fig.~\ref{ProbDist}) do drive up the uniformity, since there is a vanishing number of configurations further to the extremes, even if the peaks themselves also arise quite close to the center.
On the other hand, the smaller discontinuities in the shapes of the MPG distributions are due to fluctuations inherent in the numerical sampling on which the estimation is based and the histogram binning: $n=20000$ sampled vectors are divided among 131 values of $F_{+}$ which are divided among 41 bins. 

It is thus possible to construct groups that are internally strong but also non-uniform,
which exhibit high and significant eigenvalues but low and non-significant eigenvector entropies.
Such groups would not be recognized as authentic by the analysis applied in Sec.~\ref{Empir_2}, although they are more plausible as manifestations of real-world ideologies than internally-uniform groups.
This is clearly illustrated by the mixed prototypes scenario, which builds on theoretical considerations from social science, while also exhibiting properties that are generically compatible with empirical data~\cite{Babeanu_2}.  

\subsection{Redundant feature elimination}\label{Feat_Elim}

We have repeatedly emphasized during previous sections that empirical cultural states exhibit arbitrary feature-feature redundancies that have to do with how the underlying survey is designed rather than with system-specific properties.
As shown in previous work~\cite{Valori, Stivala, Babeanu_1}, such redundancies are often visible at the level of the feature-feature correlation matrix.
Although correlations between features can be due either to groups or to redundancies (see Sec.~\ref{Theor_1}), for some datasets, subsets of obviously redundant features may be easily identified by inspecting the correlation matrix. 
This allows for eliminating these obvious redundancies and investigating the stability of structural modes under this operation.
Stable structural modes are much more likely to be due to system-specific groups. 

\begin{figure}
\centering
	\includegraphics[width=8.5cm]{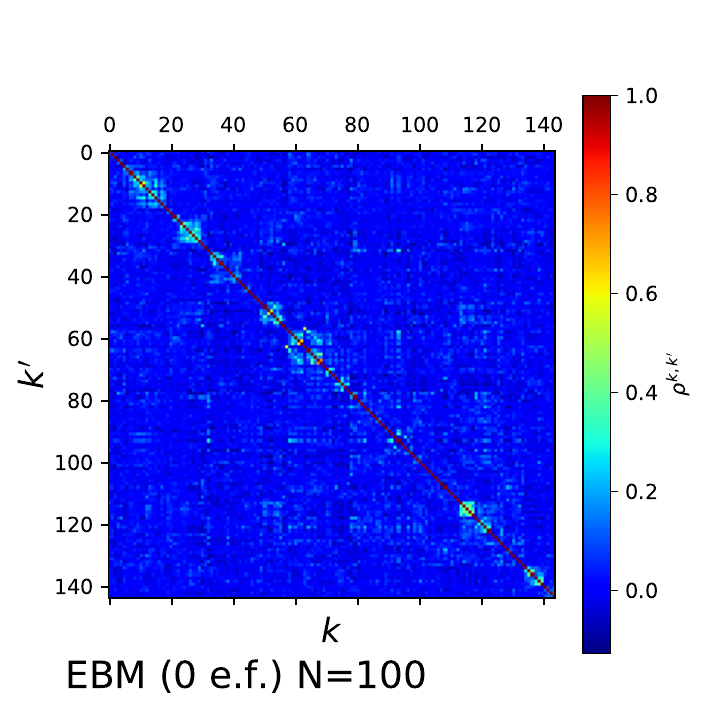}
	\caption{Matrix of feature-feature correlations for empirical data.
	Each square shows the evaluated correlation for a pair $(k,k')$ of the $F=144$ features available 
	for the same Eurobarometer (EBM) data (with $N=100$) used in Fig.~\ref{EntVal_EBM} -- no eliminated features (0 f.e.).
}
\label{CorMat_EBM}
\end{figure} 

This section focuses on the Eurobarometer (EBM) dataset, for which the redundancies between features are most obvious.
We start by illustrating this with Fig.~\ref{CorMat_EBM}, showing the matrix of correlations between the $F=144$ features, 
based on the $N=100$ cultural state used above. 
Notice the block-diagonal form of the matrix, signaling the presence of multiple blocks of consecutive features that are highly correlated with each other.
The features within each block actually correspond to survey questions that are concearned with different aspects of the same topic.
For instance, within the block corresponding to features $k,k' \in \overline{9,19}$, the items measure attitudes with respect to $11$ policy proposals that were part of the Maastricht Treaty, which aimed at enhancing integration within the European Union.
Thus, these blocks are clearly due to survey-dependent redundancies between features. 

It is worth mentioning at this point that the matrix entries in Fig.~\ref{CorMat_EBM} are not computed based on the standard, 
Pearson correlation formula, since this is not appropriate when both nominal and ordinal variables are present. 
Instead, the values are based on a variant of Pearson correlation that uses the feature-level similarity values associated to different pairs of cultural vectors, rather than using actual traits attained by different vectors. 
Formally, for a pair of features $(k,k')$, this modified correlation reads:

\begin{equation}
	\rho^{k,k'} = \frac{\sigma^{k,k'}}{\sqrt{\sigma^{k,k}\sigma^{k',k'}}},
\end{equation}
where the associated variance-covariance matrix is given by: 

\begin{equation}
	\sigma^{k,k'} = \langle s_{ij}^k s_{ij}^{k'} \rangle_{i,j \in \overline{1,N}}^{i < j} - \langle s_{ij}^k \rangle_{i,j \in \overline{1,N}}^{i < j} \langle s_{ij}^{k'} \rangle_{i,j \in \overline{1,N}}^{i < j},
\end{equation}
where $s_{ij}^k$ is the similarity associated to a single feature $k$ -- its expression is visible upon eliminating the averaging over $\overline{1,F}$ in Eq.~\eqref{CultSim} -- and in all cases the averaging is performed over all distinct pairs $(i,j)$ of cultural vectors.
This modified correlation is effectively identical to that used in Refs.~\cite{Babeanu_1,Valori}, 
where it is formulated in terms of (feature-level) cultural distances $d^k_{ij}$ instead of similarities $s^k_{ij}$, 
formulation which is mathematically equivalent, due to the simple, linear relationship ($d^k_{ij} = 1-s^k_{ij}$) between distances and similarities. 
Because of the unconventional formulation, the correlation values in this section cannot be directly compared to those in Sec.~\ref{Theor_1}, Sec.~\ref{Theor_2} and Sec.~\ref{Mixed_Protos}, which rely on a conventional Pearson formulation, which is appropriate for a cultural space built entirely from binary features.

The next step is to eliminate features from the EBM dataset so that the redundancy blocks in Fig.~\ref{CorMat_EBM} are no longer present.
We describe here a deterministic procedure/algorithm that does this in a sequential way, so that a specified number of features are eliminated one by one. 
Let the (dynamic) set of features $\mathcal{F}$ initially contain the integer labels of all features: $\mathcal{F} = \overline{1,F}$.
At each step, the feature $k^* \in \mathcal{F}$ that is ``the most correlated'' is identified, according to the following criterion:  

\begin{equation}
    \label{pick_feature}
    k^* =   \underset{k \in \{k_1,k_2\}}{\argmax}\ 
            \left[ \underset{
                k' \in \mathcal{F} - \{k_1,k_2\}
            }{\max}\ \rho^{k,k'} \right],
\end{equation}
where $k_1$ and $k_2$ are jointly defined by:

\begin{equation}
    \label{pick_pair_features}
    (k_1,k_2) = \underset{(k,k')}{\argmax}\ \rho^{k,k'},
\end{equation}
where $k, k' \in \mathcal{F}$ and $k < k'$. 
Feature $k^*$ is then eliminated from the dynamical set: $\mathcal{F} := \mathcal{F} - \{k^*\}$.
The procedure continues with the next step, unless the desired number of eliminated features (n.e.f.) has already been achieved.
In a less formal language, at each step in the iteration, from the set of surviving features $\mathcal{F}$,
one identifies the pair of distinct features $(k_1, k_2)$ exhibiting the largest correlation -- Eq.~\eqref{pick_pair_features}.
Among these two features, one eliminates the one that exhibits the largest correlation with any other surviving feature different from $k_1$ and $k_2$  -- Eq.~\eqref{pick_feature}. 
Note that one can think of other criteria for identifying ``the most correlated'' feature $k^*$  at each step in the algorithm, 
since each feature will generally exhibit a different correlation value with each of the other features in $\mathcal{F}$.
The criterion Eq.~\eqref{pick_feature} and Eq.~\eqref{pick_pair_features} represents a pragmatical, greedy-type approach that we believe is suitable for the current analysis.

\begin{table*}
\begin{center}
  \begin{tabular}{ C{1.0cm} || C{1.5cm} | C{1.5cm} || C{1.5cm} | C{1.5cm} || C{1.5cm} | C{1.5cm} || C{1.5cm} | C{1.5cm} }
    \hline
    n.e.f. & $z(\lambda_2)$ & $z(H_2)$ & $z(\lambda_3)$ & $z(H_3)$ & $z(\lambda_4)$ & $z(H_4)$ & $z(\lambda_5)$ & $z(H_5)$   \\ \hline \hline
    0    &   70.38   &   -6.24   &   21.37   &   -1.68   &   7.54    &   -5.25   &   5.89    &   -7.17   \\ 
    10   &   62.48   &   -8.05   &   20.42   &   -1.45   &   5.49    &   -1.61   &   3.17    &   -7.09   \\
    20   &   55.08   &   -9.44   &   17.20   &   -1.34   &   3.31    &   -5.89   &   1.19    &   -7.64   \\
    30   &   47.35   &   -10.80  &   13.68   &   -3.06   &   2.65    &   -3.71   &   0.71    &   -6.61   \\
    40   &   40.86   &   -9.44   &   8.23    &   -3.47   &   0.48    &   -3.16   &   -0.07   &   -6.99   \\
    50   &   33.91   &   -10.40  &   6.13    &   -2.89   &   0.78    &   -0.40   &   -0.58   &   -6.74   \\
    60   &   30.00   &   -11.67  &   5.58    &   -3.66   &   0.50    &   -1.11   &   -2.14   &   -3.24   \\
\end{tabular}
	\caption{Robustness of structural modes under redundant feature elimination. 
	The first column shows the number of eliminated features (n.e.f.). 
	Each of the following 4 pairs of columns corresponds to one of the structural modes in Fig.~\ref{EntVal_EBM}.
    The 2 columns show, respectively, 
	the eigenvalue z-score $z(\lambda_l)$ and the eigenvector entropy z-score $z(H_l)$ associated to the $l$'th empirical eigenvalue.
    The z-scores are computed with respect to the r-random null model, from which $n=1000$ random matrices are sampled for each number of eliminated features.
    The calculations are based on the Eurobarometer (EBM) data with $N=100$ used in Fig.~\ref{EntVal_EBM}.
}
	\label{TabPars_FeatElim}
\end{center}
\end{table*}

\begin{figure}
\centering
	\includegraphics[width=8.5cm]{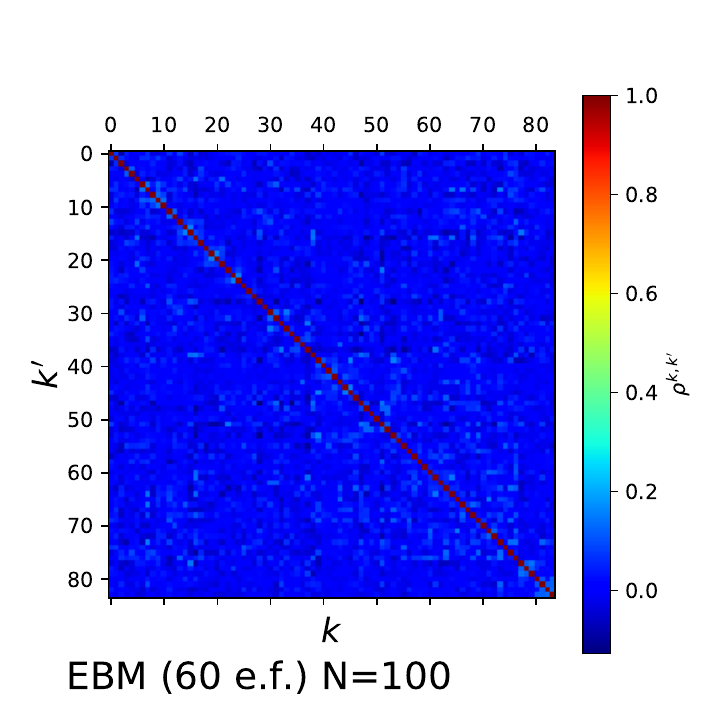}
	\caption{Matrix of feature-feature correlations for empirical data after extensive elimination of redundant features.
	Each square shows the evaluated correlation for a pair $(k,k')$ of the $F-60=84$ features available 
	for the Eurobarometer (EBM) data (with $N=100$) used in Fig.~\ref{EntVal_EBM},
	after the $60$ most redundant features are eliminated (60 e.f.).
}
\label{CorMat_EBM_FeatElim}
\end{figure}

Table~\ref{TabPars_FeatElim} illustrates the behavior of interesting EBM eigenmodes upon gradually increasing the number of features eliminated with the above procedure.
The focus is on the eigenmodes associated to eigenvalues $\lambda_2, \lambda_3, \lambda_4, \lambda_5$, which are the EBM structural modes (see Fig.~\ref{EntVal_EBM}) when the n.e.f (first column) is still zero (first line). 
The table shows the statistical significance (z-scores) of the eigenvalue and eigenvector entropy associated to these modes, computed with respect to the r-random null model.
In terms of eivenvalue z-scores, 
one can see that $\lambda_5$ and $\lambda_4$ become compatible with r-randomness when n.e.f. reaches a value of $20$ and $40$ respectively,
while $\lambda_2$ and $\lambda_3$ remain incompatible with restricted randomness even when n.e.f. reaches $60$ --
note that the block diagonal form of the correlation matrix is no longer recognizable after eliminating $60$ features, as shown by Fig.~\ref{CorMat_EBM_FeatElim}.
In terms of eigenvector entropy, as expected, all four modes remain compatible with r-randomness, as indicated by the negative values of the associated z-scores. 
These results suggest that the two weakest EBM structural modes ($\lambda_4, \lambda_5$) are artifacts of feature redundancies, while the two strongest ones ($\lambda_2, \lambda_3$) are signatures of authentic grouping tendencies, although they (consistently) fail to exhibit any eigenvector uniformity.

\begin{figure}
\centering
	\includegraphics[width=8.5cm]{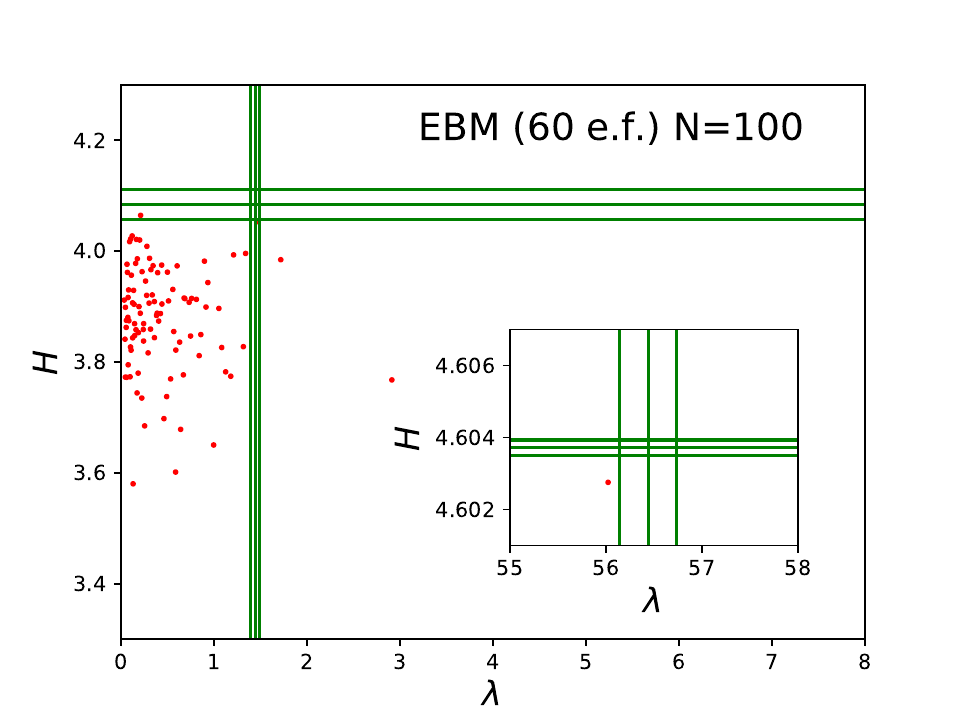}
	\caption{Empirical eigenvalues and eigenvector entropies after extensive elimination of redundant features.
	The figure illustrates the same type of eigenpair analysis as Fig.~\ref{EntVal_EBM}, 
	for the same Eurobarometer (EBM) data with $N=100$ cultural vectors used, 
	but after eliminating the $60$ most redundant of the $F=144$ features available initially (60 e.f.).
}
\label{EntVal_EBM_FeatElim}
\end{figure}

At this point, it is essential that the statistical significance of the two, stronger structural modes is still clear after all the obvious feature redundancies are eliminated. 
It is much less important that their eigenvalue significance decreases to a certain extent under feature elimination,
which is actually compatible with one would expect from authentic structural modes.
On one hand, this has to do with a decrease of the eigenvalues which is conceivably due to the fact that some features that are eliminated also store information about authentic cultural groups, despite exhibiting strong redundancies with other features (note that feature elimination is not carried out based on noisiness considerations).
On the other hand, this has to do with an increase of the upper boundary of the random bulk -- in this case, the r-random expectation for $\lambda_2$ -- 
when reducing $F$ while keeping $N$ constant, 
which is compatible with naive extrapolations from the (much better understood) behavior of random correlation matrices. 
Both aspects seem jointly responsible for the effective decrease in ``discrimination power'' encoded by the decrease of eigenvalue z-scores:
the presence of both effects is confirmed by Fig.~\ref{EntVal_EBM_FeatElim}, 
which shows smaller empirical $\lambda_2$ and $\lambda_3$ and higher r-randomness $\lambda_2$ than Fig.~\ref{EntVal_EBM} (before feature elimination).

It is important to emphasize that the feature-elimination procedure used in this section is mostly meaningful for the current context and purpose: 
using EBM data, it aids the idea that structural modes identified by random matrix theory can be authentic,
even if they do not exhibit an eigenvector uniformity that is higher than the null model expectation. 
The procedure does not qualify as a general approach for validating structural modes, since many datasets affected by redundancies might not exhibit an obvious, block-diagonal structure of the feature-feature matrix.
Moreover, if the survey designer has a priori intuition about and interest in the cultural groups that exist in the society that is being measured, 
the variables might actually exhibit (due to deliberate design or unconscious bias) a grouping into subsets that are associated to different cultural groups, 
which could induce a block-diagonal structure in the feature-feature matrix.
However, the features in a block would then retain much valuable information about a system-specific group, so eliminating them would be counterproductive. 
In such a situation, the authentic-vs-artifactual question translates to a question of whether the latent construct behind a certain block is well aligned with a systemic grouping tendency or not.
Although very interesting, such problems and questions are beyond the purpose of the current study. 

\section{Discussion}\label{Disc}

This was the first study where empirical matrices of cultural similarity between individuals were analyzed from a random matrix perspective, 
allowing for a separation of structurally irrelevant eigenmodes from the structurally relevant ones. 
The statistical significance of the latter, here referred to as ``structural modes'', was demonstrated in Sec.~\ref{Empir_1}, 
using a detailed numerical approach of explicitly sampling configurations from three null models. 
Among these three, the ``restricted randomness'' model, first proposed here, was concluded to be the most appropriate for later use. 
Restricted randomness enforces, in a flexible way, the non-uniformity inherent in each cultural feature, as this is assumed to be mostly a consequence of experimental design rather than a consequence of system-specific structure. 
This null model thus reproduces well the leading eigenvalue of the empirical matrix, which is interpreted as the ``global mode''. 
By using this null model, meaningful empirical structure is implicitly associated to inhomogeneities present in the cultural space distribution~\cite{Babeanu_1, Babeanu_2}
that cannot be expressed in terms of the feature-level inhomogeneities. 

A central question for the rest of the study was whether the structural modes identified in Sec.~\ref{Empir_1} are just signatures of redundancies between cultural features or, more interestingly, signatures of cultural groups.
The former hypothesis was based on the idea that some of the items in the questionnaire are semantically related to each other. 
The latter hypothesis was based on the idea of coexistence, within the geographical region from which the empirical data was obtained, 
of several types of individuals, where each type could correspond, for instance, to a certain political affiliation,
assuming that each affiliation comes along with a certain set of values, opinions or beliefs.
Even more interesting was the possibility that structural modes correspond to groups that form around cultural prototypes~\cite{Stivala, Babeanu_2} associated to a small number of universal ``rationalities'' or ``ways of life''~\cite{Thompson}.
This hypothesis had been shown to be compatible with some generic structural properties of culture, provided that prototype mixing is in place~\cite{Babeanu_2}. 
Note that individuals that are strongly associated to one cultural group do not need to know each other and not even to be close to each other in terms of geography or social network, but just to have a significant ideological overlap. 

We approached the ``groups or redundancies?'' question by designing, in the simplest possible setting, two probabilistic toy models that implement the ``redundancies'' scenario and the ``groups'' scenario (see Sec.~\ref{Theor_1}), 
named ``FCI'' (Sec.~\ref{FCI}) and ``S2G'' (Sec.~\ref{S2G}) respectively. 
The FCI model only enforces feature-feature couplings, in a manner that does not introduce any unintended assumption, by means of a maximum-entropy formulation~\cite{Jaynes}.
This is meant to ``simulate'' an overall level of pairwise overlap between the questions of a hypothetical survey, assuming that the hypothetical system from which the answers are obtained is otherwise maximally random. 
We have shown that there is a non-vanishing correlation interval (Sec.~\ref{FCIvS2G}) for which the S2G model induces a bimodal $F_{+}$-distribution, while the FCI model induces a unimodal $F_{+}$-distribution, where $F_{+}$ is a 1-dimensional projection of the cultural space that is very informative for that context. 
We argued that, for this interval, the groups of S2G are obvious, while the feature-feature couplings of FCI too weak to imitate such groups -- 
when peaks are present, most vectors sampled from the distribution can be unambiguously assigned to one of two categories, based on their $F_{+}$ value.
The boundaries of this interval are well defined, via the symmetry breaking phase transition of S2G on the low-correlation side and the one of FCI on the high-correlation side. 

This correlation interval was exploited (Sec.~\ref{Theor_2}) for understanding how the presence or absence of groups becomes visible via spectral analysis. 
For both FCI and S2G, one eigenvalue was shown to become increasingly separated from the random bulk when increasing the level of correlation between features. 
However, this increasing trend was shown to be, up to statistical errors arising from finite sampling, exactly the same for the two models, 
even for the above-mentioned correlation region.
This showed that the presence of deviating eigenvalues in empirical data is not a certain signature of group structure.
The difference between the two scenarios became visible when calculating the uniformities of the eigenvectors by means of ``eigenvector entropy'' (inspired by Ref.~\cite{Patil}, where it is called ``information entropy'').
There is one eigenvector uniformity that, for an increasing level of correlation, becomes increasingly separated from the random bulk. 
This increasing trend is significantly different for the two models, while starting in an abrupt way and replicating well, for each model, the phase transition expected on theoretical grounds.
Thus, for the interesting correlation region, S2G shows a deviating eigenvector uniformity, while FCI does not. 
This suggested that empirical eigenmodes corresponding to authentic groups should exhibit not only an eigenvalue that is significantly higher than the null model expectation, 
but also an eigenvector uniformity that is significantly higher than the null model expectation.  

This motivated a more detailed investigation of empirical data in Sec.~\ref{Empir_2},
which showed that all empirical eigenvalues that are significantly higher than what can be expected based on restricted randomness 
are associated to eigenvector uniformities that are not significantly higher than what can be expected based on the same null model. 
This suggests that empirical deviating eigenvalues are artifacts of arbitrary redundancies between features, 
which are known to exist in the datasets that we use~\cite{Babeanu_1}.

Based on the results in Sec.~\ref{Empir_2}, one may have even been tempted to reject altogether the existence of cultural groups, 
along with the ``cultural prototypes'' hypothesis previously showing promising results~\cite{Stivala, Babeanu_2} -- where each prototype is associated to one group.
However, Ref.~\cite{Babeanu_2} had clearly shown that this hypothesis is structurally compatible with empirical data only when prototype ``mixing'' is enforced:
cultural vectors associated to different individuals are random hybrids of the prototype vectors.
In turn, this idea goes along with a high amount of stochastic freedom for the strengths with which each vector couples to the different prototypes.
As a consequence, the ``mixed prototypes generation'' (MPG) can simultaneously give rise to a relatively broad ``spectrum of vectors'', from those
that are very biased towards one of the prototypes, to those that are highly balanced combinations of all the prototypes and anything in between.
In turn, this leads to groups that are internally non-uniform, where each group contains vectors that are arbitrarily central, arbitrarily peripheral and anything in between.
This is at odds with groups induced by S2G (which does not incorporate mixing),
whose vectors are highly localized around a specific separations from group cores, as illustrated Sec.~\ref{FCIvS2G}.
One can thus say that S2G groups are internally uniform, unlike those of MPG.

As argued in Sec.~\ref{Non_Unif_Group}, internally non-uniform groups appear likely to induce structural modes that do not exhibit significant eigenvector entropy/uniformity, just like the empirical ones (Sec.~\ref{Empir_2}).
This has been demonstrated in Sec.~\ref{Mixed_Protos}, which made use of the MPG model, adapted to the binary-feature setting used for FCI and S2G:
the deviating eigenmode capturing the group structure induced by two mixing prototypes has an eigenvector entropy which is entirely compatible with expectations based on restricted randomness. 
The eigenvector entropy departs from null model expectations only when the free MPG parameter attains values that correspond to a very high (compared to FCI and S2G) feature-feature correlation levels $C$.
This can be well understood in terms of the shape of the $F_{+}$ distribution (Fig.~\ref{ProbDist_MPG}) induced by MPG:
compared to S2G (Fig.~\ref{ProbDist}), this distribution shows much weaker decays when approaching the $F_{+} = 0$ and the $F_{+} = F$ endpoints, 
for any given $C$ that is reasonably small.
For MPG, probability peaks associated to significant eigenvector uniformity are formed directly on the two ($F_{+} = 0$ and $F_{+} = F$) endpoints when the correlation level is high enough.
This is unlike S2G and FCI, which develop probability peaks closer to the middle of the $F_{+}$ axis for smaller $C$ values 
and gradually approach the extremes as $C$ increases. 
Below the critical $C$ value of MPG, the vectors composing each of the two groups have highly different levels of ``centrality'' within the group, 
leading to highly diverse entries in the structural eigenvector, which thus exhibits low entropy. 
Complementary to Sec.~\ref{Mixed_Protos}, Sec.~\ref{Feat_Elim} provided data-driven indications that some of the structural modes identified in empirical data are due to internally non-uniform groups rather than to feature-feature redundancies.
This made use of the Eurobarometer cultural set, which allows for easy identification of blocks of highly redundant features. 
Upon sequentially eliminating such features, two of the structural modes were shown to be highly robust. 

In relation to the MPG-based analysis in Sec.~\ref{Mixed_Protos}, one could be tempted to interpret the absence of two, well defined peaks (like those of S2G) in the $F_{+}$ distribution (Fig.~\ref{ProbDist_MPG}) as an absence of groups below the critical $C$ value.
Indeed, the results of Sec.~\ref{Theor_1} and Sec.~\ref{Theor_1} may seem to implicitly suggest a (re)definition of groups via peaks in the $F_{+}$ distribution.
We argue that this interpretation is not appropriate. 
On one hand, even for low $C$ value, the MPG distribution gives significantly more emphasis to the extremes of the $F_{+}$ axis than either S2G or FCI. 
On the other hand, $F_{+}$ values that are closer to these extremes carry much fewer possible configurations which are, on average, more similar to each other.
This means that MPG generally induces a relatively high density in cultural space around these extremes (the prototypes), 
effectively providing ``hard cores'' for its groups, which otherwise have relatively ``soft external boundaries''.
By contrast, S2G induces groups with ``hard external boundaries'' and ``hollow cores'', which are thus easily recognizable via eigenvector uniformity.
Thus, MPG comes with a somewhat different meaning for ``groups'' and ``group structure'' than that implicit in S2G,
which is conceptually more compatible with what one would expect from cultural groups in the real world, 
regardless of whether they are centered on universal rationalities or on more contextual, ephemeral ideological movements,
as explained at the beginning of Sec.~\ref{Non_Unif_Group}.

One may also wonder, from a modeling perspective, whether internally non-uniform groups strictly require the mixing prototypes mechanism and whether the latter unavoidably induces the former.  
On one hand, it appears conceptually very hard to implement the mixing mechanism in a manner that induces uniform groups or that does not induce any groups at all -- at least without introducing highly arbitrary/implausible assumptions.
On the other hand, it appears possible to define alternative procedures that are capable of generating non-uniform groups without making (explicit) use of mixing.
Exploring such alternatives and their empirical validity is beyond the purpose of this study, 
which did not aim at further empirical validation of the mixing ingredient, 
but just at using it as an easy, accessible way (due to availability from previous research) of generating internally non-uniform groups. 

The fact that this study used multidimensional sociological data, while heavily relying on eigenvalue decomposition,
may raise the question of how the approach here is different from traditional social science research using principal component analysis~\cite{Dunteman}.
Although principal component analysis heavily relies on eigenvalue decomposition, 
in a social science context, the former most often implies a decomposition of the matrix of covariances or correlations between the variables,
while this study focuses on the matrix of similarities between individuals. 
This actually makes the approach here conceptually more similar to clustering methods~\cite{Kaufman}, which aim at identifying group structure, while providing an optimal clustering of the given set of individuals. 
However, these methods do not attempt to decompose the similarity matrix and remove the irrelevant eigenmodes. 
In fact, following the approach of Ref.~\cite{MacMahon}, the sum of the similarity matrix contributions associated to the structural modes identified here can be interpreted as a (modified) modularity matrix, 
which could provide a new method for clustering individuals via maximization of what one may call ``spectral modularity''. 
Since this automatically eliminates the noise components and the common trend encoded in the global mode, such a method should be able to disentangle clusters that are not recognized by previous approaches. 
However, such a method might also be sensitive to false positive cluster splittings, due to structural modes possibly being artifacts of feature redundancies, as shown in this study 
(at this point, it is not clear whether this is also a problem for the method in Ref.~\cite{MacMahon}, intended for matrices of correlations between time series). 
There is certainly a need of finding new, eigenmode-dependent criteria for distinguishing feature redundancy artifacts from authentic groups that are internally non-uniform, criteria that are generic enough to be used with empirical data.
This would be valuable regardless of whether or not the structural modes are used for spectral modularity maximization. 
These are some of the aspects left for future research.

\section{Conclusion}\label{Conc}

This study examined cultural structure from a new angle, relying on certain notions from random matrix theory. 
This provided a filtering procedure for matrices of cultural similarity between individuals,
which eliminates, in a statistically robust way, the structurally-irrelevant components.
Much effort was dedicated to the interpretation of the remaining, structurally-relevant components. 
On one hand, these may be a consequence of redundancies between cultural variables, mainly encoding information about the experimental setup. 
On the other hand, they may be a consequence of a modular organization of culture, thus encoding information about cultural groups. 
We have shown that it makes sense to conceptually distinguish between a ``redundancies scenario'' and a ``groups scenario'', 
as well as between internally uniform and internally non-uniform groups, even when cultural variables take very few, discrete values.
Although we have been able to exclude internally uniform groups as a structural mechanism,
being able to distinguish between redundancies and internally non-uniform groups requires further research.
This would allow, on one hand, to reject or accept the idea that culture has a modular structure,
on the other hand to increase the reliability of the procedure explored here,
for the purpose of identifying subtle groups in discrete, multivariate data.

{\bf Acknowledgements:} 

The author acknowledges insightful discussions with Diego Garlaschelli, Assaf Almog, Marco Verweij, Maroussia Favre, Jason Roos, Pieter Schoonees, Santo Fortunato and Vincent Traag, as well as financial support from the Netherlands Organization for Scientific Research (NWO/OCW), in particular via grant number 314-99-400.

\bibliographystyle{unsrt}
\bibliography{Paper}{}



\appendix

\section{Proof of positive semidefiniteness}
\label{App_Proof_PSD}

This section presents a proof of the fact that cultural similarity matrices of a ``Hamming-Manhattan'' type, as defined via Eq.~\eqref{CultSim}, are all positive semidefinite.
Mathematically, the statement may be written as: $S \ge 0$ for any real vector $\vec{w}$, where $S$ is the scalar quantity defined in Eq.~\eqref{Sandwich}.
It is of great use to first reformulate the statement in terms of feature-level similarities.
Specifically, by inserting Eq.~\eqref{CultSim} in Eq.~\eqref{Sandwich} one finds that:
\begin{equation}
    \label{SandwichSplit}
    S = \frac{1}{N}\sum_{k=1}^{F} \left( \sum_{i=1}^N \sum_{j=1}^N w_i s^k_{ij} w_j \right),
\end{equation}
with:
\begin{equation}
    \label{FeatCultSim}
    s^k_{ij} =  f_{\text{nom}}^k \delta(x_i^k, x_j^k) + (1-f_{\text{nom}}^k) \left( 1 - \frac{|x_i^k - x_j^k|}{q^k-1} \right)
\end{equation}
denoting one of the $F$ single-feature similarity matrices.
It becomes clear that any aggregate similarity matrix $(s_{ij})_{i,j \in \overline{1,N}}$ is positive semidefinite
if and only if any single-feature matrix $(s^k_{ij})_{i,j \in \overline{1,N}}$ is positive semidefinite. 
The sufficiency of the latter condition results from a simple averaging over feature level inequalities of the form: $\vec{w}^T s^k \vec{w} \ge 0$ to obtain $S \ge 0$, based on Eq.~\eqref{SandwichSplit}.
On the other hand, the necessity results from the observation that, in general, an aggregate similarity matrix may use any feature configuration, including any single-feature configuration.

To complete the proof, we show that any single-feature similarity matrix is positive semidefinite,
first for the nominal (Hamming) case in Sec.~\ref{App_Proof_PSD_Nom}, second for the ordinal (Manhattan) case in Sec.~\ref{App_Proof_PSD_Ord}.
These two sections make exclusive use of the ``$s_{ij}$'' notation for entries of single-feature similarity matrices, instead of the ``$s^k_{ij}$'' notation.
This simplification is also reflected by the use of ``$x_i$'' instead of ``$x^k_i$'' when denoting the entry of cultural vector $i$ for the respective feature. 

\subsection{Nominal single-feature similarity}
\label{App_Proof_PSD_Nom}

In order to prove that a similarity matrix $(s_{ij})_{i,j\in{\overline{1,N}}}$ constructed from one, nominal feature is positive semidefinite, we show that $S \ge 0$ for any real vector $\vec{w}$, where $S$ is the scalar quantity defined in Eq.~\eqref{Sandwich}.
This translates to:
\begin{equation}
    \label{CondPosSemDefNom_1}
    \sum_{i=1}^N w_i^2 + 2 \sum_{i=1}^{N-1} \sum_{j=i+1}^N w_i w_j \delta(x_i,x_j) \ge 0,
\end{equation}
after having used the fact that $s_{ii} = 1$ and that $s_{ij} = \delta(x_i,x_j)$, resulting from Eq.~\eqref{FeatCultSim}. 

It is important to note that the nominal feature induces a clustering (or partition) $\mathcal{O}$ of the set of vectors $\overline{1,N}$, 
so that all vectors picking a specific trait belong to a certain cluster (or part) $O$.
Using this observation, together with the definition of the $\delta$-function, one may rewrite Eq.~\eqref{CondPosSemDefNom_1} as: 
\begin{equation}
    \label{CondPosSemDefNom_2}
    \sum_{O\in\mathcal{O}} \left[ \sum_{i\in O} w_i^2 + \sum_{i\in O} \sum_{j\in O-\{i\}} w_i w_j \right] \ge 0,
\end{equation}
which can be further reduced to:
\begin{equation}
    \label{CondPosSemDefNom_3}
    \sum_{O\in\mathcal{O}} \left[ \left(\sum_{i\in O} w_i \right)^2 \right] \ge 0,
\end{equation}
which is a valid statement, since the left-hand side is a sum of non-negative numbers.
This concludes the proof for the the nominal (Hamming), single-feature case.

\subsection{Ordinal single-feature similarity}
\label{App_Proof_PSD_Ord}

In order to prove that a similarity matrix $(s_{ij})_{i,j\in{\overline{1,N}}}$ constructed from one, ordinal feature is positive semidefinite, we show that an extension of Sylvester's criterion, namely Prussing's criterion~\cite{Prussing} is satisfied:
that all principal minors are non-negative.
First, note that all principal minors of order 1 correspond to the diagonal elements and are thus equal to 1.
Thus, the proof focuses on higher order principal minors.
These are essentially determinants of smaller similarity matrices,
based on the same ordinal feature and on subsets of cultural vectors sampled from the larger set associated to the larger matrix.
Thus, the proof reduces to showing that the determinant of any ordinal, single-feature similarity matrix with $N \ge 2$ elements is nonzero. 

Based on Eq.~\eqref{CultSim}, such a similarity matrix can be written as:
\begin{equation}
    s_{ij} = 1 - \frac{|x_i - x_j|}{q-1},
\end{equation}
which makes it clear that the entire matrix is specified by the relative positioning of $N$ rational numbers $x_i/(q-1)$ within the $[0,1]$ interval.
Let $x_{i'}/(q-1)$ denote the same rational numbers but sorted for increasing values, so that: $x_{i'} \le x_{i'+1}, \forall i' \in \overline{1,N-1}$.
This amounts to permuting the rows and columns of $s$ in the same way, leaving the value of its determinant $\det(s)$ unchanged.
After this operation, the determinant may be conveniently written in terms of the (Manhattan) distance values $d_{i'} = 1 - s_{i'(i'+1)}$ associated to pairs of consecutive numbers, in the following way:
\begin{equation}
    \det(s) =
    \begin{vmatrix}
        1 & 1 - d_1 & 1 - d_1 - d_2 & ... \\
        1 - d_1 & 1 & 1 - d_2 & ... \\
        1 - d_1 - d_2 & 1 - d_2 & 1 & ... \\
        ... & ... & ... & ...
    \end{vmatrix}.
\end{equation}
This can be brought to an upper-diagonal form by applying further row and column operations that conserve the determinant (up to a minus sign).
Specifically, the following steps are taken:
\begin{itemize}
    \item subtract row $i+1$ from row $i$ for every $i \in \overline{1,N-1}$
    \item add column $N$ to every column $j \in \overline{1,N-1}$
    \item bring row $N$ to the top (by exchanging it with rows $N-1$ to $1$), while producing a factor of $(-1)^{N-1}$
\end{itemize}
to obtain:
\begin{widetext}
    \begin{equation}
        \det(s) = (-1)^{N-1}
        \begin{vmatrix}
            2 - \sum_{l=1}^{N-1} d_l & 2 - \sum_{l=2}^{N-1} d_l & 2 - \sum_{l=3}^{N-1} d_l & ... & 2 - d_{N-1} & 1 \\
            0 & -2d_1 & -2d_1 & ... & -2d_1 & -d_1 \\
            0 & 0 & -2d_2 & ... & -2d_2 & -d_2 \\
            ... & ... & ... & ... & ... & ... \\
            0 & 0 & 0 & ... & -2d_{N-2} & -d_{N-2} \\
            0 & 0 & 0 & ... & 0 & -d_{N-1}
        \end{vmatrix}
    \end{equation}
\end{widetext}
which evaluates to the product of the diagonal elements:
\begin{equation}
    \label{DetPrincMinorFinal}
    \det(s) = 2^{N-2} \left( 2 - \sum_{i=1}^{N-1} d_i \right) \prod_{j=1}^{N-1} d_j,
\end{equation}
where $d_i \ge 0, \forall i\in\overline{1,N}$ (non-negativity of Manhattan distances) and $\sum_{i=1}^{N-1} d_i \le 1$ (the sum is equal to the difference between the highest and the lowest of the $x_i/(q-1)$ values, which are constrained to the $[0,1]$ interval).
Thus, $\det(s) \ge 0$,
which concludes the proof for the ordinal (Manhattan), single-feature case.

\section{The fully-connected Ising (FCI) model}
\label{App_Theor_FCI}

This section gives the details behind the mathematical expressions in Sec.~\ref{FCI}, which introduced the fully-connected Ising model. 
Deriving the probability distribution $p$ associated to this model follows the maximum-entropy approach introduced by Ref.~\cite{Jaynes}.
This crucially relies on the Shannon entropy, which is a functional of the probability distribution: 
\begin{equation}
	\label{entropy}
  H[p] = - \sum_{\vec{S}} p_{\vec{S}} \log{p_{\vec{S}}},
\end{equation}
where $\vec{S}$ denotes a generic spin configuration with $F$ spins on a fully-connected lattice, 
or a generic cultural vector with $F$ binary cultural features whose possible traits are marked as ``$-1$'' and ``$+1$''.
The value of the functional $H$ is maximized subject to two constraints, one related to the normalization of the probability distribution over the set of possible configurations:
\begin{equation}
	\label{constr1}
  \sum_{\vec{S}} p_{\vec{S}} = 1, 
\end{equation}
the other related to enforcing, on average, a certain amount $K$ of alignment:
\begin{equation}
	\label{constr2_1}
  \sum_{a<b} \sum_{\vec{S}} S_a S_b p_{\vec{S}} = K,
\end{equation}
namely the average number of pairs of similarly labeled traits within a given configuration $\vec{S}$,
where the first summation is over all distinct pairs of distinct features (or lattice sites).
The maximization is done using the Lagrange multipliers technique for Eqs. \eqref{entropy}, \eqref{constr1}, \eqref{constr2_1}, which implies that one should find the extrema of the following functional: 
\begin{multline}
	\label{Lagrange}
  L[p] = H[p] - \lambda_0 \left( \sum_{\vec{S}} p_{\vec{S}} - 1 \right) \\
							- \lambda \left( \sum_{a<b} \sum_{\vec{S}} S_a S_b p_{\vec{S}} - K \right),
\end{multline}
where $\lambda_0$ and $\lambda$ are free parameters associated to the two constraints. 
By taking partial derivatives of Eq.~\eqref{Lagrange} with respect to each $p_{\vec{S}}$ and further manipulations, one finds the following probability distribution:
\begin{equation}
	\label{probDist}
	p_{\vec{S}} = \frac{1}{Z(-\lambda)}\exp{\left[-\lambda \sum_{a<b} S_a S_b \right]},
\end{equation}
where $Z(-\lambda)$ is a normalization factor, known in statistical physics as the ``partition function'': 
\begin{equation}
	\label{partFunc}
	Z(-\lambda) = \sum_{\vec{S}} \exp{\left[-\lambda \sum_{a<b} S_a S_b \right]},
\end{equation}
where one can replace the coupling parameter $-\lambda$ with $\mu > 0$ (whose positive value favors alignment as opposed to anti-alignment, which corresponds to ferromagnetism)
and re-express the sum over configurations $\vec{S}$ as a sequence of sums over the possible traits of each feature $S_{k}$, leading to:
\begin{equation}
	Z(\mu) = \prod_{k=1}^{F} \left( \sum_{S_k = \pm1} \right) \exp{\left[\mu \sum_{a=1}^{F-1}  \sum_{b=a+1}^{F} S_a S_b \right]}.
\end{equation}
In the exponent of this expression, there are $F(F-1)/2$ terms, out of which $F_{+}(F-F_{+})$ are equal to $-1$, while the other are equal to $+1$. 
Based on this, after further manipulations and after taking advantage of symmetries, the partition function can be expressed as as: 
\begin{widetext}
\begin{equation}
	\label{partFunc_mu_F}
	Z(\mu) = \sum_{F_{+} = 0}^{F} \frac{F!}{F_{+}!(F-F_{+})!} \exp{\left[\frac{\mu}{2} \left( (2F_{+}-F)^2 - F \right) \right]},
\end{equation}
\end{widetext}
where the combinatorial factor (binomial coefficient) before the exponential function counts the number of configurations with the same number $F_{+}$ of $+1$ traits.
In a way rather analogous to the partition function, the double summation in the exponent of Eq.~\eqref{probDist} can also be eliminated.
After multiplication with the combinatorial factor, this leads to Eq.~\eqref{prob_mu_F}, which gives the probability of having a configuration with $F_{+}$ spins up. 

On the other hand, using Eq.~\eqref{partFunc}, Eq.~\eqref{constr2_1} can be written as: 
\begin{equation}
	\label{K_from_deriv}
	K = - \frac{\partial(\log(Z(-\lambda)))}{\partial\lambda} = \frac{\partial(\log(Z(\mu)))}{\partial\mu},
\end{equation}
while the correlation between features/spins $a$ and $b$ is:
\begin{equation}
	\label{C_ab_gen}
	C_{ab} = \frac{\langle S_a S_b\rangle - \langle S_a \rangle \langle S_b \rangle}{\sqrt{\langle S_a^2 \rangle - \langle S_a \rangle^2} \sqrt{\langle S_b^2 \rangle - \langle S_b \rangle^2}},
\end{equation}
where $\langle Q \rangle = \sum_{\vec{S}} Q_{\vec{S}} p_{\vec{S}}$ is the expected value of quantity $Q$ with respect to the statistical ensemble. 
However, one can easily show, using Eq.~\eqref{partFunc_mu_F} that $\langle S_a^2 \rangle = 1$ and that $\langle S_a \rangle = \langle S_b \rangle = 0$, 
so $C_{ab} = \langle S_a S_b\rangle = \sum_{\vec{S}} S_a S_b p_{\vec{S}}$, which combined with Eq.~\eqref{constr2_1} leads to $\sum_{a<b} C_{ab} = K$.
But due to symmetry, the expected correlation $C_{ab}$ is the same for all pairs $(a,b)$, so:
\begin{multline}
	\label{C_ab_mu}
	C_{ab} = C(\mu, F) = \frac{2}{F(F-1)} K = \\
	= \frac{2}{F(F-1)} \frac{\partial(\log(Z(\mu)))}{\partial\mu},
\end{multline}
for any pair $(a,b)$, which can also be written in the form shown by Eq.~\eqref{C_mu} -- Eq.~\eqref{K_from_deriv} was used for the last transformation in Eq.~\eqref{C_ab_mu}.

One should expect that $C(0.0) = 0.0$ (null correlations for null coupling), which based on Eq.~\eqref{C_mu}, implies that the following identity holds:
\begin{equation}
	\sum_{F_{+} = 0}^{F} \frac{(F-2)! \left( (2F_{+}-F)^2 - F \right)}{F_{+}!(F-F_{+})!} = 0,
\end{equation}
which, after substitution of $F_+$ with $k$ and of $F$ with $N$ and some further manipulations leads to the following combinatorial identity: 
\begin{equation}
	\sum_{k = 0}^{N} {N \choose k} \left( (2k-N)^2 - N \right) = 0
\end{equation}
which can be shown to hold using the expressions for the binomial expansion and for the first and second moments of a binomial distribution with the probability parameter set to $0.5$.

\section{The symmetric two-groups (S2G) model}
\label{App_Theor_S2G}

This section provides the mathematical derivations of the important mathematical formulas related to the symmetric two-group model, introduced in Sec.~\ref{S2G}.
The derivations are based on the model description there. 

First, we proove Eq.~\eqref{prob_nu_F}.
On one hand, the probability that a cultural vector meant to be part of group $+1$ is assigned to a configuration with $F_{+}$ traits $+1$ is:
\begin{equation}
    \label{p++}
	p_{+}^{+}(\nu,F,F_{+}) = \frac{F!}{F_{+}!(F-F_{+})!} (1-2\nu)^{F_{+}} (2\nu)^{F-F_{+}},
\end{equation}
which is a binomial distribution with probability $1-2\nu$ for the $+1$ possibility and $2\nu$ for the $-1$ possibility. 
On the other hand, the probability that a configuration meant to be part of group $-1$ has $F_{+}$ traits $+1$ is:
\begin{equation}
    \label{p-+}
	p_{+}^{-}(\nu,F,F_{+}) = \frac{F!}{F_{+}!(F-F_{+})!} (2\nu)^{F_{+}} (1-2\nu)^{F-F_{+}},
\end{equation}
which is the same binomial distribution, but with inverted probabilities. 
Since the two groups are by construction equally likely, the combined probability of all configurations with $F_{+}$ traits $+1$ is:
\begin{equation}
	\label{prob_nu_F_prelim}
	p(\nu,F,F_{+}) = \frac{1}{2} p_{+}^{+}(\nu,F,F_{+}) + \frac{1}{2} p_{+}^{-}(\nu,F,F_{+}).
\end{equation}
Inserting Eq.~\eqref{p++} and Eq.~\eqref{p-+} in Eq.~\eqref{prob_nu_F_prelim} leads to Eq.~\eqref{prob_nu_F}.

Second, we prove Eq.~\eqref{C_nu}.
The correlation coefficient of any two features $a$ and $b$ is given by Eq.~\eqref{C_ab_gen}, which, for symmetry reasons similar to the case of the FCI model, simplifies to:
\begin{equation}
	\label{C_ab_nu_prelim}
	C_{ab}(\nu) = \sum_{\vec{S}} S_a S_b p_{\vec{S}}(\nu) = C(\nu).
\end{equation}
Moreover, the probability attached to any configuration $\vec{S}$ can be written as: 
\begin{equation}
	\label{caseSplit}
	p_{\vec{S}}(\nu) = \frac{1}{2}\left(p_{\vec{S}}^{-}(\nu) + p_{\vec{S}}^{+}(\nu)\right),
\end{equation}
where $p_{\vec{S}}^{-}(\nu)$ and $p_{\vec{S}}^{+}(\nu)$ are the probabilities of configuration $\vec{S}$, conditional on whether it is generated for group $-1$ or for group $+1$ respectively. 
In turn, these probabilities can be factorized in terms of feature-level probabilities of traits:
\begin{equation}
	\label{factorization}
	p_{\vec{S}}^{-}(\nu) = \prod_{a=1}^{F} p_{S_a}^{-}(\nu), ~~~~~~~ p_{\vec{S}}^{+}(\nu) = \prod_{a=1}^{F} p_{S_a}^{+}(\nu),
\end{equation}
because once the group is chosen, each trait $S_a$ (with possible values $-1$ and $+1$) is chosen independently at the level of the respective feature $a$.
By inserting Eq.~\eqref{factorization} in Eq.~\eqref{caseSplit} and the result in Eq.~\eqref{C_ab_nu_prelim}, 
by carrying out appropriate algebraic manipulations, 
while making use of the fact that $\sum_{\vec{S}} = \prod_{a=1}^F (\sum_{S_a})$ 
and of the fact that $p_{S_a = -1}^{-/+}(\nu) + p_{S_a = +1}^{-/+}(\nu) = 1.0$, 
one obtains: 
\begin{multline}
	\label{C_nu_prelim}
	C(\nu) = 	\frac{1}{2} \left[ p_{--}^{-}(\nu) - p_{-+}^{-}(\nu) - p_{+-}^{-}(\nu) + p_{++}^{-}(\nu) \right] + \\
				 +	\frac{1}{2} \left[ p_{--}^{+}(\nu) - p_{-+}^{+}(\nu) - p_{+-}^{+}(\nu) + p_{++}^{+}(\nu) \right],
\end{multline}
where, for instance, $p_{-+}^{-}(\nu)$ is the probability that trait $-1$ is chosen for one of the features and that trait $+1$ is chosen for the other feature,
conditional on the given configuration being generated for group $-1$. 
Based on the model description in Sec.~\ref{S2G}, one can see that:
\begin{align}
	p_{--}^{-}(\nu) = p_{++}^{+}(\nu) &= (1-2\nu)^2, \\
	p_{++}^{-}(\nu) = p_{--}^{+}(\nu) &= (2\nu)^2, \\
	p_{-+}^{-}(\nu) = p_{+-}^{+}(\nu) &= (1-2\nu)(2\nu), \\
	p_{+-}^{-}(\nu) = p_{-+}^{+}(\nu) &= (2\nu)(1-2\nu). 	
\end{align}
By plugging these in Eq.~\eqref{C_nu_prelim}, after simple algebraic manipulations, one obtains Eq.~\ref{C_nu}.

\section{The structure of the FCI and S2G models}
\label{App_lambda_1-3}

\begin{figure*}
\centering
	\subfigure[\label{Val:1}]{\includegraphics[width=8.5cm]{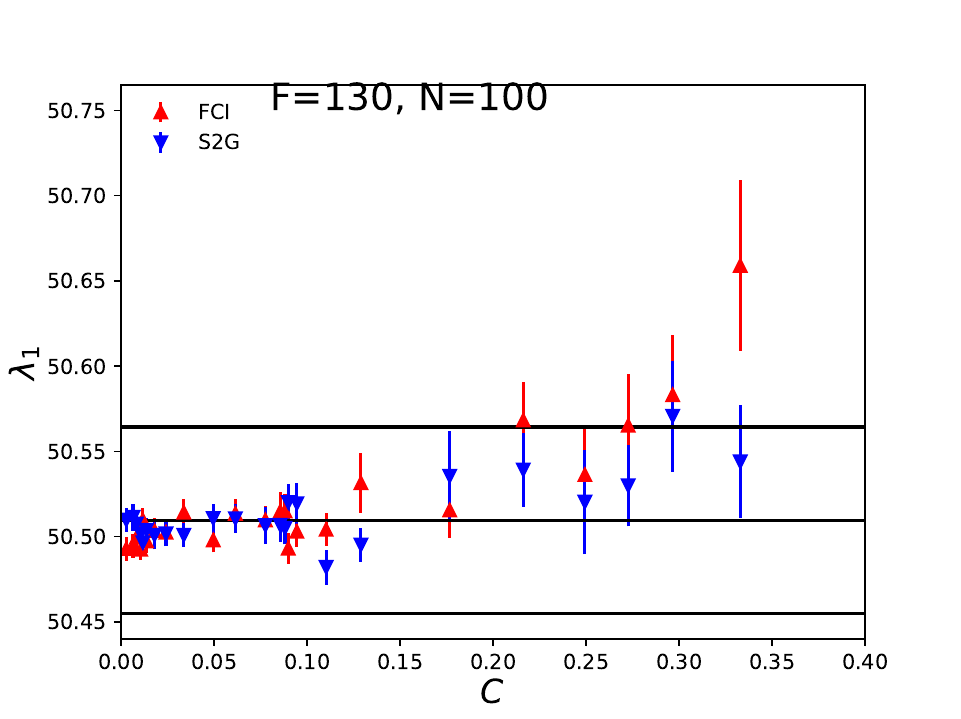}}
 	\subfigure[\label{Val:3}]{\includegraphics[width=8.5cm]{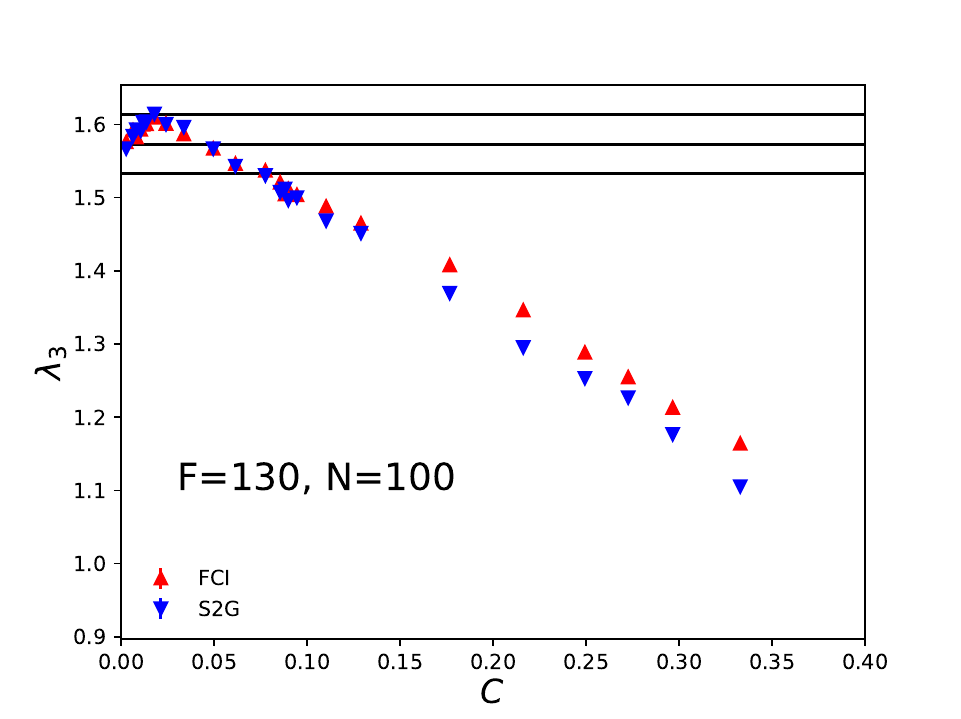}}
	\caption{Behaviour of largest and third-largest eigenvalues $\lambda_1$ and $\lambda_3$.
	The figure shows how $\lambda_1$~\subref{Val:1} and $\lambda_3$~\subref{Val:3} depend on the correlation level $C$ for the fully-connected Ising (FCI, red, upward triangles) and for the symmetric two-groups (S2G, blue, downward triangles) models. 
	For each $C$ value, for each of the two models, an averaging is performed over 80 sets of cultural vectors independently sampled from the respective ensemble --
	the vertical bar associated to each point shows the interval spanned by one standard mean error on each side of the mean. 
	The black, horizontal lines show, for comparison, the mean $\lambda_1$~\subref{Val:1} and mean $\lambda_3$~\subref{Val:3} based on uniform randomness, along with the width of the $\lambda_1$ and the $\lambda_3$ distributions -- one standard deviation on each side --
	where the calculations are based on 60 sets of cultural vectors generated via uniform randomness 
	-- these lines do not imply that, for uniform randomness, the correlation $C$ (which actually vanishes by construction) can be arbitrarily large. 
}
\label{Val}
\end{figure*} 

This section shows that the structure implicit in cultural states generated with either the FCI or the S2G model is mostly captured by only one eigenpair of the similarity matrix, 
so that there is at most one structural mode.
Specifically, as the correlation level is increased for the FCI and the S2G models, there is only one eigenvalue -- the subleading eigenvalue $\lambda_2$ -- that becomes separated from the random bulk, 
while becoming significantly larger than the upper boundary of the bulk that is expected based on uniform randomness. 
The behavior of $\lambda_2$ has already been presented in Fig.~\ref{Val2}.
The results shown here, via Fig.~\ref{Val}, are complementary to those shown in Fig.~\ref{Val2}, which uses the same format, 
while focusing on the behavior of $\lambda_1$ in Fig.~\ref{Val:1} and on the behavior of $\lambda_3$ in Fig.~\ref{Val:3}.
Note that $\lambda_1$, associated to the global mode, remains statistically compatible with the null model as the level of correlation is increased, for both FCI and S2G.
On the other hand, $\lambda_3$ decreases, while becoming, for large enough $C$, significantly smaller than the upper boundary of the bulk predicted by uniform randomness. 
All this shows that the structure FCI and S2G is mostly captured by the eigenpair of $\lambda_2$, which becomes increasingly stronger as the correlation level increases. 
This appears to be a consequence of the fact that each model is controlled by one parameter, while all the non-uniformity of the associate probability distribution is captured by one dimension, 
namely the $F_{+}$ axis of Fig.~\ref{ProbDist}. 

\end{document}